# Thermodynamics of Element Volatility and its Application to Planetary Processes

**Paolo A. Sossi**

*Institut de Physique du Globe de Paris, 1 rue Jussieu, F-75005, Paris, France*

*sossi@ipgp.fr*

**Bruce Fegley, Jr.**

*Planetary Chemistry Laboratory, Department of Earth and Planetary Sciences and McDonnell Center for the Space Sciences, Washington University, St. Louis, Missouri, USA*

*bfegley@wustl.edu*

## 1.0. Introduction

Despite its importance in geological sciences, our understanding of interactions between gas and condensed phases (comprising solids and liquids) remains clouded by the fact that, often, only indirect evidence remains for their occurrence. This arises from the tendency for the vapour phase to escape from the condensed phase with which it interacts, owing to its much lower density and thus greater volume. For a gas that is sufficiently tenuous that interactions do not occur between its constituent molecules, this relationship is quantified in the ideal gas law (Clapeyron 1834):

$$PV = nRT \tag{1}$$

where $P$ is the total pressure exerted by the gas, $V$ its volume, $n$ is the number of moles, $R$ the gas constant (8.3145 $Jmol^{-1}K^{-1}$, Horstmann, 1873) and $T$ the absolute temperature. One mole of an ideal gas at 273.15 K and $10^5$ Pa (standard temperature and pressure for gases)[1] has a molar volume of 22,711 $cm^3$/mol, $10^3$ × greater than typical silicate liquids or minerals. As a result, vaporisation processes in nature are often informed by chemical and textural evidence remaining in the condensed phase. Such evidence points to the influence of vapour-condensed phase interaction at a variety of scales and geological settings, both in our solar system and beyond (Nagahara, this volume). Although also important in terrestrial processes such as volcanism (see Symonds and Reed 1993; Henley and Seward, this volume) and impact events (Koeberl 1986; King et al., this volume), we present in this chapter an overview of how elements can record gas-condensed phase interactions as they apply to planetary processes that include, but are not limited to:

1) Degassing of planetary magma oceans (section 3.0.)
2) Atmospheres on existing bodies (section 3.0.)

[1] Standard pressure was changed by IUPAC in 1982 from 760 mm Hg barometer (Partington 1950; Ewing et al. 1994)

3) Evaporation on small planetary bodies (section 4.0.)
4) Evaporation during giant impacts (section 4.0.)

To constrain the conditions under which these processes occur necessitates thermodynamic data of gaseous- and condensed species. Therefore, a review of our knowledge and lack thereof, is also presented (section 2.0.). Much of our information on thermodynamic quantities of gaseous species comes from spectrometric measurements, undertaken from the 1950s onwards, largely on simple (binary and ternary) systems (e.g., see Brewer 1953; Margrave 1967; Lamoreaux et al. 1987, and section 2.0.). This information is now readily available in compilations of thermodynamic data such as JANAF (Chase 1998) or IVTAN (Glushko et al. 1999). However, the application of these data to more complex, natural geological systems is less straightforward.

Instead, for lack of any other comprehensive volatility scale, 50% nebular condensation temperatures, $T_c$, (Grossman and Larimer 1974; Lodders 2003), which describe the temperature at which half of the abundance of an element is condensed in the solar nebula, are frequently used to discuss elemental volatility in planetary environments. However, these $T_c$ are strictly relevant only at the pressure and temperature conditions of a solar-composition protoplanetary accretion disk (*e.g.*, the solar nebula). Namely, at low total pressures ($P_T \lesssim 10^{-2}$ bar), solar or close to solar metallicity[2] (i.e., $P_T \sim P_{H2} + P_H + P_{He}$), oxygen fugacities ~ 6 log bar units below the Iron – Wüstite (IW) buffer (due to the low $H_2O/H_2$ ratios, $5 \times 10^{-4}$; Rubin et al. 1988; Grossman et al., 2008). Boss (1998) and Woolum and Cassen (1999), to give two examples, used astronomical observations to model midplane temperatures in protoplanetary accretion disks in the planet-forming region and found temperatures <1600 K. Petrological observations of Calcium, Al-rich inclusions (CAI) with the highly fractionated group II rare-earth element (REE) pattern[3] (Boynton 1975) indicate they formed by fractional condensation (Boynton 1975; Davis and Grossman 1979) at temperatures between 1676 K, 1463 K, 1296 K at total pressures ($P_T$) of $10^{-3}$ bar, $10^{-6}$ bar, and $10^{-9}$ bar, respectively (Kornacki and Fegley 1986). At total pressures above $\approx 10^{-2}$ bar, liquids condense in place of solids (Ebel, 2004) and departures from solar composition (*e.g.,* by changes in metallicity or H-loss on a planetary body) lead to condensation temperatures different to those in a solar-composition system (Larimer and Bartholomay 1979; Lodders and Fegley 1999; Schaefer and Fegley 2010). Thus, the complexity of evaporation and condensation reactions are such that, at conditions outside this range, the relevant gas species, their equilibrium partial pressures and mechanisms driving their escape diverge greatly.

Potentially powerful tracers of the conditions under which these gas-liquid/solid processes occur are the moderately volatile elements (MVEs), defined as those that condense from a solar composition gas at temperatures between the major components (≈1320 K, Fe, Mg and Si) and troilite (≈660 K, FeS) at $10^{-4}$ bar total pressure (Palme et al. 1988). Their utility lies in their balanced partitioning between the gas and the condensed phase, such that significant quantities remain to permit their analysis. However, as these elements are typically present in trace quantities in planetary materials, knowledge of their thermodynamic properties in minerals and melts is frequently incomplete. In particular, our understanding of the mechanisms by which MVEs substitute into common high-temperature phases and of their activity coefficients in silicate liquids and minerals remain inadequate. Furthermore, how the complex chemical environments relevant to planetary systems affect the relative stability of gaseous molecules is poorly known. Thus, at present, these gaps in knowledge limit the information one can extract from the behaviour of moderately volatile- and other rock-forming elements during high-temperature planetary processes.

[2] Metallicity is the mass proportion of elements heavier than He relative to H in a star.
[3] The Group II CAI comprise ~ 38% of CAI analysed for REE (see Table 1 of Ireland & Fegley 2000) and are depleted in both the most volatile and the most refractory REE, taken as evidence of direct condensation from the solar nebula.

Here, we present an introductory thermodynamic treatment of vaporisation/condensation reactions, which serves as a framework for the application of thermodynamic models to geological processes. The two end-member styles of evaporation/condensation, equilibrium and kinetic, are described in section 2.1 and 2.2, respectively. Calculation of thermodynamic equilibrium is discussed, taking into account the effect of non-ideality in the gas phase, enthalpies and entropies of fusion and vaporisation, and activity coefficients of melt oxide species (section 2.1.1.). Information on equilibrium partial pressures of metal-bearing gas species, mainly in simple systems, is derived from Knudsen Effusion Mass Spectrometry (KEMS) (e.g., see Copland and Jacobsen, 2010) and other experimental techniques (transpiration, torsion effusion; e.g., see Margrave, 1967), though alternative methods, such as *ab-initio* calculations (Langhoff and Bauschlicher, 1988), CALPHAD modelling (Saunders and Miodownik, 1998) and free-energy minimisation programs (*e.g.*, HSC chemistry, Roine, 2002; IVTANTHERMO, Belov et al. 1999) are becoming increasingly widespread (section 2.1.2.). A primer on the geological applications of these techniques, and the thermodynamic data they provide, including evaporation of simple-, complex- and natural oxide systems, are presented in sections 2.1.3 to 2.1.6.

The kinetic treatment (section 2.2.1.) is based on the Maxwell-Boltzmann kinetic theory of gases, which is then applied to derive the Hertz-Knudsen-Langmuir (HKL) equation used to calculate evaporation rates from a surface (Chapman and Cowling 1970; and references therein). Using this equation, extraction of thermodynamic data may also be achieved, though less precisely, from free or Langmuir evaporation experiments, which are performed in furnaces under vacuum or in a controlled atmosphere (section 2.2.2.). In this case, the HKL equation serves as the basis for predicting element loss as a function of time, temperature, and equilibrium partial pressures for element evaporation from minerals and melts (sections 2.2.3 – 2.2.5). Much of the information of the evaporation of trace elements in geologically relevant systems comes from these methods (Wulf et al. 1995; Sossi et al., 2016; Norris and Wood 2017; section 2.2.5). These experiments highlight the differences between nebular condensation and evaporation of silicates under post-nebular conditions (Visscher and Fegley, 2013).

At lower temperatures, higher pressures, or in systems that have not lost their complement of major volatiles, the formation of metal hydroxide, -halide, -carbonate, and/or -sulfate gas species becomes potentially important (section 3.0.). The volatility of metals in $H_2O$ (or steam) atmospheres is enhanced relative to their volatilities in the binary metal-oxygen system (Schaefer et al. 2012; Fegley et al. 2016) and speciation of an element in gas phase may change as a function of pressure, temperature and fugacity of other volatiles (sections 3.2 – 3.5). As such, volatile loss of these species may engender depletions in the complementary condensed phase that are distinct from evaporation under volatile-poor conditions.

This information is reconciled together to summarise and interpret volatile element depletions observed in the terrestrial planets in section 4.0. Here, we give an overview of how volatile element depletion in planetary materials can be quantified, and how they relate to the processes, both physical and chemical, that occur during the formation of meteorites and subsequently the planets (section 4.1.). We discuss that the conditions under which the planets form differ from those present during the condensation of the solar nebula, and, as a result, the mechanisms responsible for volatile depletion leave imprints on their composition distinct from those found in chondritic meteorites[4] (section 4.2.). Though the Earth is depleted in volatile elements with respect to chondritic meteorites, its current mass and atmospheric temperature effectively impedes atmospheric escape (Zahnle & Catling 2017), except in special cases, namely for H, $^{3}$He, $^{4}$He (Hunten and Donahue 1976; Torgesen 1989), such that this depletion likely occurred early in its history (section 4.3.). By comparison, the Moon and the asteroid 4-Vesta are even more impoverished than the Earth in these volatile elements, which is partially attributable to their much smaller masses, the composition of their source materials, and their individual thermal histories (section 4.4.).

This chapter is intended as an introduction and as a reference to the thermodynamic approaches, methods and existing data relevant to evaporation/condensation processes of metal oxides and their application to planetary processes. The reader is encouraged to select the sections that are most relevant to them, and to delve into the references cited, which are themselves not exhaustive. Finally, throughout this chapter, we highlight promising avenues of future research in the field that may help in gaining a more quantitative understanding of gas-condensed phase interactions in planetary science.

[4] Chondritic meteorites – the most common meteorites – are undifferentiated rocks containing metal, silicate, and sulfide, so called for the millimetre-sized spherical objects named 'chondrules' that comprise them.

## 2.0. Thermodynamics of evaporation/condensation of metal oxides in simple, complex and natural systems

Evaporation is the (solid, liquid) to gas phase transition and condensation is the reverse process. Unless otherwise noted we do not distinguish between sublimation (solid to gas) and vaporisation (liquid to gas). Metals are taken to broadly to encompass metals, metalloids, and semiconductors that remain in metallic form, form oxides and silicates, or combine with other ligands such as C, N, P, OH, chalcogens (S, Se, Te) or halogens (F, Cl, Br, I) in planetary materials. The cases of silicates, oxides and metals (liquids and solids), common in planets and chondrites, are considered henceforth, but other compounds are discussed where relevant. The review covers virtually all naturally-occurring elements with an emphasis on the moderately volatile. The MVE occur on the inner right-hand side of the periodic table; transition metals in groups 11 and 12 (Cu, Zn, Ag, Cd), metals and metalloids of groups 13 and 14 (*e.g.* Ga, Sn, Pb, Sb) in addition to the alkali metals (Li, K, Na, Rb, Cs).

A description of the basic mechanics of evaporation are presented, followed by some of the ways in which evaporation rates and equilibrium partial pressures can be determined experimentally. Finally, examples are presented of natural materials in which volatile element loss by evaporation has been recorded, and the implications this has for their geologic history.

## 2.1. Equilibrium evaporation/condensation

### *2.1.1. General treatment of equilibrium evaporation/condensation*

Evaporation can occur congruently or incongruently. The majority of metals in condensed phases, evaporate congruently (*e.g.,* Brewer 1953; Ackermann and Thorn 1961; Lamoreaux et al. 1987; Table 1), where a condensed component, such as a metal-oxide, M-O, evaporates to one (associative) or more (dissociative) species in the gas phase, with the same overall stoichiometry. Congruent dissociative evaporation involving oxygen can be written generally:

$$\left(M^{x+n}\frac{x+n}{2}O\right)(s,l) = \left(M^{x}\frac{x}{2}O\right)(g) + \frac{n}{4}O_2(g) \qquad (2)$$

where $M$ is the metal species, $x$ its formal charge, and $n$ the number of electrons in the redox equilibrium. The metal is present in metallic form if $x+n=0$, whereas if $x+n>0$, then an oxide compound is stable in at least the more oxidised of the two phases. The latter case typifies evaporation of silicate material, where metals are present as $\left(M^{x+n}\frac{x+n}{2}O\right)$ components in the condensed phase, and $n$ is positive such that the metal becomes more volatile under reducing conditions as $\left(M^{x}\frac{x}{2}O\right)$. However, when $n$ is negative, oxidising conditions promote vaporisation of the metal- or metal-oxide component in the condensed phase.

Incongruent evaporation is characterised by reactions in which $(x+n)(s,l) \neq (x+n)(g)$. That is, the M/O ratio is greater- or less than unity in the vapour phase with respect to the condensed phase. Incongruence is manifested in a change in stoichiometry of the condensed phase, which may be small and result in defect formation (*e.g.,* ZnO, Kodera et al 1968), or larger and lead to the formation of a new phase (as we discuss later). We do not discuss defect chemistry further and refer the reader to Kröger (1964). Incongruent evaporation involving the production of oxygen only can be written with Equation (2) where $\left(M^{x}\frac{x}{2}O\right)$ is also condensed. For cases in which a metal-bearing gas species is produced, incongruent evaporation, for all $x \geq n$, may be expressed:

$$2\left(M^{x}\frac{x}{2}O\right)(s,l) = \left(M^{x+n}\frac{x+n}{2}O\right)(s,l) + \left(M^{x-n}\frac{x-n}{2}O\right)(g) \tag{3}$$

As per congruent evaporation, if $n$ is > 0, the speciation of the element in the gas is more reducing, and vice-versa.

Speciation of metals in oxides, silicate liquids or minerals can be combined with thermodynamic data for stable gas species (*e.g.,* Lamoreaux et al. 1987) to write an equilibrium constant for a given volatilisation reaction:

$$K_{(2)} = \frac{f\left(M^{x}\frac{x}{2}O\right)f(O_2)^{n/4}}{a\left(M^{x+n}\frac{x+n}{2}O\right)} \tag{4}$$

The $a$ and $f$ terms are the activity and fugacity of a species. The fugacity of a gas is the product of the partial pressure and fugacity coefficient $\phi$ of the gas,

$$f_i = \phi_i p_i \tag{5}$$

The fugacity coefficient is equal to unity for an ideal gas and is either < 1 or > 1 for a non-ideal (real) gas. The fugacity coefficient is unity to a very good first approximation for equilibrium evaporation reactions with total vapour pressures ≤ 1 bar. For example, the 2$^{nd}$ virial coefficient for pure $O_2$ gas (Levelt Sengers et al. 1972) at one bar total pressure is within 0.03% of unity from 500 to 2000 K. The total vapour pressures in the equilibrium evaporation experiments discussed here are ≤ 1 bar and thus ideal gas behaviour can safely be assumed, and hence the approximation $f = p$ (i.e., $\phi = 1$) can be made.

The activity of a pure condensed solid or liquid is unity at a partial pressure equal to that of the saturated vapour, $p_{i,\,sat}$,

$$a_i = \frac{p_i}{p_{i,sat}} \tag{6}$$

The activity of a pure condensed phase is simply the ratio of its partial pressure to its saturated vapour pressure at the same temperature. If equilibrium evaporation takes place at high total pressure, e.g., due to a high pressure exerted by an inert gas, the effect of total pressure on activity of the condensed phase(s) has to be considered using the thermodynamic relationship

$$RTlna_i = \int_{P^\circ}^{P} V_m dP \tag{7}$$

Where $V_m$ is the molar volume of the condensed phase and the integration limits are the saturated vapour pressure of the pure phase ($P^o$) and the total pressure of the system (P). This relation is known as the Poynting correction and expresses the effect of total pressure on vapour pressure.

The equilibrium constant for reaction (2) is related to the standard Gibbs Free Energy of the reaction via the reaction isotherm

$$\Delta G = \Delta G^o + RTlnK \tag{8}$$

At equilibrium $\Delta G = 0$ and we have the familiar relationship

$$\Delta G^o_{(2)} = -RTlnK_{(2)} \tag{9}$$

The standard Gibbs Free energy of an evaporation reaction is given by the Gibbs – Helmholtz Equation,

$$\Delta G^o = \Delta H^o - T\Delta S^o \tag{10}$$

The standard Gibbs energy, enthalpy, and entropy of formation are all functions of temperature, and the Gibbs – Helmholtz equation can be rewritten in terms of the heat capacities of the product and reactant species,

$$\Delta G^o_T = \Delta H^o_{T,ref} + \int_{T,ref}^{T} \Delta C^o_P \, dT - T\left[\Delta S^o_{T,ref} + \int_{T,ref}^{T} \frac{\Delta C^o_P}{T} dT\right] \tag{11}$$

The reference temperature (T,ref) is generally 298.15 K but any other convenient value can also be used. The $\Delta C_P^\circ$ is the difference between the heat capacity of the compound and its constituent elements in their reference states with the appropriate stoichiometric coefficients. This expression can be considerably simplified using the Gibbs energy function defined (typically at a reference temperature of 298.15 K) as

$$\frac{G^o_T - H^o_{298}}{T} = \frac{(H^o_T - H^o_{298})}{T} - S^o_T \tag{12}$$

Substituting this back into the expression for $\Delta G^\circ_T$ gives

$$-RlnK_{(2)} = \frac{\Delta G_T^o}{T} = \Delta\left[\frac{(G_T^o - H_{298}^o)}{T}\right] + \frac{\Delta H_{298}^o}{T} \quad (13)$$

The Gibbs energy functions for a given compound tabulated in the JANAF Tables or other thermodynamic data compilations generally take solid-state phase changes (polymorphs, $solid_1$ → $solid_2$) and fusion (solid → liquid) into account, but if this is not the case then the Gibbs energy function of the liquid species must be calculated using an equation such as

$$\Delta_{fus}G_T^o = \Delta_{fus}H_{T_m}^o + \int_{T_m}^{T} \Delta C_P^o\, dT - T\left[\Delta_{fus}S_{T_m}^o + \int_{T_m}^{T} \frac{\Delta C_P^o}{T}\, dT\right] \quad (14)$$

The melting point is $T_m$, $\Delta_{fus}H°_{Tm}$ and $\Delta_{fus}S°_{Tm}$ are the standard molar enthalpy and entropy of fusion, and $\Delta C°_P$ is the heat capacity difference between the solid and liquid oxide. Rewriting in terms of Gibbs energy functions this equation becomes

$$\frac{\Delta_{fus}G_T^o}{T} = \Delta\left[\frac{\left(G_T^o - H_{T_m}^o\right)}{T}\right] + \frac{\Delta_{fus}H_{T_m}^o}{T} \quad (15)$$

If the $\Delta C°_P$ is not known and cannot be estimated, it is common to use the approximation

$$\frac{\Delta_{fus}G_T^o}{T} = \frac{\Delta_{fus}H_{T_m}^o}{T} - \Delta_{fus}S_{T_m}^o = \frac{\Delta_{fus}H_{T_m}^o}{T} - \frac{\Delta_{fus}H_{T_m}^o}{T_m} \quad (16)$$

Assuming ideality, it is clear from Equation (4) that the relative stabilities of the gas/condensed phase will depend on i) the $fO_2^{(n/4)}$ and ii) the equilibrium constant, *K*, as given by the standard state entropy, enthalpy and volume change of the reaction. However, mixing of elements in silicate melts are non-ideal and the activity of a component is related to its mole fraction via an activity coefficient $\gamma_i$

$$a_i = \gamma_i X_i \quad (17)$$

Use of thermodynamic properties obtained on pure compounds is only applicable to vaporisation reactions from silicate minerals or melts if the non-ideality of the metal oxides in the condensed phase are considered (see *sections 2.1.4. to 2.1.6.*)

### *2.1.2. Equilibrium evaporation/condensation experimental techniques*

Techniques for acquiring thermodynamic data for evaporation reactions relevant to the geosciences include Knudsen Cell Effusion Mass-Spectrometry (KEMS), transpiration and torsion effusion, where partial pressures or gas species are detected by various spectroscopic methods (*e.g.,* infra-red, Raman, and photoelectron spectroscopies), weight-loss or -gain, and mass spectrometry. Detailed descriptions of each of these techniques can be found in the reviews on high-temperature vapours by Margrave (1967), Hastie (1981) and Wahlbeck (1986), and only a brief primer is given here.

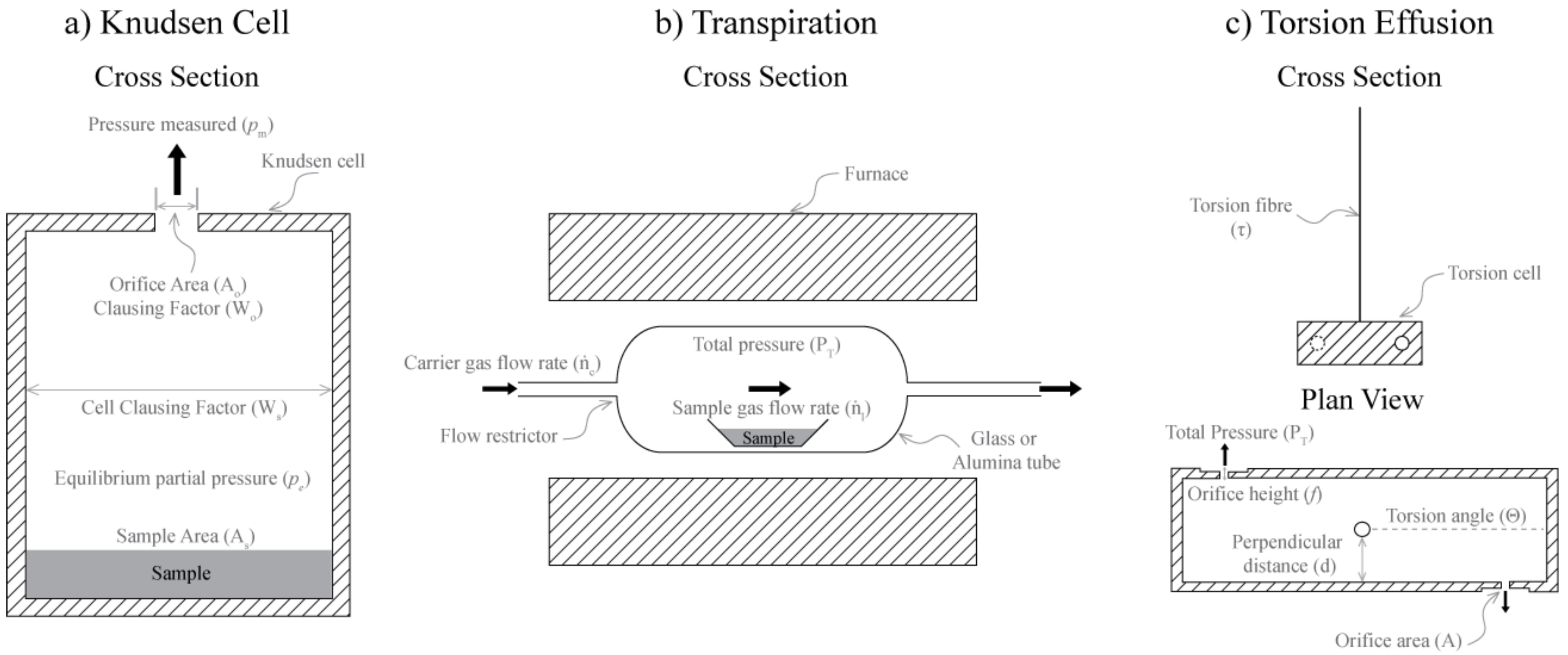

***Fig. 1.*** *Schematic illustrations of the three types of apparatus used for measuring equilibrium vapour pressures. Only the essential components of each are represented here. a) Knudsen cell. See Chatillon et al. (1975); Copland and Jacobson (2010); Wetzel et al. (2012) b) Transpiration apparatus. See Merten and Bell (1967); Wahlbeck (1986); Holstein (1993). c) Torsion effusion cell apparatus. See Pratt and Aldred (1959); Munir and Searcy (1964); McCreary and Thorn (1968); Viswanadham and Edwards (1975); Wahlbeck (1986).*

The most widely-used technique is KEMS, which is predicated upon chemical and thermal equilibration of gaseous species with each other and with the condensed phase during heating of sample material inside a Knudsen cell in a vacuum (typically $10^{-4}$ to $10^{-10}$ bar) due to the small size of the cell and of the pinhole orifice relative to the large mean free path of gases at such low pressures. Thus, a Knudsen cell approximates a closed system. The gas, which has chemically and thermally equilibrated with sample of surface area $A_s$, effuses out of the pinhole orifice of area $A_o$. The equilibrium vapour pressure inside the cell ($p_e$) relative to that calculated by mass loss via the HKL equation ($p_m$) is given by the Whitman-Motzfeldt equation,

$$p_e = p_m \left[1 + \frac{W_o A_o}{A_s}\left(\frac{1}{\alpha_e} + \frac{1}{W_c} - 2\right)\right] \quad (18)$$

where $W_o$ and $W_c$ are the Clausing factors for the orifice and cell body, respectively (Fig. 1a). The orifice therefore should ideally have a diameter and thickness that are infinitesimally small relative to the cell dimensions, $A_s$. A portion of the effusing gas is then ionised and the resulting positively-charged ions are separated according to their mass/charge ratio in quadrupole time-of-flight or magnetic sector mass spectrometers (*e.g.*, Drowart and Goldfinger, 1967; Margrave 1967; Cater 1970; Copland and Jacobson 2010).

Merten and Bell (1967) describe the transpiration method in which either an inert or reactive gas flows over a sample held in the isothermal hot zone of a tube furnace at gas flow rates such that the weight loss is independent of gas flow rate (see Fig 4 of Belton and Richardson 1962; Fig. 1b). The partial pressure of an evaporating species, $p_i$, is given by:

$$p_i = \frac{\dot{n}_i}{\dot{n}_i + \dot{n}_c} P_T \quad (19)$$

where $\dot{n}_i$ and $\dot{n}_c$ are the molar flow rates of the evaporating species and carrier gas, respectively. At low flow rates, $\frac{\dot{n}_i}{\dot{n}_i+\dot{n}_c}$ becomes large and $\dot{n}_i$ is partially controlled by gas-phase diffusion, whereas $\dot{n}_i$ is unable to reach equilibrium values at high flow rates due to kinetic effects. Transpiration experiments have provided thermodynamic data for number of metal oxide, -chloride, -hydroxide, and -oxyhydroxide gases and have clarified vaporisation behaviour of numerous metals and their oxides (e.g., Darken and Gurry, 1946; Kitchener and Ignatowicz, 1951; Gilbert and Kitchener, 1956; Ackermann et al., 1960; Alcock and Hooper, 1960; Cubicciotti, 1960; Meschi et al., 1960; Caplan and Cohen, 1961; Shchukarev et al., 1961; Cordfunke and Mayer, 1962; Alexander et al., 1963; Altman, 1963; Berkowitz et al., 1960; Berkowitz-Mattuck and Büchler, 1963; Belton and McCarron, 1964; Stafford and Berkowitz, 1964; Belton and Jordan, 1965; Faktor and Carasso, 1965; Kim and Belton, 1974; Sano and Belton, 1974; Tetenbaum 1975; Sasamoto et al., 1979; Matsumoto and Sata, 1981; Rau and Schnedler, 1984; Rego et al., 1985; Hashimoto, 1992; Holstein 1993a,b; Hashizume et al., 1999; Jacobson et al. 2005; Meschter et al. 2013; Myers et al., 2016; Nguyen et al., 2017).

The torsion effusion method, as first developed by Volmer (1931), and summarised by Freeman (1967), is predicated on the relationship between torsional force exerted on a cell suspended at its midpoint, by a molecular vapour effusing horizontally from two antipodal orifices ($A_1$ and $A_2$) which displace the cell by a perpendicular distance $d_1$ and $d_2$ to an angle Θ against the restoring force of the fibre with a torsion constant τ (Fig. 1c). The total pressure, $P_T$, is given by momentum loss,

$$P_T = \frac{2\tau\Theta}{(A_1 d_1 f_1 + A_2 d_2 f_2)} \tag{20}$$

where *f*s are the force recoil factors that account for the finite dimensions of the orifice. Some early works yielded data for As (Rosenblatt and Birchenall, 1960), Fe halides (Sime and Gregory, 1960) and Ga (Munir and Searcy, 1964). Later studies (Viswanadham and Edwards, 1975; Edwards, 1981; Botor and Edwards, 1985; Govorchin, 1986; Edwards et al., 1989) combine weight loss measurements by Knudsen effusion with torsion effusion approaches in the same apparatus as a means of verifying vapour pressure measurements. Several groups have measured vapor pressures of metal oxides via torsion effusion or Langmuir effusion (e.g., Hildenbrand et al. 1963, Peleg and Alcock 1966, Alcock & Peleg 1967, Piacente et al. 1972, Hildenbrand & Lau 1993).

These techniques were used extensively in the 1950s to 1980s to measure thermodynamic quantities of materials relevant to military applications, and leading to the construction of the JANAF (Chase 1998) and IVTAN (Glushko et al. 1999) databases in the US and former USSR. Some of this work was also relevant to the ceramic and steelmaking industries for understanding phase equilibria (Muan and Osborn, 1965). More recent studies have combined experimental measurements with *ab-initio* calculations that enable prediction of geometries, vibrational frequencies, and free energies of formation of molecular species (Opila et al., 2007; Gong et al., 2009; Nguyen et al., 2017; Xiao and Stixrude, 2018) and represent a fruitful future direction.

*2.1.3. Evaporation of binary metal oxides ($MO_x$)*

Thermodynamic data for refractory oxides and their evaporation products were instrumental in understanding the major element mineralogy, trace element abundances, and rare earth element (REE) abundance patterns in Ca, Al-rich inclusions (CAIs) in the Allende CV3 chondrite and other meteorites (e.g., Grossman 1972, 1973, Grossman and Larimer 1974, Boynton 1975, 1978, Davis and Grossman 1979, Fegley and Palme 1985, Kornacki and Fegley 1986, Ireland and Fegley 2000). The same data are now useful for modelling atmosphere – surface interactions on hot rocky exoplanets such as CoRoT-7b and Kepler-10b (e.g., Schaefer and Fegley 2009; Kite et al 2016) and chemistry in the proto-lunar disk (Visscher and Fegley 2013).

Table 1 lists the metal bearing gases in the saturated vapour over ~ 80 binary metal oxides, both volatile and refractory, based on experimental measurements and/or thermodynamic calculations.

**Bold type** indicates the most abundant metal-bearing gas; others are listed in descending order. Monatomic and/or diatomic oxygen are also present in all cases but are not listed. The temperature range of each study and the melting point of each oxide are also listed, e.g., 2327 K for $Al_2O_3$.

Many oxides evaporate congruently such as $Al_2O_3$, CdO, $Ga_2O_3$, $GeO_2$, MnO, NiO, $SiO_2$, $SnO_2$, $Ti_3O_5$, $V_2O_3$, $WO_3$ (T≤1550 K). In these cases evaporation involves the production of oxygen and is dissociative, in which the gaseous species each have different stoichiometry than the condensed phase, but the bulk composition of the vapour is the same. Some oxides evaporate mainly as gases with the same stoichiometry as the solid or liquid oxide (congruent associative evaporation), *e.g.*, $As_4O_6$, $B_2O_3$, BaO, $Cs_2O$, $OsO_4$, $P_4O_{10}$, $Re_2O_7$, $Sb_4O_6$, and $Tl_2O$, but even in these cases other metal-bearing gases with different stoichiometry are present in non-negligible amounts (Table 1). However several oxides do evaporate incongruently with different metal/oxygen ratios in the saturated vapour and the condensed phase, which may vary with temperature. Examples of this behaviour include Fe oxides, TiO, $TiO_2$, $Ti_2O_3$ VO, $V_2O_4$, $V_2O_5$, CuO, $Ag_2O$, $La_2O_3$, $Ce_2O_3$, $Y_2O_3$, $WO_2$, $WO_3$ (T>1550 K), $ThO_2$, and $UO_2$ (Ackermann and Rauh 1963, 1971a,b; Ackermann et al. 1956, 1963, 1970; Farber et al. 1972b; Kodera et al. 1968). In addition, Taylor and Hulett (1913) and Otto (1964, 1965, 1966a, b) measured manometrically the equilibrium $O_2$ dissociation pressures over HgO (mercuric oxide), $Mn_2O_3$ (manganic oxide), $MnO_2$ (pyrolusite), $Ag_2O$, and $PbO_2$ (plumbic oxide). With the exception of HgO, which vaporises to Hg (g) + $O_2$ (g), the total vapour pressure for these oxides, as well as $Fe_{1-x}O$, $Fe_3O_4$, $Fe_2O_3$ (Chizikov et al. 1971) and CuO (Kodera et al., 1968) is effectively the $O_2$ pressure.

Figure 2 shows vapour pressure curves for some selected metal-oxide systems listed in Table 1. The interested reader should consult the references cited for details about the metal-oxide vapour pressure measurements. Early summaries of vaporisation studies of oxides can be found in Brewer and Mastick (1951), Brewer (1953), Ackermann and Thorn (1961), Ackermann et al. 1961, Margrave (1967), Brewer and Rosenblatt (1969), Pedley and Marshall (1983), Lamoreaux and Hildenbrand (1984), and Lamoreaux et al. (1987). Margrave (1967) is a superb book that is still relevant today.

**Table 1.** The metal bearing gases in the saturated vapour over 77 pure oxides.

| Oxide | Temperature (K) | Gas Species | References |
|---|---|---|---|
| α-$Al_2O_3$(s,l) | 2309 – 2605[a] | **Al**, AlO, $Al_2O$, | [a]Brewer & Searcy 1951 |
| 2327 K | 1500 – 1800[b] | $AlO_2$[d,g], $Al_2O_2$ | [b]Porter et al. 1955a; |
| | 2188 – 2594[c] | | [c]Drowart et al. 1960 |
| | 1900 – 2600[d] | | [d]Farber et al. 1972a |
| | 2300 – 2600[e] | | [e]Chervonnyi et al 1977 |
| | 2450[f] | | [f]Paule 1976 |
| | 2200 – 2318[g] | | [g]Ho & Burns 1980 |
| | 2105 – 2800[h] | | [h]Milushin et al 1987 |
| | 3500 – 4700(l)[i] | | [i]Hastie et al. 2000 |
| | 2300 – 2600[j] | | [j]Chervonnyi 2010 |
| $As_2O_3$(s,l) | 367 – 429[a] | **$As_4O_6$** >> $As_4O_5$, | [a]Behrens & Rosenblatt 1972 |
| arsenolite | 400 – 1700[b] | $As_4O_4$, $As_2O_3$, AsO | [b]Brittain et al. 1982 |
| 551 K | 375 – 415[c] | | [c]Kazenas & Petrov 1997 |
| $B_2O_3$(l) | 1300 – 1500[a] | **$B_2O_3$**, $B_2O_2$, BO, B | [a]Inghram et al. 1956 |
| 723 K | 1410 – 1590[b] | | [b]Hildenbrand et al. 1963 |
| | 1200 – 1440[c] | | [c]Büchler & Berkowitz-Mattuch 1963 |
| | 1170 – 1400[d] | | [d]Jacobson & Myers 2011 |
| BaO(s) | 1200 – 1500[a] | **BaO**, Ba, $Ba_2O_3$, | [a]Claasen & Veenemans 1933 |
| 2246 K | 1150 – 1390[b] | $Ba_2O_2$, $Ba_2O$ | [b]Pelchowitch 1954 |
| | 1530 – 1758[c] | | [c]Inghram et al. 1955 |
| | 1365 – 1917[d] | | [d]Newbury et al. 1968 |
| | 1340 – 1660[e] | | [e]Hilpert & Gerads 1975 |
| BeO(s) | 1900 – 2400[a] | **Be**, $(BeO)_3$, $(BeO)_4$, | [a]Chupka et al. 1959 |
| 2851 K | 2300 – 2500[b] | BeO, $(BeO)_2$, $(BeO)_5$, $(BeO)_6$, $Be_2O$[b] | [b]Theard & Hildenbrand 1964 |
| $Bi_2O_3$(s,l) | 1003 – 1193[a] | **Bi**, BiO, $Bi_4O_6$, $Bi_2$, | [a]Sidorov et al. 1980 |
| 1098 K | 987 – 1017[b] | $Bi_2O_3$ $Bi_3O_4$, $Bi_2O_2$, $Bi_2O$ | [b]Oniyama & Wahlbeck 1998 |
| CaO(s) | 2080 – 2206[a] | **Ca**, CaO | [a]Farber and Srivastava 1976 |
| 3172 K | 1500 – 1900[b] | | [b]Sata et al 1982 |
| | 1923 – 2227[c] | | [c]Samoilova & Kazenas 1995 |

| Oxide | Temperature (K) | Gas Species | References |
|---|---|---|---|
| CdO(s) | 1150 – 1374[a] | **Cd** >> CdO | [a]Gilbert & Kitchener 1956 |
| 1755 K, dec. | 873 – 1056[b] | | [b]Kodera et al. 1968 |
| | 1014 – 1188[c] | | [c]Miller 1975 |
| | 886 – 1090[d] | | [d]Behrens and Mason 1981 |
| | 980 – 1334[e] | | [e]Kazenas et al. 1984a |
| $CeO_2$(s) | 1825 – 2320[a] | **$CeO_2$**, CeO >> Ce | [a]Ackermann & Rauh 1971b |
| 2753 K | 1736 – 2067[b] | | [b]Piacente et al. 1973 |
| | 1700 – 2200[c] | | [c]Marushkin et al. 2000 |
| $Ce_2O_3$(s) | 1825 – 2320[a] | **CeO**, $CeO_2$ >> Ce | [a]Ackermann & Rauh 1971b |
| 2523 K | 1850 – 2050[b] | | [b]Marushkin et al. 2000 |
| CoO(s) | 1578 – 1744[a] | **Co**, CoO | [a]Grimley et al. 1966 |
| 2103 K | 1617 – 1818[b] | | [b]Kazenas and Tagirov 1995 |
| $Cr_2O_3$(s) | 1839 – 2059[a] | **Cr**, CrO, $CrO_2$, $CrO_3$ | [a]Grimley et al. 1961 |
| 2603 K | 1690 – 2020[b] | | [b]Chizikov et al. 1972 |
| $CrO_3$(s) | 423 – 523[a] | **$(CrO_3)_n$**, | [a]McDonald and Margrave 1968 |
| 470 K | 440 – 480[b] | (n = 3-5), $Cr_xO_{3x-2}$ | [b]Kazenas & Samoilova 1995 |
| $Cs_2O$(s,l) | 443 – 1000[a] | **$Cs_2O$**, Cs, $Cs_2O_2$, | [a]Tower 1969 |
| 768 K | 1070 – 1170[b] | CsO | [b]Nicolosi et al. 1979 |
| | 298 – 2500[c] | | [c]Glushko et al. 1982 |
| | 620 – 1100[d] | | [d]Lamoreaux & Hildenbrand 1984 |
| $Cu_2O$(s) | 1280 – 1450[a] | **Cu** | [a]Kazenas et al. 1969 |
| 1517 K | 1210 – 1452[b] | | [b]Kodera et al. 1968 |
| CuO(s) | 873 – 1173 | | Kodera et al. 1968 |
| $Dy_2O_3$(s) | 2432 – 2637[a] | **DyO**, Dy | [a]Ames et al. 1967 |
| 2680 K | 2000 – 2500[b] | | [b]Panish 1961b |
| $Er_2O_3$(s) | 2492 – 2687[a] | **ErO**, Er | [a]Ames et al. 1967 |
| 2691 K | 2000 – 2500[b] | | [b]Panish 1961b |
| $Eu_2O_3$(s) | 1984 – 2188[a] | Eu, EuO[a,b] | [a]Ames et al. 1967 |
| 2623 K | 1950 – 2350[b] | **EuO** >> Eu[c] | [b]Panish 1961a |
| | | | [c]Dulick et al 1986 |

| Oxide | Temperature (K) | Gas Species | References |
|---|---|---|---|
| $Fe_{1-x}O$(s) | 1776 (l)[a] | **Fe** > FeO | [a]Brewer & Mastick 1951 |
| wüstite | 1773 – 1867 (l)[b] | | [b]Washburn 1963 |
| 1650 K | 1600 – 1750 (s,l)[c] | | [c]Chizikov et al. 1971 |
| FeO(l)[c] | 1543 – 1573[d] | | [d]Shchedrin et al. 1978 |
| | 1930 – 1950 (l)[e] | | [e]Kazenas and Tagirov 1995 |
| $Fe_3O_4$(s) | 1400 – 1600[a] | **Fe**[b] | [a]Chizikov et al. 1971 |
| 1870 K | 1373 – 1673[b] | | [b]Shchedrin et al. 1978 |
| $Fe_2O_3$(s,l) | 1030 – 1200[a] | **Fe**[b], FeO[b], $FeO_2$[b] | [a]Chizikov et al. 1971 |
| 1730 K dec in 1 atm $O_2$ | 1639 – 1764[b] | | [b]Hildenbrand 1975 |
| $Ga_2O_3$(s) | 1073 – 1273[a] | **$Ga_2O$**, Ga, GaO | [a]Frosch & Thurmond 1962 |
| 2080 K | 2068[b] | | [b]Burns 1966 |
| | 1013 – 1841[c] | | [c]Shchukarev et al. 1969 |
| $Gd_2O_3$(s) | 2000 – 2500[a] | **GdO** >> Gd | [a]Panish 1961b |
| 2693 K | 2350 – 2590[b] | | [b]Messier 1967 |
| | 2400 – 2550[c] | | [c]Ames et al. 1967 |
| | 2100 – 2350[d] | | [d]Alcock & Peleg 1967 |
| $GeO_2$(s) | 758 – 859[a] | **GeO**, Ge, $Ge_2O$, $(GeO)_2$, $(GeO)_3$ | [a]Jolly & Latimer 1952 |
| 1388 K | 1313 – 1373[b] | | [b]Shimazaki et al. 1957 |
| | 744 – 1373[c] | | [c]Drowart et al. 1965b |
| | 1213 – 1347[d] | | [d]Kazenas et al. 1973 |
| | 965 – 1144[e] | | [e]Rau & Schnedler 1984 |
| $HfO_2$(s) | 2246 – 2600[a] | **HfO** > $HfO_2$ >> Hf | [a]Panish & Reif 1963 |
| 3073 K | 2150 – 2500[b] | | [b]Alcock & Peleg 1967 |
| | 2200 – 2900[c] | | [c]Belov & Semenov 1985 |
| $Ho_2O_3$ (s) | 2487 – 2711[a] | **HoO**, Ho | [a]Ames et al. 1967 |
| 2686 K | 2000 – 2500[b] | | [b]Panish 1961b |
| $In_2O_3$(s) | 1290 – 1490[a] | **$In_2O$**, In, InO | [a]Shchukarev et al. 1961 |
| 2186 K | 1269 – 1539[b] | | [b]Burns et al. 1963 |
| | 1300 – 1500[c] | | [c]Burns 1966 |
| | 1539 – 1631[d] | | [d]Shchukarev et al. 1969 |
| | 873 – 1223[e] | | [e]Valderraman and Jacob 1977 |
| | 1322 – 1520[f] | | [f]Gomez et al 1982 |

| Oxide | Temperature (K) | Gas Species | References |
|---|---|---|---|
| $K_2O$ (s,l) | [a]1087 – 1186 | **K**, $K_2O$, KO, $K_2O_2$[d], | [a]Ehlert 1977 |
| 1013 K[b] | [b]1037 – 1184 | $K_2$[d], $KO_2$[d] | [b]Simmons et al. 1977 |
| | [c]580 – 825 | | [c]Byker et al. 1979 |
| | [d]298 – 3000 | | [d]Lamoreaux and Hildenbrand 1984 |
| $La_2O_3$ (s) | 2230 – 2440[a] | **LaO** >> La | [a]Walsh et al 1960 |
| 2586 K | 2234 – 2441[b] | | [b]Goldstein et al. 1961 |
| | 1778 – 2427[c] | | [c]Ackermann & Rauh 1971a |
| $Li_2O$(s) | 1000 – 1600[a] | **Li** ~ **$Li_2O$** > LiO > | [a]Berkowitz et al. 1959 |
| 1711 K | 1494 – 1669[b] | $Li_3O$[d,e], $Li_2O_2$[d,e] | [b]Hildenbrand et al. 1963 |
| | 1447 – 1662[c] | | [c]White et al. 1963 |
| | 1300 – 1700[d] | | [d]Kudo et al. 1978 |
| | 1225 – 1507[e] | | [e]Kimura et al. 1980 |
| $Lu_2O_3$ (s) | 2615 – 2700[a] | **LuO** >> Lu | [a]Ames et al. 1967 |
| 2763 K | 2000 – 2500[b] | (LuO/Lu ~ 4)[c] | [b]Panish 1961b |
| | 2700[c] | | [c]Lopatin & Shugurov 2014 |
| MgO(s) | 1950[a] | **Mg**, MgO | [a]Porter et al 1955b |
| 3100 K | 1720 – 1960[b] | | [b]Alcock & Peleg 1967 |
| | 2020 – 2160[c] | | [c]Farber & Srivastava 1976 |
| | 1573 – 1973[d] | | [d]Sata et al. 1978 |
| | 1820 – 1983[e] | | [e]Kazenas et al 1983 |
| MnO(s) | 1767[a] | **Mn**, MnO, $MnO_2$[c] | [a]Brewer & Mastick 1951 |
| 2115 K | 1618 – 1815[b] | | [b]Kazenas et al. 1984b |
| | 1858 – 1999[c] | | [c]Hildenbrand & Lau 1994 |
| | 1535 – 1782[d] | | [d]Matraszek et al. 2004 |
| $MoO_2$(s) | 1481 – 1777[a] | **$MoO_3$**, $(MoO_3)_2$, | [a]Burns et al., 1960 |
| 1373 K dec. | 2262 – 2466[b] | $MoO_2$, $(MoO_3)_3$, MoO[b] | [b]DeMaria et al. 1960 |
| $MoO_3$(s) | 800 – 1000[a] | **$(MoO_3)_3$**, $(MoO_3)_n$ | [a]Berkowitz et al. 1957b |
| 1074 K | 980 – 1060[b] | n=4,5 | [b]Ackermann et al. 1960 |
| $Na_2O$(s) | 987 – 1114[a] | **Na**, $Na_2O$, NaO, | [a]Hildenbrand and Murad 1970 |
| 1405 K | 1144 – 1304[b] | $NaO_2$[b] | [b]Piacente et al 1972 |
| | 984 – 1056[c] | | [c]Hildenbrand and Lau, 1993 |
| | 873 – 1003[d] | | [d]Popovic et al. 2012 |

| Oxide | Temperature (K) | Gas Species | References |
|---|---|---|---|
| NbO (s,l)<br>2210 K | 1773 – 2473 (s,l)[a]<br>2032 – 2125[b]<br>1971 – 2175[c] | **$NbO_2$ ~ NbO** | [a]Shchukarev et al. 1966<br>[b]Kamegashira et al. 1981<br>[c]Matsui & Naito 1982 |
| $NbO_2$ (s,l)<br>2175 K<br>2188 K[b] | 1773 – 2473[a]<br>1963 – 2323[b] | **$NbO_2$**, NbO | [a]Shchukarev et al. 1966<br>[b]Kamegashira et al. 1981 |
| $Nb_2O_5$ (s,l)<br>1785 K | 1726 – 2271 | **$NbO_2$** | Matsui & Naito 1983 |
| $Nd_2O_3$ (s)<br>2593 K | 2250 – 2400[a]<br>2255 – 2408[b]<br>1950 – 2350[c]<br>2155 – 2485[d] | **NdO** >> Nd | [a]Walsh et al. 1960<br>[b]Goldstein et al. 1961<br>[c]Panish 1961a<br>[d]Tetenbaum 1975 |
| NiO(s)<br>2228 K | 1782 – 1816[a]<br>1575 – 1709[b]<br>1400 – 1570[c]<br>1587 – 1730[d] | **Ni**, NiO | [a]Brewer & Mastick 1951<br>[b]Grimley et al. 1961b<br>[c]Kodera et al. 1968<br>[d]Kazenas and Tagirov 1995 |
| $OsO_4$ (l)<br>314 K | 1100 – 1750 | **$OsO_4$**, $OsO_3$ >> $OsO_2$ | Grimley et al. 1960 |
| $P_4O_{10}$(s)<br>699 K | 333 – 923[a]<br>313 – 773[b]<br>370 – 423[c] | **$P_4O_{10}$** >> $P_4O_9$, $P_4O_8$, $P_4O_7$, $P_4O_6$ | [a]Hashizume et al. 1966<br>[b]Muenow et al. 1970<br>[c]Kazenas & Petrov 1997 |
| PbO(s)<br>1160 K | 1000 – 1150[a]<br>1010 – 1105[b]<br>810 – 1040[c] | **PbO**, Pb, $(PbO)_n$ (n = 2 to 6) | [a]Drowart et al. 1965a<br>[b]Kazenas & Petrov 1996<br>[c]Kobertz et al. 2014 |
| $Pr_2O_3$ (s)<br>2573 K | 1950 – 2350[a]<br>Calculated[b] | **PrO** >> Pr | [a]Panish 1961a<br>[b]Dulick et al. 1986 |
| $Rb_2O$ (s,l)<br>900 K | 1020 – 1135[a]<br>298 – 3000[b] | **Rb**, $Rb_2$, $Rb_2O$, RbO, $Rb_2O_2$ | [a]Nicolosi et al. 1979<br>[b]Lamoreaux and Hildenbrand 1984 |
| $Re_2O_7$(s)<br>570 K | 327 – 463 | **$Re_2O_7$**, $ReO_3$, $ReO_2$, $Re_2O_6$, $Re_2O_5$ | Skinner and Searcy, 1973 |
| $Sb_2O_4$(s)<br>1524 K dec. | 1090 | **$Sb_4O_6$**, $Sb_3O_4$, $Sb_2O_3$, Sb, $Sb_2O_2$ | Asryan et al. 2003 |

| Oxide | Temperature (K) | Gas Species | References |
|---|---|---|---|
| $Sb_2O_3$(s,l) | 742 – 829 (cubic)[a] | **$Sb_4O_6$**[a,b,c] | [a]Hincke 1930 |
| orthorhombic[a,b] | 742 – 917 (orthorh)[a] | (+ $Sb_4O_5$, $Sb_3O_4$, | [b]Behrens and Rosenblatt 1973 |
| 928 K | 929 – 1073 (liq)[a] | $Sb_2O_2$, SbO)[c] | [c]Semenov et al. 1983 |
| cubic[a,c] | 627 – 732 (orthorh)[b] | | |
| 843 K $T_{trans}$ | 580 – 730 (cubic)[c] | | |
| $Sc_2O_3$ (s) | 2555 – 2693[a] | **ScO**, Sc | [a]Ames et al 1967 |
| 2762 K | 2481 – 2740[b] | | [b]Belov et al. 1987 |
| | 2521 – 2768[c] | | [c]Rostai & Wahlbeck 1999 |
| $SiO_2$ | 1840 – 1951[a] | **SiO**, $SiO_2$ >> Si | [a]Brewer & Mastick 1951 |
| cristobalite | 1200 – 1950[b] | | [b]Porter et al. 1955c |
| 1996 K | 1600 – 1754[c] | | [c]Nesmeyanov & Firsova 1960 |
| $SiO_2$(l)[g] | 1823 – 1983[d] | | [d]Nagai et al. 1973 |
| | 1833 – 1877[e] | | [e]Zmbov et al. 1973 |
| | 1887 – 1988[f] | | [f]Kazenas et al. 1985 |
| | 1800 – 2200[g] | | [g]Shornikov et al. 1998 |
| $Sm_2O_3$ (s) | 2333 – 2499[a] | **SmO**, Sm | [a]Ames et al. 1967 |
| 2613 K | 1950 – 2350[b] | | [b]Panish 1961a |
| $SnO_2$(s) | 1254 – 1538[a] | **SnO**, $Sn_2O_2$[a,c] | [a]Colin et al. 1965 |
| 1903 K | 1315 – 1660[b] | | [b]Hoenig and Searcy, 1966 |
| | 1296 – 1672[c] | | [c]Kazenas et al. 1996 |
| | 1160 – 1450[d] | | [d]Zimmermann et al. 1999 |
| SrO(s) | 1575 – 1750[a] | **Sr ~ SrO** | [a]Pelchowitch 1954 |
| 2805 K | 2100[b] | | [b]Porter et al. 1955b |
| | 2010 – 2102[c] | | [c]Farber and Srivastava 1976 |
| | 1905 – 1996[d] | | [d]Samoilova & Kazenas 1994 |
| $Ta_2O_5$(s,l) | 2019 – 2314[a] | **$TaO_2$**, TaO >> Ta | [a]Inghram et al. 1957 |
| 2145 K | 1573 – 2073[b] | | [b]Kofstad 1964 |
| | 1924 – 2268[c] | | [c]Krikorian & Carpenter 1965 |
| | 1880 – 2525[d] | | [d]Smoes et al. 1976 |
| | 2043 – 2262[e] | | [e]Kazenas et al. 1994 |
| $Tb_2O_3$(s) | 2000 – 2500 | **TbO** >> Tb | Panish 1961b |
| 2682 K | | | |

| Oxide | Temperature (K) | Gas Species | References |
|---|---|---|---|
| $ThO_2$(s,l) | 2050 – 2250[a] | **$ThO_2$**, ThO >> Th | [a]Shapiro 1952 |
| 3650 K | 2170 – 2400[b] | | [b]Wolff & Alcock 1962 |
| | 2000 – 3000[c] | | [c]Ackermann et al. 1963 |
| | 2170 – 2500[d] | | [d]Peleg & Alcock 1966 |
| | 2400 – 2800[e] | | [e]Ackermann & Rauh 1973 |
| | 1782 – 1940[f] | | [f]Murad & Hildenbrand 1974 |
| | 2390 – 2950[g] | | [g]Belov & Semenov 1979 |
| | 5140 – 6230 (l)[h] | | [h]Joseph et al. 2002 |
| TiO(s,l) | 1840[a] | **TiO**, Ti, $TiO_2$ | [a]Berkowitz et al. 1957d |
| 2023 K | 2023 – 2153[b] | | [b]Gilles et al 1968 |
| $Ti_2O_3$(s,l) | 2194[a] | **TiO ~ $TiO_2$** >> Ti | [a]Berkowitz et al. 1957d |
| 2115 K | 1675 – 1768[b] | | [b]Wu & Wahlbeck 1972 |
| $Ti_3O_5$ (s,l) | 1870 – 2000[a] | **$TiO_2$ ~ 2×TiO** | [a]Hampson & Gilles 1971 |
| 2050 K | 1837 – 2040[b] | | [b]Wahlbeck and Gilles 1967 |
| | 2097 – 2143[c] | | [b]Gilles et al 1968 |
| $TiO_2$(rutile,l) | 1881[a] | **$TiO_2$ ~ TiO** >> Ti | [a]Berkowitz et al. 1957d |
| 2116 K | 2193[b] | | [b]Gilles et al 1968 |
| | 1850 – 2540[c] | | [c]Semenov 1969 |
| $Tl_2O$(s,l) | 620 – 770[a] | **$Tl_2O$**, >> $Tl_3O$, $Tl_4O_2$ | [a]Cubicciotti 1969, 1970 |
| 852 K | 820 – 940[b] | | [b]Holstein 1993a |
| $Tl_4O_3$ (s,l) | ~ 773[a] | **$Tl_2O$** | [a]Wahlbeck & Myers 1997 |
| 990 K | 920 – 1080[b] | | [b]Holstein 1993a |
| $Tl_2O_3$(s) | 915 – 1030[a] | **$Tl_2O$**, Tl, TlO | [a]Shchukarev et al. 1961 |
| 1107 K | 803 – 948[b] | | [b]Cubicciotti & Keneshea 1967 |
| | 657 – 773[c] | | [c]Wahlbeck et al. 1991 |
| | 880 – 1060[d] | | [d]Holstein 1993a |
| $Tm_2O_3$ (s) | 2450 – 2641[a] | **Tm**, TmO | [a]Ames et al. 1967 |
| 2683 K | 2000 – 2500[b] | | [b]Panish 1961b |
| $UO_2$ (s,l) | 1600 – 2800[a] | **$UO_2$**, $UO_3$, UO | [a]Ackermann et al. 1956 |
| 3123 K | 2200 – 2800[b] | | [b]Ohse 1966 |
| | 2025 – 2343[c] | | [c]Matsui & Naito 1985 |
| | 2500 – 5000 (s,l)[d] | | [d]Joseph et al. 1996 |
| | 4715, 5708 (l)[e] | | [e]Joseph et al. 2002 |
| | 2800 – 3400 (s,l)[f] | | [f]Pflieger et al 2011 |

| Oxide | Temperature (K) | Gas Species | References |
|---|---|---|---|
| $U_3O_8$(s) | 1230 – 1700[a] | **$UO_3$** | [a]Ackermann et al. 1960 |
| 2010 K | 840 – 1450[b] | | [b]Ackermann & Chang 1973 |
| (in 1 kb He) | | | |
| VO(s) | 1680 – 1950[a] | **VO**, V >> $VO_2$ | [a]Berkowitz et al. 1957a |
| 2063 K | 1803 – 1999[b] | | [b]Banchorndhevakul et al. 1986 |
| $V_2O_3$(s) | 2270[a] | **VO ~ $VO_2$**, V | [a]Farber et al. 1972b |
| 2340 K | 1914 – 2182[b] | | [b]Banchorndhevakul et al. 1985 |
| | 1839 – 2105[c] | | [c]Wang et al. 2010 |
| $V_2O_4$(s,l) | 1673 – 2273[a] | **$VO_2$**, VO, $V_4O_{10}$, V | [a]Frantseva & Semenov 1969 |
| 1818 K | 2063 – 2270[b] | | [b]Farber et al 1972b |
| $V_2O_5$(s,l) | 1680 – 1950[a] | **$V_4O_{10}$**, $V_4O_8$, $V_6O_{14}$, | [a]Berkowitz et al. 1957a |
| 943 K | 1003 – 1205[b] | $V_6O_{12}$, $V_2O_4$ | [b]Farber et al. 1972b |
| $WO_3$(s) | 1368 – 1492[a] | **$W_3O_9$**, $W_4O_{12}$, $W_5O_{15}$ | [a]Berkowitz et al. 1957c |
| 1745 K | 2188 – 2475[b] | $W_2O_6$, $WO_3$, $WO_2$, | [b]DeMaria et al 1960 |
| | 1300 – 1600[c] | WO,$W_3O_8$, $W_3O_9$ | [c]Ackermann & Rauh 1963 |
| | 1400 – 3150[d] | (over W)[b,d] | [d]Schissel & Trulson 1966 |
| $Y_2O_3$ (s,l) | 2500 – 2700[a] | **YO** >> Y | [a]Walsh et al. 1960 |
| 2712 K | 2492 – 2697[b] | | [b]Ames et al. 1967 |
| | 2170 – 2500[c] | | [c]Alcock & Peleg 1967 |
| | 1400 – 2100[d] | | [d]Ackermann et al. 1970 |
| | 1600 – 2700[e] | | [e]Ackermann & Rauh 1973 |
| | 3500 – 5000 (l)[f] | | [f]Hastie et al. 2000 |
| $Yb_2O_3$ (s) | 2000 – 2500[a] | **Yb**, YbO | [a]Panish 1961b |
| 2707 K | 2371 – 2626[b] | | [b]Ames et al. 1967 |
| | 2000 – 2500[c] | | [c]Alcock & Peleg 1967 |
| | 2290 – 2580[d] | | [d]Stolyarova et al. 2014 |
| ZnO(s) | 1229 – 1259[a] | **Zn** >> ZnO[c] | [a]Anthrop and Searcy 1964 |
| 2248 K | 1321 – 1632[b] | | [b]Hoenig 1964 |
| | 1023 – 1473[c] | | [c]Kodera et al. 1968 |
| | 1200 – 1540[d] | | [d]Kazenas et al. 1984 |
| $ZrO_2$ (s) | 1700 – 2500[a] | **ZrO** ~ **$ZrO_2$** >> Zr | [a]Chupka et al. 1957 |
| 2950 K | 1800 – 2390[b] | | [b]Peleg & Alcock 1966 |
| | 2060 – 2875[c] | | [c]Ackermann et al. 1975 |
| | 1790 – 2430[d] | | [d]Murad & Hildenbrand 1975 |
| | 2600 – 2820[e] | | [e]Belov et al. 1981 |

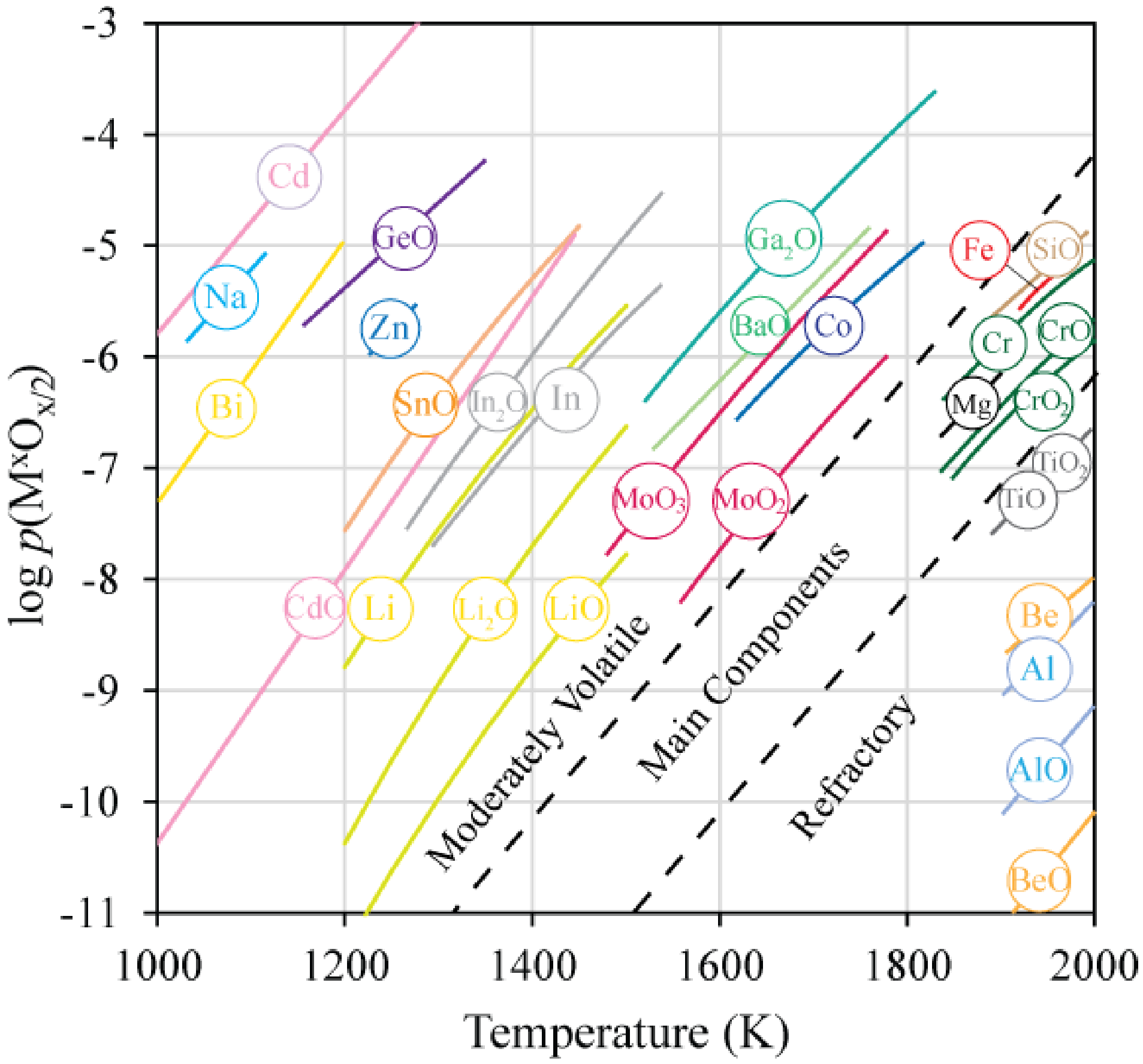

***Fig. 2.*** *Equilibrium partial pressures of major metal-bearing gas species (listed in circles) over selected pure oxides during their congruent evaporation in a vacuum. Not shown is $Tl_2O$(s) where logp($Tl_2O$) = -3 at 870 K (Cubicciotti 1970); CdO(s), Behrens and Mason 1981; $Na_2O$(s), Hildenbrand and Murad 1970; $Bi_2O_3$(s), Sidorov et al. 1980; $GeO_2$(s) Drowart et al, 1965; ZnO(s), Anthrop and Searcy 1964; $SnO_2$(s), Zimmermann et al. 1999; $In_2O_3$(s), Burns et al. 1963; $Li_2O$, Kimura et al. 1980; $Ga_2O_3$(s), Shchukarev et al. 1969; BaO(s), Inghram et al. 1955; $MoO_2$(s), Burns et al. 1960; CoO(s), FeO(s) Kazenas and Tagirov 1985; $SiO_2$(s), Kazenas et al. 1985; $Cr_2O_3$(s), Grimley et al. 1961; MgO(s), Kazenas et al. 1983; $TiO_2$(s), Semenov 1969; BeO(s), Chupka et al. 1959; $Al_2O_3$(s), Farber et al. 1972.*

### *2.1.4. Activity measurements in binary metal oxide - silicate systems ($MO_x$ – $SiO_2$)*

The application of Knudsen effusion, KEMS, torsion effusion, or transpiration measurements to natural geologic compositions has so far been limited, with a few exceptions, and has instead been used extensively to study binary glass-forming melts (borates, germanates, phosphates, and silicates; Stolyarova 2001). Electromotive force measurements, solubility measurements, and partitioning of a solute between two phases have also been used to measure activities in binary silicate melts. Kubaschewski and Alcock (1979) describe numerical calculation of activities and activity coefficients from all these methods. Several of the papers listed below also give good descriptions of the underlying principles. The most activity data are available for $Na_2O$-$SiO_2$ and $K_2O$-$SiO_2$ melts: (e.g., Kröger and Sörström 1965; Charles 1967; Frohberg et al. 1973; Eliezer et al., 1978a; Sanders and Haller 1979; Plante 1981; Rego et al. 1985; Zaitsev et al. 1999; Zaitsev et al. 2000). Some data are available for CaO-$SiO_2$, FeO-$SiO_2$, and MgO-$SiO_2$ melts (e.g., Schuhmann and Ensio 1951;

Bodsworth 1959; Distin et al 1971; Zaitsev and Mogutnov 1995; 1997; Zaitsev et al. 2006). Apparently only a few studies exist for vaporisation of $Li_2SiO_3$ (s,l) (Penzhorn et al. 1988; Asano and Nakagawa 1989). They show Li is the major Li-bearing gas with other $LiO_x$ gases being significantly less abundant. Kim & Sanders (1991) computed Redlich – Kister parameters for the $Rb_2O$ – $SiO_2$ and $Cs_2O$ – $SiO_2$ binaries.

*2.1.5. Activity measurements for multicomponent silicate systems*

Carmichael et al (1977) and Bottinga and Richet (1978) are the pioneering studies on computations of activity coefficients in multicomponent silicates. A large amount of subsequent work was done on thermochemical properties of silicate glasses and melts (e.g., Stebbins et al 1984; Richet and Bottinga 1986; Lange and Navrotsky 1992, 1993). Most work on activity measurements in multicomponent silicate systems focused on alkali-bearing aluminosilicates (e.g., (Eliezer et al 1978b; Hastie et al. 1982; Kassis and Frischat 1981; Rammensee and Fraser 1982; Fraser et al. 1983, 1985; Chastel et al. 1987; Fraser and Rammensee 1987) with applications to granitic systems. There is also a large body of metallurgical literature in which the activities of silica and other major rock-forming oxides have been measured, generally at one or a few isotherms, e.g., the silica activity measurements of Kay and Taylor (1960) for the $CaO$–$Al_2O_3$–$SiO_2$ system which is relevant to the mineralogy of Ca, Al-rich inclusions; $FeO$-$MgO$-$SiO_2$ melts (Plante et al. 1992) and the CMAS system (Rein and Chipman, 1965), that has been applied to model the phase equilibria of terrestrial basalts (Osborn and Tait, 1952; Presnall et al., 1978), as well as olivine (Dohmen et al., 1998; Costa et al. 2017).

Aside from the major oxide components (*i.e.,* $SiO_2$, $TiO_2$, $Al_2O_3$, FeO, $FeO_{1.5}$, CrO, $CrO_{1.5}$, MgO, CaO, $NaO_{0.5}$, $KO_{0.5}$, $LiO_{0.5}$, $PO_{2.5}$ and $H_2O$; Chipman 1948; Fincham and Richardson 1954; Rein and Chipman 1965; Charles 1967; Carmichael et al. 1977; Ryerson, 1985; Ghiorso and Sack 1995; O'Neill, 2005), whose activities can be constrained by phase equilibria (*e.g.,* buffering of $a$SiO$_2$ of a silicate liquid by olivine and orthopyroxene, Kushiro, 1975), and a handful of trace elements (*e.g.,* Mo, W, Co, Mn, Ni, Cu, In, Pb; Hirschmann and Ghiorso 1994; O'Neill and Eggins 2002; O'Neill et al., 2008; Wood and Wade 2013), the nature of metal oxide species (and of their activity coefficients) in complex silicate melts are poorly known and should be an experimental target in future. Table 2 lists determinations for some metal oxide activity coefficients in complex silicate melts. These should only be used as guide and the interested reader is referred to the relevant paper for further information as to how the activity coefficients were calculated and how they vary with melt composition, temperature, and oxygen fugacity.

**Table 2.** Metal oxide activity coefficients of 28 species in complex silicate melts.

| Oxide | Temperature (K) | Composition | Activity Coefficient | References |
|---|---|---|---|---|
| $AlO_{1.5}$ | 1573 – 1773 | CMAS (An-Di) | 0.28 – 0.37 | Ghiorso and Sack 1995 |
| $AsO_{1.5}$ | 1573 | FCS | 0.03 – 0.56 | Chen and Jahanshahi 2010 |
| $BiO_{1.5}$ | 1573 | FCS | 0.3 – 0.5 | Paulina et al. 2013 |
| CaO | 1873 | CMAS (0.55 < Basicity < 0.65) | ≈0.001 – 0.15 | Beckett 2002 |
| CoO | 1573 – 1873[a]<br>1673[b] | (F)CMAS[a]<br>CMAS[b] | 0.5 – 2.2[a]<br>0.8 – 1.8[b] | [a]Holzheid et al. 1997<br>[b]O'Neill and Eggins 2002; O'Neill and Berry, 2006 |
| CrO | 1773 | CAS | 1.9 – 7.2 | Pretorius and Muan 1992 |
| $CrO_{1.5}$ | 1773 | CAS | 23 – 42 | Pretorius and Muan 1992 |
| $CuO_{0.5}$ | 1573[a]<br>1923[b] | (F)CMAS[a]<br>CMAS[b] | 9 – 11[a]<br>3.5[b] | [a]Holzheid and Lodders 2001<br>[b]Wood and Wade 2013 |
| FeO | 1600[a]<br>1573 – 1873[b]<br>1673[b]<br>1923[c] | SKAnF[a]<br>(F)CMAS[b]<br>CMAS[c, d] | 0.6 – 2.3[a]<br>1.2 – 2.1[b]<br>0.9 – 1.8[c]<br>≈0.5 – 2.5[d] | [a]Doyle 1988<br>[b]Holzheid et al. 1997<br>[b]O'Neill and Eggins 2002<br>[c]Wood and Wade 2013 |
| GeO | 1423 – 1523 | FCAS | 1.44 – 2.55 | Yan and Swinbourne 2013 |
| $GeO_2$ | 1573 | FCMS | 0.24 – 1.50 | Shuva et al. 2016 |
| $InO_{1.5}$ | 1923 | CMAS | 0.02 | Wood and Wade 2013 |
| $KO_{0.5}$ | 1573 | KFCAS*<br>* with $KAlSiO_4(s)$ | $6.3\times10^{-5}$ – $7.1\times10^{-4}$ | Hastie et al. 1981 |
| MgO | 1873 | CMAS (0.55 < Basicity < 0.65) | ≈0.25 – 4 | Beckett 2002 |
| MnO | 1823 – 1923[a]<br>1873[b] | FCMS(P)[a]<br>CAS[b] | 0.7 – 3.5[a]<br>0.5 – 7.2[b] | [a]Suito and Inoue 1984<br>[b]Ohta and Suito 1995 |

| Oxide | Temperature (K) | Composition | Activity Coefficient | References |
|---|---|---|---|---|
| $MoO_2$ | 1673[a] | CMAS[a, b] | 67 – 227[a] | [a]O'Neill and Eggins 2002 |
| | 1923[b] | | ≈5 – 100[b] | [b]Wood and Wade 2013 |
| $MoO_3$ | 1673 | CMAS | 0.1 – 0.8* *log$\gamma$$MoO_3$ = 1.49(log$\gamma$$MoO_2$)-3.59 | O'Neill and Eggins 2002 |
| $NaO_{0.5}$ | 1623[a] | CMAS[a] | [a]$10^{-3}$ | [a]Mathieu et al. 2008 |
| | 1673[b] | CMS[b] | [b]$8.4\times10^{-4}$ – $2.2\times10^{-3}$ | [b]Mathieu et al. 2011 |
| NiO | 1573 – 1873[a] | (F)CMAS[a] | 1.3 – 3.5[a] | [a]Holzheid et al. 1997 |
| | 1673[b] | CMAS[b, c] | 2.0 – 4.3[b] | [b]O'Neill and Eggins 2002; |
| | 1923[c] | | ≈1.5 – 3.0[c] | O'Neill and Berry, 2006 |
| | | | | [c]Wood and Wade 2013 |
| $PO_{2.5}$ | 1823 – 1923 | CMFS | $10^{-6}$ – $10^{-10}$ | Turkdogan (2000) |
| PbO | 1923 | CMAS | 0.22 | Wood and Wade 2013 |
| $SiO_2$ | 1573 – 1773 | CMAS (An-Di) | 0.9 – 1.1 | Ghiorso and Sack 1995 |
| SnO | 1573 | FCS | 5 – 6.2 | Takeda and Yazawa 1989 |
| $TiO_2$ | 1573 – 1773 | CMAS | 1.5 – 1.7 | Ghiorso and Sack 1995 |
| $V_2O_3$ | 1900 | CMAS | 1.2 – 20.8 | Wang et al. 2010 |
| | 2000 | CMAS | 0.3 – 5.2 | |
| | 2100 | CMAS | 0.05 – 1.65 | |
| $WO_2$ | 1673 | CMAS | 7.4 – 45* *log$\gamma$$WO_2$ = log$\gamma$$MoO_2$-2.20 | O'Neill et al. 2008 |
| $WO_3$ | 1673[a] | CMAS[a, b] | 0.02 – 0.3[a]* | [a]O'Neill et al. 2008 |
| | 1923[b] | | *log$\gamma$$WO_3$ = log$\gamma$$MoO_3$-1.42 ≈0.001 – 0.05[b] | [b]Wood and Wade 2013 |
| ZnO | 1673 – 1823 | CAS | 0.25 – 0.58 | Reyes and Gaskell 1983 |

***Notes:*** A = $Al_2O_3$; An = Anorthite; C = CaO; Di = Diopside; F = FeO; K = $K_2O$; M = MgO; P = $P_2O_5$; S = $SiO_2$

Attempts to quantify the effects of melt and mineral composition on activity coefficients commonly apply single-value melt parameter models such as the relative proportion of Non-Bridging Oxygens to Tetrahedral cations (NBO/T; *e.g.,* Mysen and Virgo, 1980; Kohn and Schofield, 1994) or optical basicity, which describes the tendency for the sum of the species in a melt to donate electrons (*e.g.,* O'Neill and Eggins, 2002; Burnham and O'Neill, 2016), and then typically only in CMAS, CaO-MgO-$Al_2O_3$-$SiO_2$, (±$Na_2O$, $TiO_2$) systems. More complex formulations, such as the combined linear-quadratic fit of Wood and Wade (2013) are empirical and generally more tractable in these simple systems. Similarly, understanding the underlying mechanisms controlling the activity coefficients of melt components via quasi-chemical Gibbs Free energy approaches (Pelton et al., 2000; Björkvall et al., 2000; Schmid-Fetzer et al., 2007; Glibin and King, 2015) remain restricted to binary and ternary systems, though free-energy minimisation programs in multi-component silicate melts (*e.g.*, MELTS; Ghiorso and Sack, 1995) and *ab-initio* simulations hold promise for modelling the properties of complex silicate melts (Guillot and Sator, 2007). As such, quantifying the evaporation/condensation of trace elements (notably, the moderately volatile elements) remains an empirical exercise for lack of a large of body of concerted experimental and theoretical work in complex, geological systems.

Once the metal oxide activities are known the partial vapour pressures can be computed using computer codes that take non-ideality of the melt into account and also simultaneously solve gas phase and gas-melt chemical equilibria (e.g., the FactSage code; Bale et al. 2002).

*2.1.6. Natural Systems*

Sample return from the Moon following the Apollo missions precipitated a push to understand the depletions of volatile metals (particularly Na and K) in a wide variety of lunar lithologies (O'Hara et al. 1970; Ringwood and Essene 1970; Baedecker et al. 1971). To aid in interpretation of these distinctive chemical signatures from terrestrial basalts, a small body of evaporation studies on lunar compositions, mainly using KEMS, were undertaken, largely in the early 1970s (De Maria and Piacente 1969; Balducci et al., 1971; De Maria et al. 1971; Naughton et al. 1971; Gooding and Muenow 1976; Markova et al. 1986). These authors report measured vapour pressures of volatile elements as a function of temperature. Imperative in these experiments is to measure vapour pressures over a finite range corresponding to the initial vaporisation of the element, lest its concentration in the condensed phase falls such that its evaporation rate decreases and the data obtained no longer refer to a single well-defined composition. The measured equilibrium vapour pressures $p\left(M^{x}\frac{x}{2}O\right)$ are instructive for simulating the composition of the atmosphere developed upon heating. However, as these elements are not evaporating from pure condensed phases, the measured $p\left(M^{x}\frac{x}{2}O\right)$ is not indicative of element volatility unless the activities of the oxides in the melt are known from some independent information (see Table 2). Much of this information is available in the metallurgical literature, e.g., from equilibrium measurements such as those of Rein and Chipman (1965), but, in many cases, has not been appropriately exploited. Rather, in order to determine volatilities of components in silicate rocks, the vapour pressure is divided by the concentration of the element initially present in the sample:

$$\log\frac{p\left(M^{x}\frac{x}{2}O\right)}{X\left(M^{x+n}\frac{x+n}{2}O\right)} = \log K - \frac{n}{4}\log f(O_2) + \log\gamma\left(M^{x+n}\frac{x+n}{2}O\right). \qquad (21)$$

This information is shown in Fig. 3, for the KEMS data of De Maria et al., (1971), acquired upon heating of lunar mare basalt 12022 between 1250 and 2500 K. The most volatile of the major silicate components are the alkalis, Na and particularly K, and hence constitute the dominant metal-bearing component of the vapour evolved from natural basalts at low temperatures. This order of evaporation is consistent with their evaporation from pure oxides of $Na_2O$ and $K_2O$ where $p$K > $p$Na, a tendency also observed in the data of Gooding and Meunow (1976) on Knudsen evaporation of a Hawaiian basalt between 1175 and 1250 K. In these compositions, alkali volatility is $10^4$ times greater than the next most volatile components measured, Fe and Mn. Both transition metals are divalent in silicate melts, occurring as $FeO_{(l)}$ and $MnO_{(l)}$ and predominantly as the monatomic gas in the vapour phase (Table 1), however, thermodynamic data for the pure Mn-O and Fe-O systems from the JANAF tables (Chase, 1998) show that $p$Mn > $p$Fe at the same $fO_2$. The evaporation of $MgO_{(l)}$ and $CrO_{(l)}$ (since $Cr^{2+}/\sum Cr \approx 0.9$ at the $fO_2$ recorded in lunar basalts; Berry et al. 2006) also occur in a similar manner, and are both more volatile than $SiO_2$. The notionally 'refractory' components, CaO, $Al_2O_3$ and $TiO_2$ (lunar glass was found to have no $Ti^{3+}$; Krawczynski et al. 2009), enter the gas phase appreciably only above 1900 K. Phosphorus, a minor element that was not measured in the experiment of De Maria et al. (1971), is also volatile (Muenow et al. 1970) and is inferred to vaporise as $PO_{(g)}$ and $PO_{2(g)}$ gas species from silicate liquids (Markova et al. 1986), though the activity coefficients of $PO_{2.5}$ in steel-making slags can be very small (Table 2).

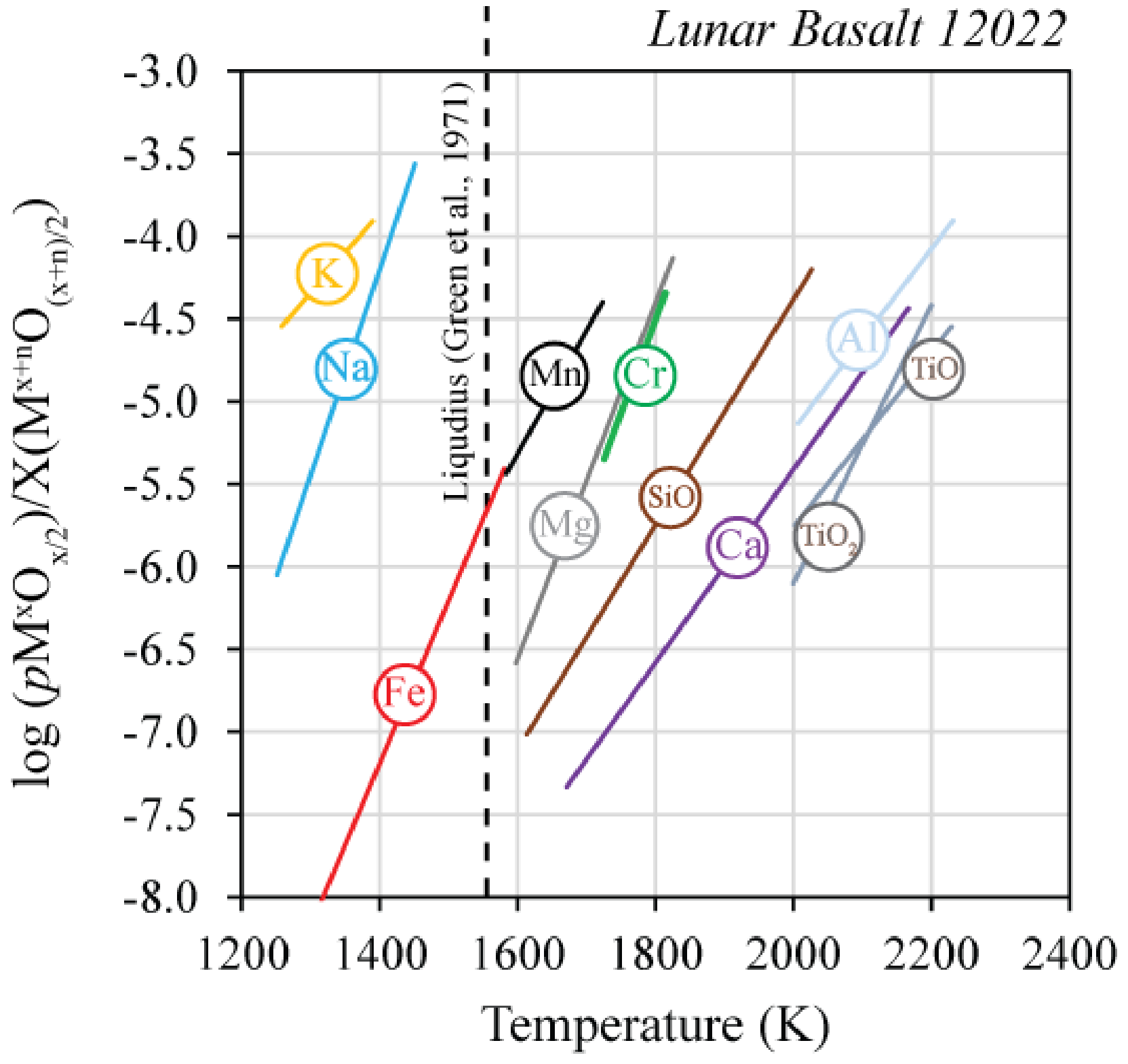

***Fig. 3.*** *The calculated equilibrium partial vapour pressures ($pM^xO_{x/2}$) of an element above Lunar Basalt 12022 relative to its initial concentration in the condensed sample ($XM^{x+n}O_{(x+n)/2}$). The speciation of the metal in the gas phase is shown in the circle. Data from DeMaria et al. (1971).*

Owing to the high volatility and abundance of the alkalis relative to other volatile metals, their vaporisation reactions determine the *f*$O_2$ of gas vaporised from silicate material at "low" temperatures where most of the major oxides remain involatile (e.g., see Schaefer & Fegley 2004a). With increasing temperature ($\gtrsim$ 2000 K) and/or with fractional vaporisation at constant temperature, the less volatile but more abundant elements Fe, Si, and Mg vaporise in larger amounts and it is their respective gas species that then define the bulk properties of the gas at super-liquidus temperatures. The evaporation reaction;

$$SiO_{2(l)} = SiO_{(g)} + \frac{1}{2}O_2 \tag{22}$$

produces an oxygen fugacity 2 log units below the Fayalite-Magnetite-Quartz (FMQ) buffer (O'Neill, 1987) or 1.5 log units above the Iron-Wüstite (IW) buffer, at temperatures above the silicate liquids, >1600 K (Nagai et al. 1973; Kazenas et al. 1985; Shornikov et al. 1998). The oxygen pressures measured in the experiments of De Maria et al. (1971) correspond to an *f*$O_2$ buffered near FMQ+0.5 over the temperature range 1396-1499 K (Fig. 4). That the *f*$O_2$ does not change appreciably over the temperature range of alkali-dominated ($K_{(g)}$, $Na_{(g)}$) at 1400 K to Fe-dominated gas by 1500 K implies that alkali evaporation also results in *f*$O_2$ close to FMQ. By contrast, evaporation of $Fo_{93}$ olivine (Costa et al. 2017), with the same three major metal-bearing gas species, $Fe_{(g)}$, $Mg_{(g)}$ and $SiO_{(g)}$ produces oxygen fugacities close to ≈FMQ-1.5 at ≈1900 K (Fig. 4). However, a clear dependence with temperature is evident over the temperature range covered, becoming more reduced relative to FMQ from 1830 K (FMQ-1.4) to 1970 K (FMQ-1.8). Extrapolating this trend linearly versus inverse temperature intersects the data of DeMaria et al. (1971) at 1450 K (Fig. 4), allowing consistency between the two datasets, and hinting that the entropy change associated with silicate vaporisation is smaller than that of the FMQ buffer. The decrease of *f*$O_2$ relative to (the extrapolated) FMQ buffer at higher temperature may be related to the dissociation of oxygen:

$$O_2 = O + O, \tag{23}$$

which proceeds to the right with increasing temperature, thereby decreasing *f*$O_2$. Nevertheless, even at temperatures in excess of 2000 K, thermodynamic calculations of evaporation of dry (H-, S-, C- and halogen-free) silicate mantles show that the *f*$O_2$ of the vapour phase is very close to that produced by evaporation of pure silica (Visscher and Fegley 2013; Fig. 4).

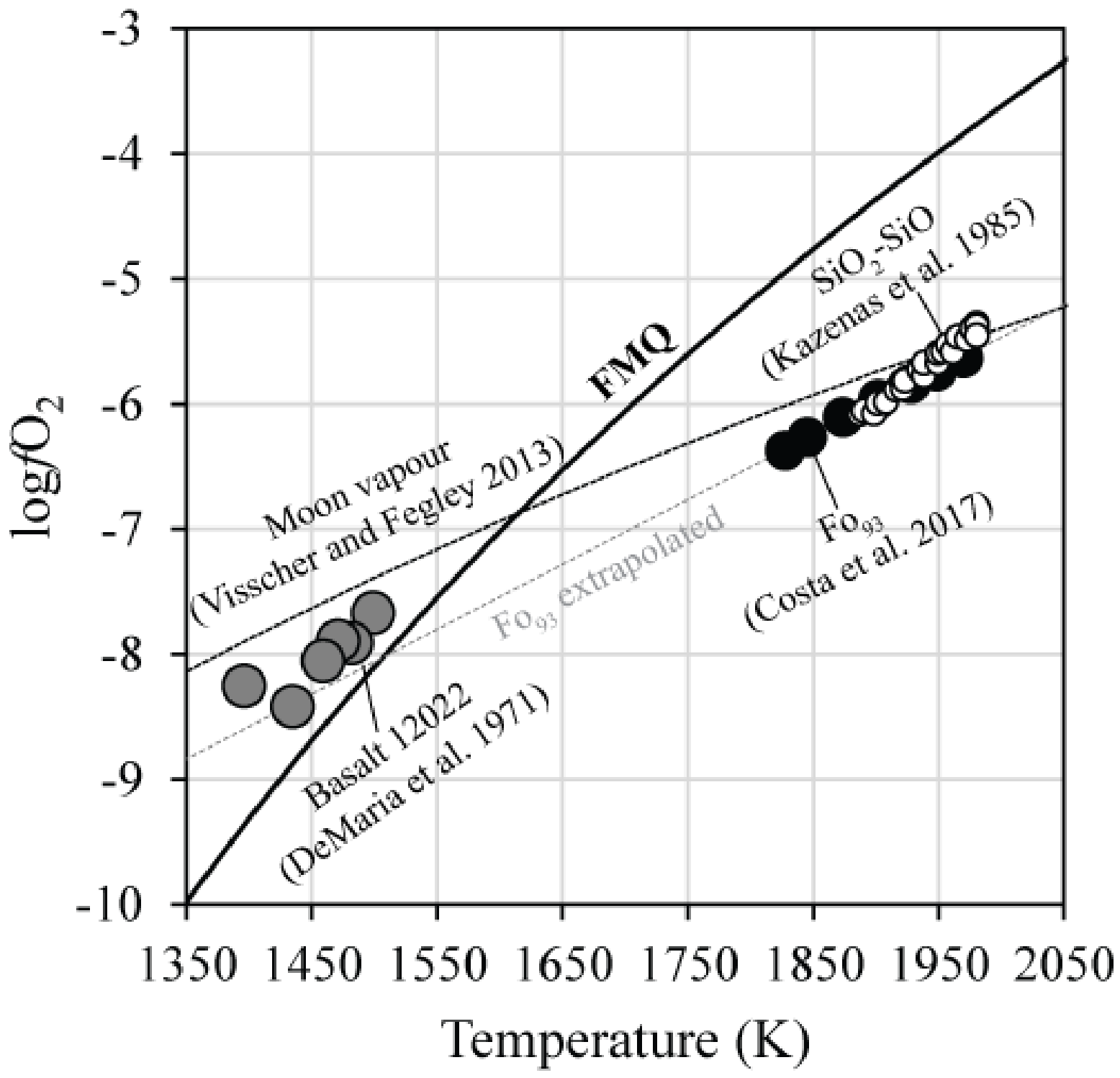

***Fig. 4.*** *Calculated log$fO_2$ as a function of temperature upon evaporation of silicate materials; lunar basalt 12022 (grey; De Maria et al., 1971), $Fo_{93}$ olivine (black, Costa et al., 2017) and pure $SiO_2$ (white, Kazenas, 1985). A model atmosphere by evaporation of a bulk silicate moon composition (Visscher and Fegley, 2013) is shown by the stippled black line. The $fO_2$ defined by the Fayalite-Magnetite-Quartz (FMQ) buffer (O'Neill 1987) is shown in black.*

Lunar basalt 12022 has a liquidus temperature of ≈1300°C (1573 K; Green et al. 1971), and hence evaporation of Na, K, and Fe occurred sub-liquidus. This consideration makes accurate extraction of thermodynamic quantities from such experiments very difficult, however, their partial pressures can be compared with those predicted from evaporation in the pure system, if the corresponding evaporation reaction is known. The activity of the M-O species in the pure phase is always equal to unity (Raoult's law). Because the gas is ideal in both the pure and complex systems, the activity coefficient of the metal oxide species in the liquid can be calculated from the ratio of the equilibrium constant (Equation (4)) of the reaction in the complex system, $K^*$, to that in the pure system, $K$ at a given temperature, pressure and $fO_2$:

$$\gamma_i = \frac{K^*}{K}. \tag{24}$$

The references cited above for activity data in binary and ternary alkali oxide melts and the results of DeMaria et al. (1971) show that alkali oxides dissolved in silicate melts are strongly non-ideal. The studies summarised in Table 1 show $Na_2O$ and $K_2O$ vaporisation proceeds primarily via the reaction

$$(Na,K)O_{0.5(l)} = (Na,K)_{(g)} + \frac{1}{4}O_2 \tag{25}$$

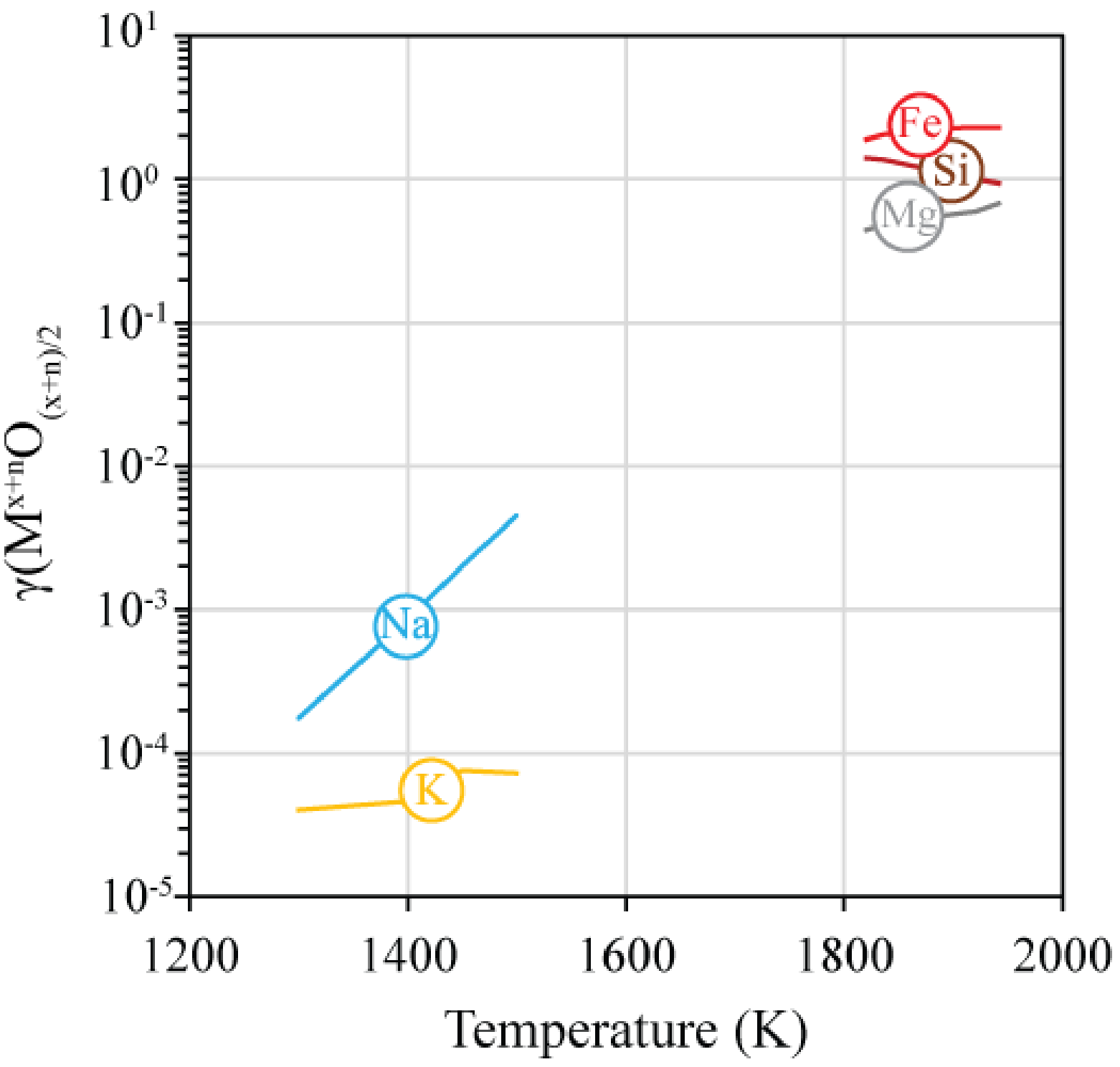

***Fig. 5.*** *Calculated activity coefficients for volatile melt species, $\gamma(M^{x+n}O_{(x+n)/2})$ as a function of temperature in the experiments of DeMaria et al. (1971) (Na, K) in a lunar basalt, and Costa et al. (2017) (Fe, Si, Mg) in olivine*.

The calculated activity coefficients of melt species from the data of De Maria et al. (1971) of $NaO_{0.5}$ and $KO_{0.5}$ are $10^{-3}$ and $5 \times 10^{-5}$, respectively (Fig. 5). Both of the calculated activity coefficients depend on the temperature, decreasing from $4.5\times10^{-3}$ at 1500 K to $1.8\times10^{-4}$ at 1300 K for $\gamma NaO_{0.5}$, while the decrease for $\gamma KO_{0.5}$ is milder, from $3.5\times10^{-5}$ to $7.2\times10^{-5}$, respectively, and both are consistent with values in Table 2. A similar decrease in γMgO with falling temperature is observed in olivine (Costa et al. 2017), whereas $SiO_2$ shows the opposite trend (Fig. 5). All major components in olivine are close to unity, but, in detail, vary by a factor of 2 (greater for Fe, smaller for Mg).

The conclusion is that equilibrium vapour pressures in liquid and solid silicates differ from those predicted from evaporation of the pure phase and require experimental data if their behaviour during evaporation of silicate materials is to be understood. Aside from the role of activity coefficients of metal oxides in silicate melts which could be informed by systematic experimental studies in over a range of compositions in complex systems, the difference may also be related to interdependent vaporisation equilibria, which often, but not always, involve production of $O_2$ and/or O gas. Thus the $O_2$ produced by vaporisation of one oxide affects the vaporisation of other oxides because of the law of mass action.

## 2.2. Kinetic Evaporation

### *2.2.1. General treatment of kinetic evaporation*

The condition for kinetic, free or Langmuir evaporation is met when the partial pressure of the vaporising species is such that the average distance travelled between collisions, the mean-free path, $\ell$, is greater than the length scale over which the evaporation occurs, where:

$$\ell = \frac{kT}{\sqrt{2}p_i\sigma} \tag{26}$$

and $k$ is Boltzmann's constant and $\sigma$ is the collisional cross-sectional area, which can be described by a Lennard-Jones 12-6 potential function that models the van der Waals forces between molecules (*e.g.*, Neufield, 1972).

The Hertz-Knudsen-Langmuir (HKL) equation describes the flux of particles striking a surface (Hertz, 1882; Knudsen 1909; Langmuir 1916; Hirth and Pound 1963; Nussinov and Chernyak 1975; Richter et al. 2002), and is derived from the Maxwell-Boltzmann kinetic theory of gases given an ideal, isothermal gas:

$$\frac{dn_i}{dt} = -A\frac{\alpha_e p_{i,sat} - \alpha_c p_i}{\sqrt{2\pi R M_i T}} \tag{27}$$

Here, $A$ is the surface area, $p_{i,sat}$ is the equilibrium partial pressure of the gas species of element $i$, $p_i$ is its actual partial pressure at the surface, $M_i$ its molar mass, $\alpha_e$ and $\alpha_c$ the Langmuir coefficients for evaporation and condensation respectively, $R$ the gas constant, $T$ the absolute temperature and $\frac{dn_i}{dt}$ the evaporation rate in mol/s.

Equilibrium conditions prevail for the condition $p_{i,sat}$ = $p_i$. However, for all other conditions, evaporation is partially kinetic in nature, reaching the end-member case for which $p_i = 0$, or pure Langmuir evaporation. This situation may physically occur for evaporation into a vacuum, or a scenario in which the gas is being transported away from the evaporation surface before it can be replenished. If this condition is demonstrated, then Equation (27) may be simplified by setting $p_i = 0$. For evaporation of a molten sphere, which approximates the geometry of the melt on the wire-loops in the widely-used experimental petrology method (*e.g.*, Tsuchiyama et al. 1981) then:

$$dn_i = -4\pi r^2 \frac{\alpha_e p_{i,sat}}{\sqrt{2\pi R M_i T}} dt. \tag{28}$$

In order to determine the extent of elemental loss (a change in concentration) experienced by a spherical body, Equation (28) is divided by the total number of moles in a sphere, $n_T^0 = \frac{4\pi r^3 \rho}{3M}$, yielding $\left(\frac{n_i}{n_T^0}\right) = X_i$, the concentration of element $i$. Equation (28) then becomes:

$$dX_i = -\alpha_e p_{i,sat} \frac{3}{r\rho} \sqrt{\frac{M_i}{2\pi RT}} dt. \tag{29}$$

Here, $r$ is the radius of the melt sphere and $\rho$ its density.

Though grounded in statistical mechanics, the applicability or otherwise of the HKL equation to experimental and natural processes remains equivocal because the assumptions used it in its derivation may be invalid. Knudsen's work on the kinetic theory of gases generated intense interest among chemists and physicists in the early 20$^{th}$ Century and led to a large number of subsequent experimental and theoretical studies that verified his ideas, *e.g.*, Loeb (1961). Much of the ambiguity stems from the meaning of the evaporation coefficient, $\alpha_e$, which is an empirical term describing the non-ideality of evaporation. Since the HKL equation describes only the number molecules striking a surface over time, Langmuir coefficients therefore express the energy required in the transformation of a condensed component into a gaseous molecule (or vice-versa), and may depend upon several factors including crystal structure, chemical composition, temperature and gas species. The HKL equation is derived from an equilibrium distribution of gas molecules at a single temperature, and is therefore applicable when the evaporating flux is null, such that the introduction of non-equilibrium coefficients *a posteriori* undermines this assumption (Ackermann et al. 1967). *Ab-initio* approaches to this inconsistency may consider a non-Maxwellian gas distribution, as attempted by Schrage (1953) whose equation yields evaporation rates double those of the HKL equation, but does not satisfy momentum- or energy conservation.

The HKL equation as written in Equation (27) assumes that the temperature in the gas is equivalent to that in the condensed phase, as pointed out by Littlewood and Rideal (1956). They investigated various organic molecules at low temperatures and found that when the temperature difference across the interface was accounted for, the agreement between $\alpha_e$ and the ideal case ($\alpha_e$ = 1) improved significantly. Persad and Ward (2016), extend this argument, though they largely focused on liquids (water, silicone oil) that have high vapour pressures at low temperatures, as well as vapour species with large dipole moments (Langmuir 1932; Wyllie 1949). An alternative law based on statistical rate theory is proposed to better account for these complexities (Persad and Ward 2016), because they give rise to conditions that diverge from those under which the HKL equation should be applied by exacerbating temperature differences between the gas and condensed phase, decreasing the mean-free path and enhancing polar interactions between molecules (see also Bond and Struchtrup, 2004).

These departures from ideal conditions are mitigated in high-temperature geological applications, and variation in $\alpha_e$ appears to be smaller, typically between 0.05-1 (Wolff and Alcock 1962; Alcock and Peleg 1967; Krönert and Boehm 1972; Srivastava and Farber 1981; Fedkin et al. 2006; Richter et al. 2011; Shornikov 2015), with those for liquids being closer to unity (Burns 1966; Fedkin et al., 2006; Safarian and Engh, 2009) which may reflect low heat of vaporisation relative to thermal energies. Schaefer and Fegley (2004a) give a convenient summary of vaporisation coefficients for several minerals and melts in their Table 10. Nevertheless, further experimental testing of the HKL equation at high temperatures is required in order to better understand its applicability and the nature of evaporation coefficients. Therefore, although the HKL equation should only strictly be applied under equilibrium conditions to ideal gases, it can provide an accurate description of Langmuir evaporation processes in controlled experiments at high temperatures, provided the caveats mentioned above are considered.

### *2.2.2. Kinetic evaporation experimental techniques*

Thermodynamic data on equilibrium partial vapour pressures of elements evaporating from natural silicate melts or solids are very scarce (*section 2.1.6.*). Information garnered on metal volatility in multicomponent systems predominantly comes from kinetic, or Langmuir evaporation experiments. As opposed to the Knudsen Cell, kinetic evaporation experiments are typically conducted in alumina tube furnaces that are capable of reaching ≈ 1800°C with $MoSi_2$ heating elements in their ≈3-5 cm 'hot zone', in which a sample is suspended on a wire loop or crucible (Fig. 6).

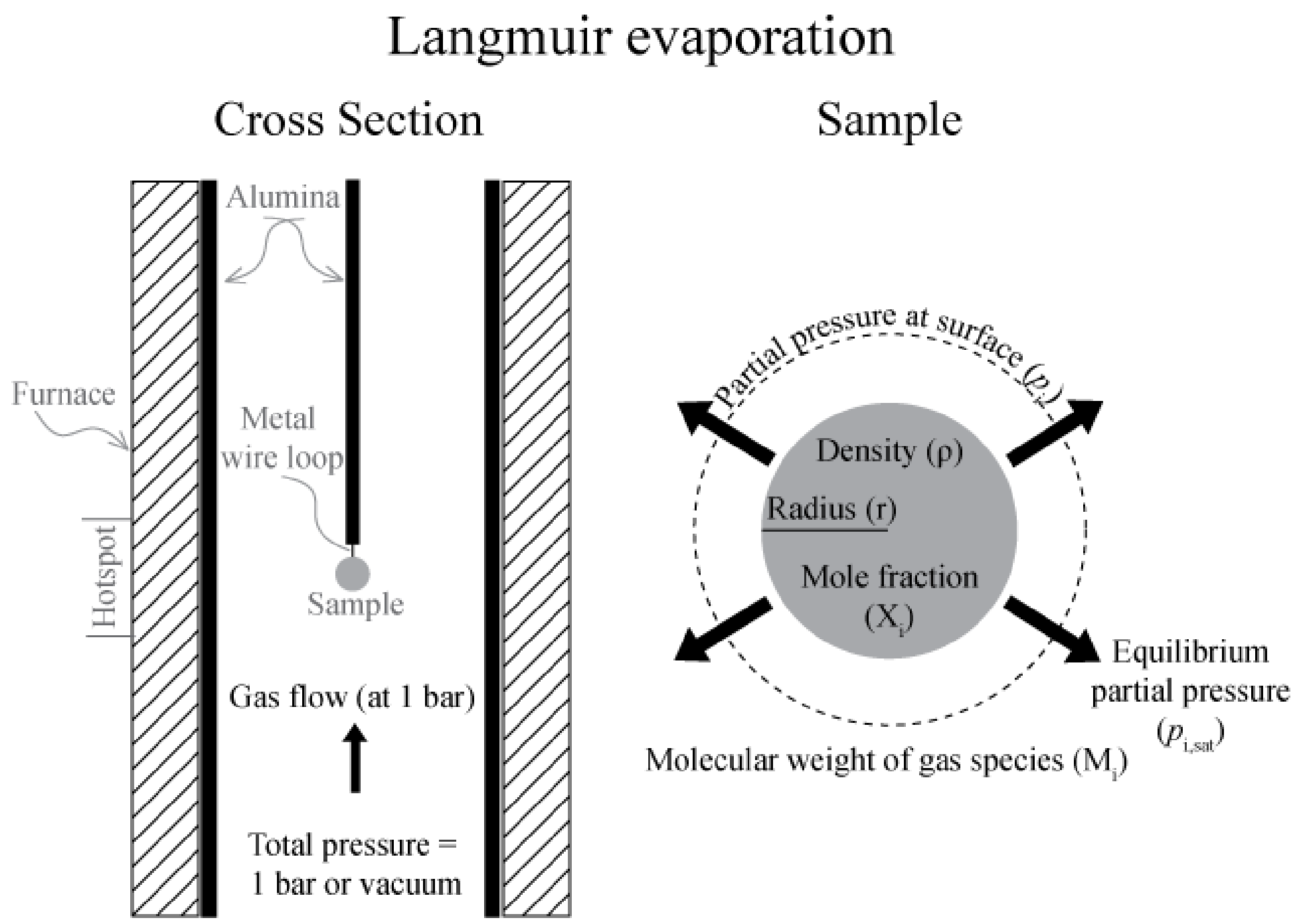

***Fig. 6.*** *Schematic illustration of an alumina tube furnace in which a sample is suspended from a metal wire loop or levitated and equilibrated at high temperatures with a gas flow or under vacuum. See Donaldson (1979); Tsuchiyama et al. (1981), Hashimoto (1983); Nagahara and Kushiro (1989); Richter et al. (2002); O'Neill and Eggins (2002); Pack et al. (2010).*

In these experiments, the temperature (≈±2°C) and pressure (1 bar) are fixed, but also the oxygen fugacity, by means of $CO/CO_2$ (*e.g.,* O'Neill, 2005) or $CO_2/H_2$ (*e.g.,* Donaldson, 1979) gas mixtures to better than ≈±0.1 log unit, depending on the mixing ratio (Huebner, 1987). Equally, experiments in this apparatus can also be conducted under vacuum (*e.g.,* Hashimoto 1983; Richter et al. 2002; Yu et al. 2003; Nagahara, this volume). Laser-heated aerodynamic levitation apparatuses, in which the atmosphere composition may also be controlled by changing the composition of the gases used to levitate the sample, are becoming increasingly employed for Langmuir evaporation experiments (Pack et al. 2010; Macris et al. 2016). This method has the advantage being able to heat samples in excess of 2000 °C *via* a $CO_2$ laser, temperatures that are amenable to the study of the vaporisation of refractory elements. The major difference is that these experiments take place in an open system, where the evolved gas is loses contact with the condensed phase, either from expansion into a vacuum or by a continual flux of gases at typical gas flow rates of 10 – 200 sccm through the furnace tube or chamber, leading to a strong dependence of the volatile content of the condensed phase with time (*e.g.,* Richter et al. 2002), see Equation (27). Furthermore, an open system means that the evolved gases are subsequently lost from the system. Therefore, vapour pressures of evaporating components must be calculated from the concentration of the element remaining in the condensed phase, by weight loss or by target collection, rather than measured as for effusion techniques.

Some of the kinetic evaporation experiments in the geochemical literature are similar to Langmuir evaporation experiments (e.g., Gibson and Hubbard 1972; Storey 1973; Donaldson 1979; Tsuchiyama et al. 1981; Hashimoto 1983), while others are similar to transpiration experiments (e.g., Jacobson et al. 2005). Langmuir vaporisation experiments for geochemically important metal oxides (e.g., $Al_2O_3$, $HfO_2$, MgO, REE sesquioxides, $SiO_2$, $ThO_2$, $ZrO_2$) are described in several papers (e.g., Wolff and Alcock 1962; Peleg and Alcock 1966; Alcock and Peleg 1967; Krönert and Boehm 1972). Sata et al. (1982) conducted both Langmuir vaporisation and transpiration experiments involving CaO. Their paper illustrates the utility of and differences between the two methods (their figures 1 & 3 and associated text). Given the uncertainty in evaporation coefficients (*section 2.2.1.*), data obtained from Knudsen effusion or other equilibrium techniques are generally preferred. As per the equilibrium experiments, the vast majority of studies in the geochemical literature deal primarily with the alkalis (Gibson and Hubbard 1972; Storey 1973; Donaldson 1979; Tsuchiyama et al. 1981; Kreutzberger et al. 1986; Tissandier et al. 1998) or the major silicate components (Hashimoto 1983; Floss et al. 1996; Richter et al. 2007).

*2.2.3. Major elements*

For a recent, comprehensive summary of the kinetic evaporation of major components (Fe, Mg, Si, Al, Ca) from silicate melts, the reader is referred to Davis and Richter (2014). In brief, as per the equilibrium case, Fe is shown to be the most volatile component of FCMAS melts (Hashimoto 1983;

(30)

Floss et al. 1996; Richter et al. 2011), in which FeO dissolved in a silicate liquid evaporates according to:

$$FeO_{(l)} = Fe_{(g)} + \frac{1}{2}O_2 \ .$$

However, work by several groups cited earlier (Table 1) shows that pure wüstite ($Fe_{1-x}O$), magnetite ($Fe_3O_4$), and hematite ($Fe_2O_3$) evaporate incongruently with lower Fe/O atomic ratios in the vapour than in the condensed phase. It is still uncertain whether or not pure molten FeO vaporises congruently; Darken & Gurry (1946) report the congruently vaporising composition has O/Fe = 1.116 but Smoes & Drowart (1984) consider that the congruently vaporising composition is stoichiometric FeO. Further work is needed to resolve this question.

In either case, Fe vaporisation and transport may result in subsequent oxidation of the silicate residue if the evolved oxygen remains behind. Initially, the next most volatile component is $SiO_2$, which volatilises as SiO, Equation (22). However, as, evaporation proceeds, the activity coefficient of $SiO_2$ decreases, such that the equilibrium partial pressure of $SiO_{(g)}$ falls below that of $Mg_{(g)}$ (Richter et al. 2002, 2007), whose vaporisation reaction occurs by dissociation of MgO to $Mg_{(g)}$ + ½$O_2$, as shown by Kazenas et al. (1983). Calcium oxide and $Al_2O_3$ behave conservatively over the range of temperatures investigated by Hashimoto (1983), up to 2000°C. Forsterite evaporation occurs congruently, whether in the liquid or solid state (Mysen and Kushiro 1988; Hashimoto 1990; Wang et al. 1999), however, this congruent vaporisation involves several gas species (dissociative), according to the reaction:

$$Mg_2SiO_4 = 2Mg_{(g)} + SiO_{(g)} + \frac{3}{2}O_2. \quad (31)$$

Forsterite-fayalite solid solutions were shown to evaporate incongruently by Nagahara and Ozawa (1994), who found the residual condensed olivine was Mg-richer post-vaporisation, resulting from preferential loss of Fe(g) relative to Mg(g), as confirmed by KEMS measurements of the vapour above $Fo_{95}$ (Costa et al. 2017). Plagioclase also evaporates incongruently, in which Na is more volatile than Ca leaving a more anorthitic residue (Nagahara and Kushiro, 1989). This behaviour is likely a general property of binary isomorphous solid-solutions with complete miscibility (i.e., cigar shaped phase diagrams). It is also possible that other melts will display azeotropic[5] behaviour upon evaporation (Heyrman et al 2004), though this has not been rigorously tested experimentally. See the chapter by Nagahara in this volume for further information on the evaporation of forsterite and of other minerals relevant to the solar nebula.

[5] An azeotrope is defined by the conditions at which the bulk composition of the vapour and the condensed phase is identical.

### *2.2.4. Alkali metals*

For the pure oxides, thermodynamic data (*e.g.*, Brewer 1953; Margrave 1967; Lamoreux and Hildenbrand 1984) indicates an order of alkali volatility (least volatile) Li < Na < K < Rb < Cs (most volatile). This sequence is not universally confirmed from evaporation of alkalis from more complex silicates with different bulk compositions, however.

Shimaoka and Nakamura (1989, 1991) and Shimaoka et al. (1994) find that, upon evaporation of fine-grained (<10 µm) chondrites, Na is more volatile than K and Rb, between 1200-1400°C at $8\times10^{-6}$ torr ($10^{-8}$ bar), with Rb evaporation marginally favoured relative to K at lower temperatures (1200°C). Although these experiments were performed sub-solidus, Donaldson (1979) also observed more rapid loss of Na than K from an alkali olivine basalt ($T_{liquidus}$ = 1232±2°C) between 1236°C and 1293°C and $\log fO_2$ between -3.3 and -11.9 (FMQ+4.5 to FMQ-4), as did Storey et al. (1973) from an Fe-rich quartz tholeiite between 1160 and 1210°C.

In a similar study, Kreutzberger et al. (1986), investigated alkali (Na, K, Rb, Cs) evaporation from An-Di melt at 1400°C, in which evaporation followed the sequence expected from the pure oxides, i.e., Na < K < Rb < Cs (Fig. 7). This hierarchy holds for evaporation in air, at the IW buffer (FMQ-3.6 at 1400°C) and under vacuum ($<10^{-6}$ bar). Furthermore, Gibson and Hubbard (1972), whose experiments on lunar basalt 12022 (20 wt. % FeO) at $3\times10^{-9}$ bar were conducted from 950°C to 1400°C, also show that $p$Rb > $p$K > $p$Na, where their volatilities become more dissimilar at lower temperatures. Hastie and Bonnell (1985) measured alkali vaporisation from NIST SRM glass and dolomitic limestone samples. They found $p$Na > $p$K for the Fe-free glass (their Fig 9), but $p$K > $p$Na for the low Fe limestone (their Fig 10).

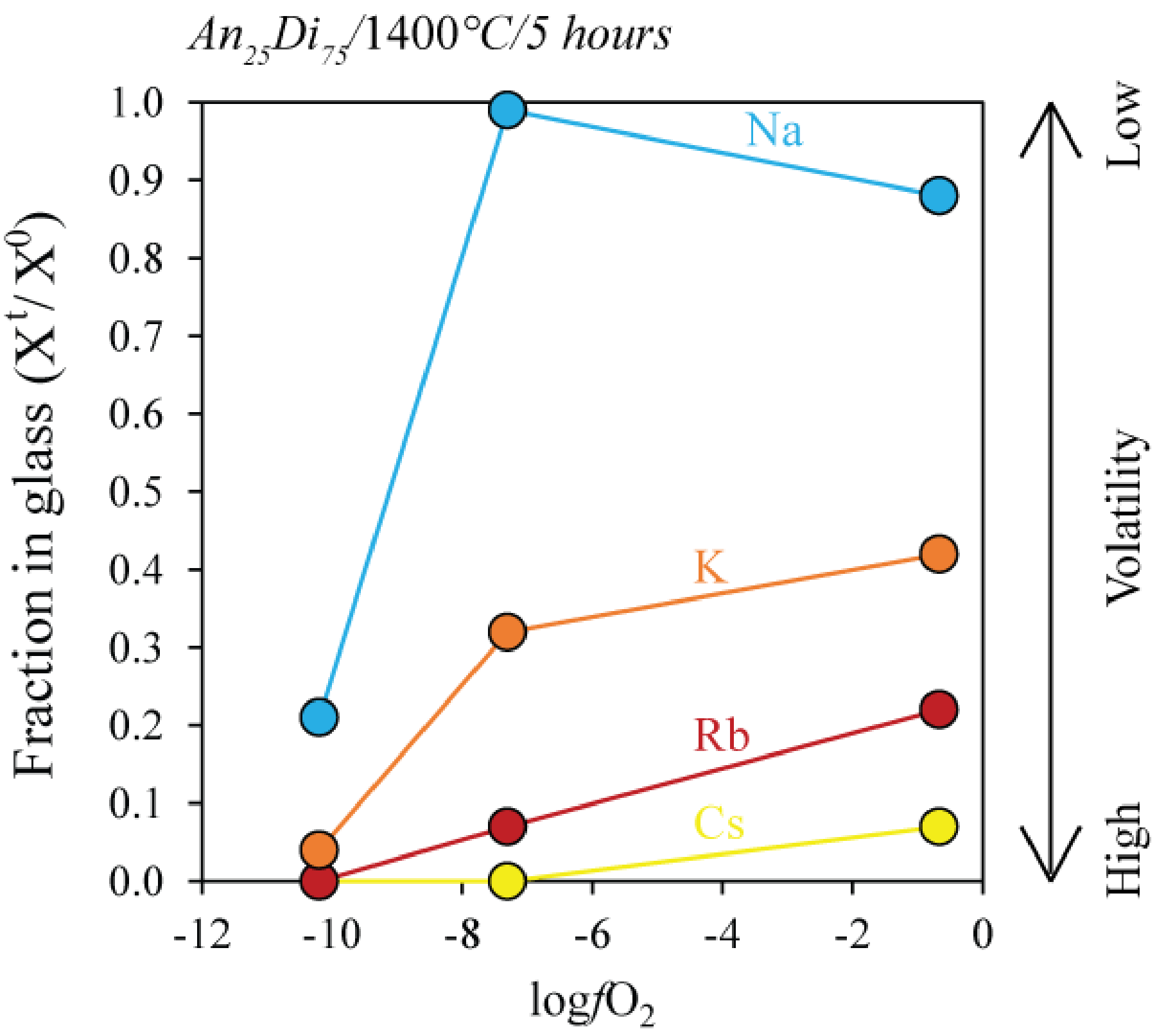

***Fig. 7.*** *The fraction remaining of a trace element after 5 hours with respect to its initial amount ($X^t/X^0$) in an $An_{25}Di_{75}$ melt as a function of log$fO_2$ after heating at 1400°C for 5 hours (Kreutzberger et al. 1986). The vacuum experiment is predicted to define a log$fO_2$ = -7.3 using the trend of log$fO_2$ with temperature for evaporation in a vacuum from Fig. 4.*

The changing order of volatility among alkalis evaporating from silicate relative to their pure oxides led Donaldson (1979) to postulate that differences in $NaO_{0.5}$ and $KO_{0.5}$ activity coefficients (*i.e.*, $\frac{\gamma_{NaO_{0.5}}}{\gamma_{KO_{0.5}}} > 1$) must act to enhance Na volatility relative to K in order to explain the data. Indeed, this behaviour occurs in binary $M_2O$-$SiO_2$ (Charles 1967; Wu et al. 1993), ternary $Na_2O$-$K_2O$-$SiO_2$ melts (Chastel et al. 1987) and multicomponent $M_2O$-bearing melts (Hastie et al., 1981; Borisov 2009; Mathieu et al. 2011). The key point from these studies is that the depression of $\gamma MO_{0.5}$ when combined with $SiO_2$- or $Al_2O_3$ + $SiO_2$-bearing melts is inversely proportional to their volatilities as pure oxides. Alkali aluminates are very stable compounds, and increasing $XAlO_{1.5}$ in the liquid causes negative deviations from ideality in the order Na < K < Rb (O'Neill 2005; Borisov 2009; Yazhenskikh et al. 2006, 2011). As illustrated by the KEMS data of De Maria et al. (1971), this does also appear to be the case in natural compositions, with Na becoming more volatile than K above 1200°C (Fig. 3). At present only limited data are available, and it therefore remains unclear as to what controls the alkali element volatility hierarchy in silicate melts and minerals. Solutions to this question require a better understanding of alkali metal activity coefficients in silicate melts as a function of *P*, *T* and *X*.

*2.2.5. Other trace elements*

Systematic studies of the evaporation behaviour of trace elements from silicate materials are exceedingly sparse, restricted to the studies of Notsu et al. (1978), Masuda and Tanaka (1979), and Wulf et al. (1995), joined recently by that of Sossi et al. (2016) and Norris and Wood (2017). Other works are generally restricted to unrefereed abstracts and include Shaffer et al. (1991); Ertel-Ingrisch and Dingwell (2010); Ertel-Ingrisch et al. (2012); Speelmans (2014), and Humayun and Crowther (2015).

Early studies on the kinetic vaporisation of natural rocks dealt predominantly with meteorites. A series of studies on heating of chondritic meteorites were published, mainly in the 1970s, by the group of M.E. Lipschutz (*e.g.,* Ikramuddin et al. 1976; Matza and Lipschutz 1977; Bart et al. 1980). These experiments provide information on the volatility sequence of trace elements during meteorite metamorphism. Although they were initially carried out in $10^{-5}$ atm $H_2$ to replicate nebular conditions, their interpretation is not straightforward for several reasons. Namely, due to the poorly constrained fugacities for oxygen, sulfur, and halogens; the relative importance of diffusion versus volatility, and uncertainty regarding the extent of equilibration during stepwise heating (Wulf et al. 1995; Schaefer and Fegley 2010). Furthermore, Wulf et al. (1995) in attempting to replicate these experiments, found that significant quantities of Na, K and Ga had entered the quartz tube containing the samples following heating. Nevertheless, experiments of this type are instructive for discerning between element mobility by diffusion and volatility, and reasonable agreement is found between volatilities and the most labile elements in their experiments. For example, In, Tl, Bi and Cd are invariably the most depleted (*e.g.,* Bart et al. 1980) and are among the most volatile in nebular environments (Lodders, 2003; Schaefer and Fegley, 2010), broadly matching their abundances in metamorphosed chondrites (Wang and Lipschutz, 2005).

Notsu et al. (1978) carried out DC arc heating experiments on the Allende CV3 chondrite liquid at 2000°C in air, as well as on a basalt (JB-1) by Mo resistance heating at 0.01 bar, and measured the residual composition after Langmuir evaporation. Masuda and Tanaka (1979) performed heating experiments in a vacuum ($\approx 5\times10^{-8}$ bar) on JB-1 heated to ≈2000°C by a W filament and measured the concentration of each element precipitated on a condensation plate. In both studies, Al, Ca, Ti, Sc and the REE behave in a refractory manner, becoming enriched in vaporisation residues. Masuda and Tanaka (1979) note that La, and particularly Eu and Yb are relatively more volatile than their neighbouring REEs, consistent with Boynton (1975). Vanadium is roughly constant, while Cr, Fe and Mn are originally refractory before decreasing in abundance beyond 20% vaporisation, being reduced to negligible levels at 90%. Magnesium only decreases in concentration after 80% vaporisation of the starting material, and is initially less volatile than $SiO_2$ but is depleted more rapidly thereafter, as also observed by Richter (2002). Sodium is the most volatile metal studied by Notsu et al. (1978), though K is similarly volatile and Rb more so in Masuda and Tanaka (1979), while Ni, Co, Ir and Au are also all quantitatively lost within 40-50% vaporisation (Notsu et al. 1978). The volatility of these ordinarily non-volatile siderophile metals comes about due to the oxidising conditions under which the experiments were performed (air). Wänke et al. (1984) also observe that the volatility of refractory siderophile elements Re, Os, W and Mo, in addition to halogens and chalcogens are enhanced in an oxidising, $H_2O$-steam atmosphere relative to a more reducing $N_2$-rich atmosphere, whereas Zn and In show the opposite behaviour. The systematics of element volatility as a function of the atmosphere under which they evaporate is expanded upon below.

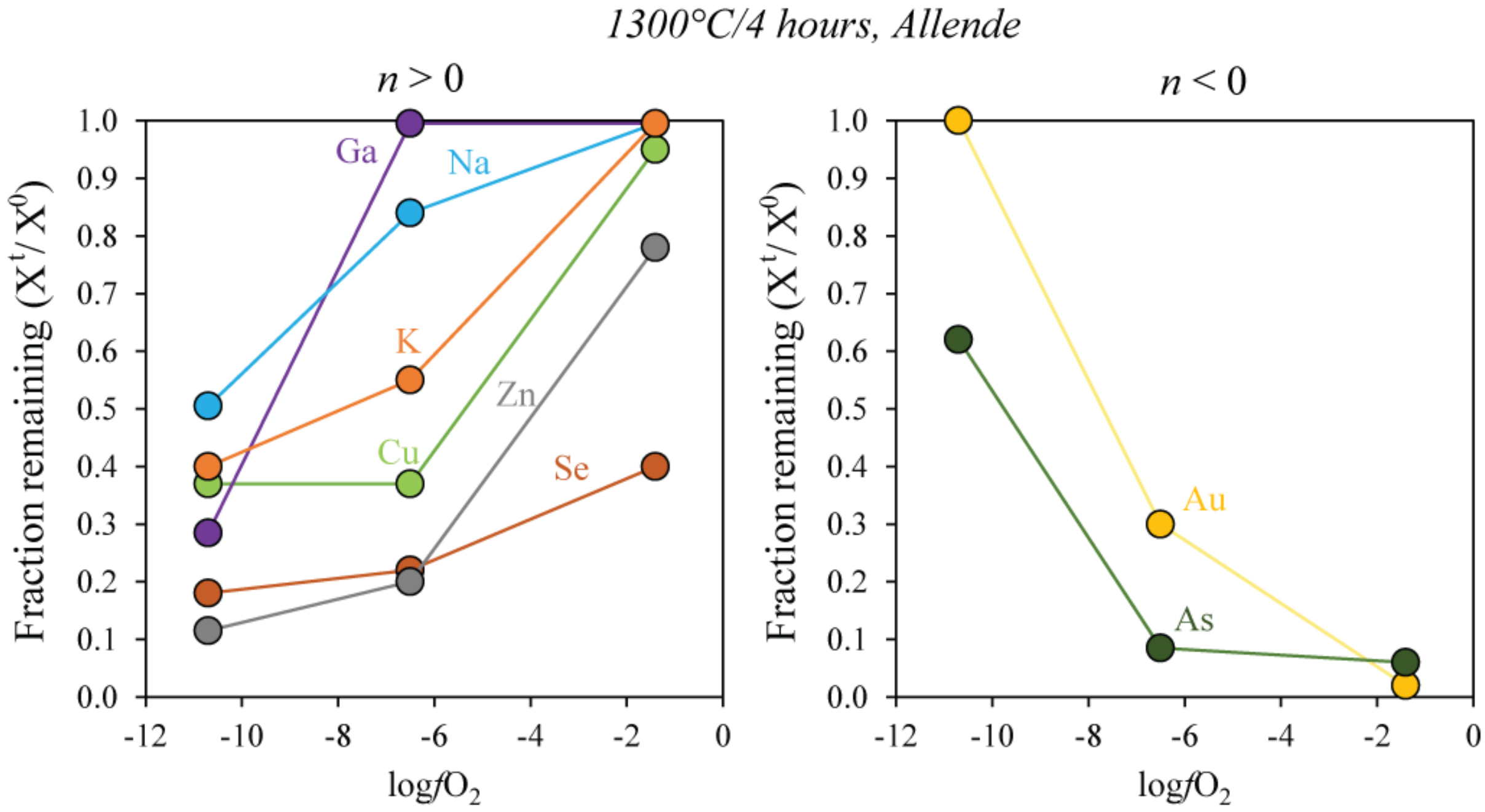

***Fig. 8.*** *The fraction remaining of a trace element after 4 hours with respect to its initial amount ($X^t/X^0$) in bulk Allende chunks as a function of log$fO_2$ after heating at 1300°C for 4 hours (Wulf et al., 1995). These conditions were chosen as equilibrium is more likely to have been reached in these higher-temperature experiments.*

Wulf et al. (1995) studied the loss of trace elements from two natural carbonaceous chondrites (Murchison and Allende) over a range of oxygen fugacities (log$fO_2$ = -16.5 to air) and temperatures (1050-1300°C; always sub-liquidus) at atmospheric pressure (Fig. 8). Wulf et al. (1995) did their work at sub-liquidus temperatures and suggested elemental volatility was controlled by a combination of equilibrium partial pressures and stability of the MVE host phases. Therefore, for trace elements locked in a phase whose decomposition depends on the thermodynamics of other major elements, these experiments give only qualitative insight into their volatility, as discussed for platinum-group metals in alloys trapped in silicates. The experiments of Norris and Wood (2017) were undertaken on a natural MORB liquid at a single temperature, 1300°C, with varying log$fO_2$ from -7 to -13 (FMQ+0.3 to IW-2.3), using a Ni stirrer to mitigate the potential for incomplete diffusive equilibration of the liquid phase during evaporation (see also Dingwell et al. 1994). Volatile loss of different elements in their experiments is not systematic as a function of time, and they deduced empirical 'volatility factors' for a constant run time (Fig. 9).

In both studies, the $fO_2$ of the gas (and silicate melt in Norris and Wood, 2017) was externally buffered by a continuous flow of $CO\text{-}CO_2$ gas mixture inside the furnace tube, a situation that diverges from natural systems, in which the $fO_2$ of the gas is defined by the partial pressures of the evolved gas species from the heated material. The telling result from Wulf et al (1995) and Norris and Wood (2017) is this: elemental volatility depends strongly on the $fO_2$ of the gas into which elements are evaporating (Fig. 8a, b; 9). This dependence is not equivalent for all elements, nor would we expect it to be so from knowledge of vaporisation stoichiometry of the pure oxides (see Table 1). For example, certain elements become *more* volatile at oxidising conditions, whereas for others, reducing conditions promote evaporation. The former group are comprised of largely highly siderophile noble metals that exist in metallic form in chondrites, notably Au, As, Re, and Os (Fig. 8b). By contrast, lithophile and chalcophile elements, such as the alkalis, Cu, Zn, Sn, Sb, Pb and Ga are more readily lost at low $fO_2$ (Fig. 8a; Fig. 9). Fegley and Cameron (1987) discussed this type of behaviour for Ce (vs. other REE) and U & Pu (vs Th).

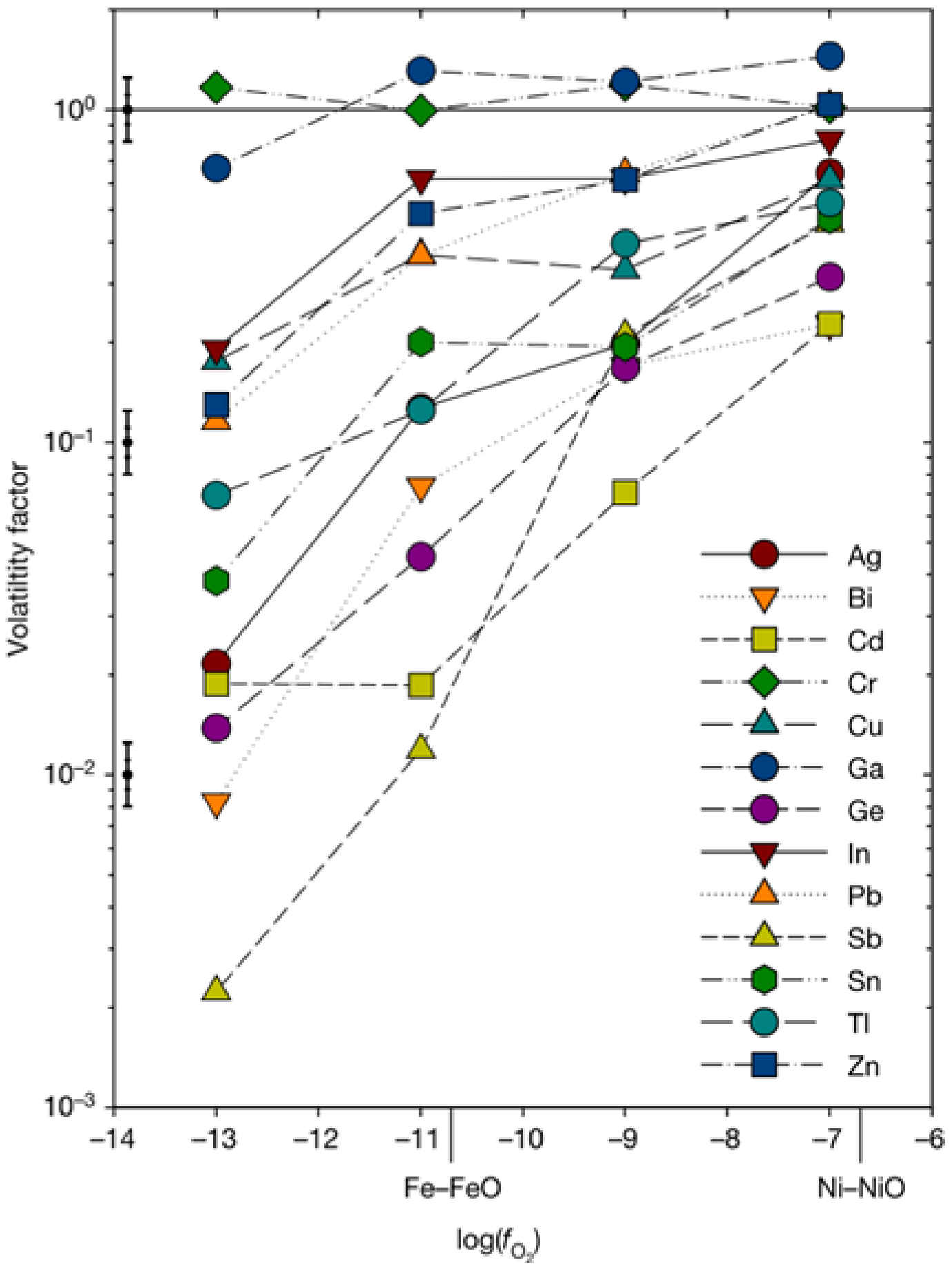

***Fig. 9.*** *The volatility factor (fraction of the element remaining the glass) for selected moderately volatile elements as a function of* $fO_2$ *at 1300°C and 60 minutes. Reproduced from Norris and Wood (2017).*

This contrasting behaviour may be understood by the reaction shown in Equation (2). For the first group of elements, *n*<0, meaning the element in the gas phase is more oxidised than in the condensed phase. Chalcophile and particularly lithophile elements mostly exist as oxides in natural rocks, and their gas species tend to be more reduced, yielding *n*>0. For many moderately volatile elements, such as Na, K, Rb, Zn, Cu, Ge, Sn, Cd, Ag, Mn, Fe, Mg, Si a single (though different) metal-bearing species predominates in both phases over a wide range of temperature and $fO_2$ (*e.g.*, Lamoreaux et al., 1987; Table 1). Some elements may exist in multiple oxidation states, in the condensed and/or the gas phase, such that several equilibria akin to Equation (2) must be written to describe the overall volatility behaviour of an element.

In this way, Sossi et al. (2016) showed that the vapour pressures of the evaporating species for a given element may be calculated from the concentration of that element remaining in the residual condensed phase (in their case, a liquid), $X(M^{x+n}\frac{x+n}{2}O)$. Re-arranging, the equilibrium constant, Equation (4) to solve for the partial pressure of the metal species yields:

$$p\left(M^{x}O_{\frac{x}{2}}\right) = \frac{K_{(2)}X\left(M^{x+n}O_{\frac{x+n}{2}}\right)\gamma\left(M^{x+n}O_{\frac{x+n}{2}}\right)}{f(O_2)^{n/4}}. \tag{32}$$

Substituting Equation (32) into Equation (29) allows it to be integrated with respect to time and concentration to yield the relative loss of the element from the sphere over a given time interval:

$$\frac{X_i^t}{X_i^0} = \exp\left(-\frac{\alpha_e\gamma\left(M^{x+n}O_{\frac{x+n}{2}}\right)K_{(2)}}{f(O_2)^{n/4}}\frac{3}{r\rho}\sqrt{\frac{M_i}{2\pi RT}}(t-t_0)\right). \tag{33}$$

Where $t_0$ is the time at which the element begins evaporating, and $X_i^t$ and $X_i^0$ its concentration at $t$ and $t_0$, respectively (Sossi et al. 2016). Tsuchiyama et al. (1981) and Humayun and Koeberl (2004) derived an expression with a similar functional form. This analysis requires implicit knowledge of the speciation of the element in at least one of the condensed- or the gas phase, and hence cannot uniquely identify its speciation. Rather, speciation must be inferred from the dependence of $\frac{X_i^t}{X_i^0}$ with $fO_2$. This analysis becomes complicated if volatiles such as water, halogens, or sulfur (important for MVE speciation at higher pressures; see section 3.0) and is thus applied only to anhydrous materials herein. As neither $\alpha_e$ nor the activity coefficient of the oxide species are known *a priori*, a new equilibrium constant, $K^*$ is defined, where $K^* = \alpha_e\gamma_i K$. It is clear that the loss of an element by kinetic evaporation still strongly depends on its equilibrium partial pressure, $p_{i,sat}$, in addition to other, non-equilibrium (or 'kinetic') factors, most notably the radius and density of the evaporating body, the time for which it is evaporating, and the square root of the molar mass of the evaporating species. To a first approximation (i.e., no concentration gradients in the evaporating material, and temperature–independent enthalpy of vaporisation and activation energy), kinetic evaporative loss depends on the temperature in an Arrhenian fashion, and hence linear arrays are anticipated in plots of $ln\frac{X_i^t}{X_i^0}$ vs. $t$. In detail, the rate constants of evaporation reactions are not exactly constant with 1/T (Benson 1960), but, predominantly, because $p_{i,sat}$ decreases as the abundance of the element in the condensed phase ($X_i^t$) decreases, evaporation rate slows with time, and linearity over time is never achieved.

In a plot of $\frac{X_i^t}{X_i^0}$ vs. log$f$O$_2$ for a given time, the data define a sigmoidal curve, where the range of log$f$O$_2$ over which the element passes from fully condensed to fully vaporised is proportional to the number of electrons ($n$) in the reaction. The more $n$ diverges from 0, the sharper this transition is. Therefore, in this way, the speciation of the element in the gas phase can be inferred from the slope of the sigmoid. An example for the case of Cu and Zn evaporation is shown below (Fig. 10).

The volatility of Cu is, at low oxygen fugacity, lower than that of Zn (compare the amount of each remaining in the glass at log$f$O$_2$ = -8, Fig. 10a, b). However, as oxygen fugacity increases, Cu becomes more volatile relative to Zn, such that, in air, the proportion of Cu and Zn remaining in the glass is equivalent. Using Equation (33) to fit the experimental data shows that Cu evaporation is consistent with an $n$ = 1 transition, whereas Zn conforms to an $n$ = 2 reaction. Given knowledge of their respective speciation in silicate melts, their volatilisation reactions are:

$$ZnO_{(l)} = Zn_{(g)} + \frac{1}{2}O_2 \quad (34)$$

and for Cu,

$$CuO_{0.5\,(l)} = Cu_{(g)} + \frac{1}{4}O_2 \quad (35)$$

The different coloured points and curves correspond to different temperatures at a given run time, and illustrate that higher temperatures result in more extensive volatile loss. This effect may be quantified by a change in the log$K$* value, which shifts the curves to the left (lower temperature) or right (higher temperature) along the log$f$O$_2$ axis. Higher values correspond to greater stability of the products (*i.e.*, the volatile species). The stoichiometry of ZnO and CuO$_{0.5}$ vaporisation from silicate melts deduced from this method is consistent with those in Table 1.

The log$K$* for each volatilisation reaction used to fit the data can be converted into an effective Gibbs Free Energy, ΔG* using Equation (9). Sossi et al. (2016) performed their experiments between 1300°C and 1550°C. For their data, the variation of $\Delta G^*$ of a volatilisation reaction against the absolute temperature, $T$ in an Arrhenius plot, results in a straight line (Fig. 11). This confirms that a reproducible, thermally-activated process is controlling elemental loss, namely, their equilibrium partial pressures.

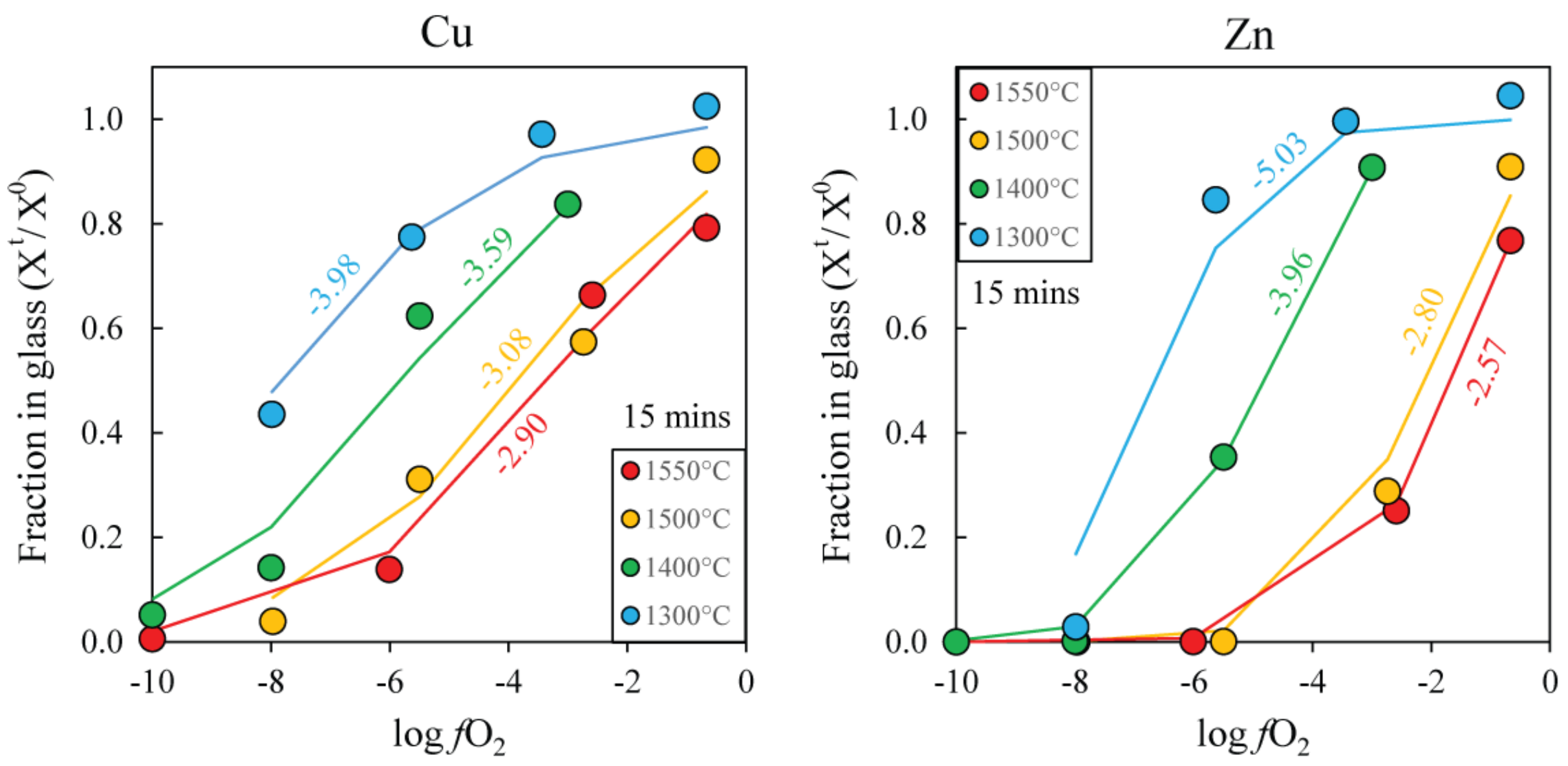

***Fig. 10.*** *The fraction remaining in the melt ($X^t/X^0$) at t = 15 minutes, where X = a) Cu and b) Zn as a function of log$fO_2$. Colours correspond to the temperature of the experiment (blue = 1300°C; green = 1400°C; yellow = 1500°C and red = 1550°C). Curves are calculated using Equation (33) and the numbers are their associated log$K^*_{(r)}$. Data from Sossi et al. (2016).*

At constant total pressure, the Gibbs – Helmoltz equation is given by Equation (10), and hence, the slope of the line is equivalent to $-\Delta S^*$ and its intercept, $\Delta H^*$, which are average values determined for the mid-point temperature over the temperature range. This is a 2nd law treatment of the vaporisation data and is analogous to the integration of the Clausius – Clapeyron Equation generally used for interpretation of vapour pressure data (Drowart and Goldfinger 1967; L'vov 2007). It implicitly assumes that $\Delta H^*$ is much greater than the product of $\Delta C°_P$ and temperature over the interval of interest, see Equation (11), such that slopes can be approximated as linear within experimental uncertainty.

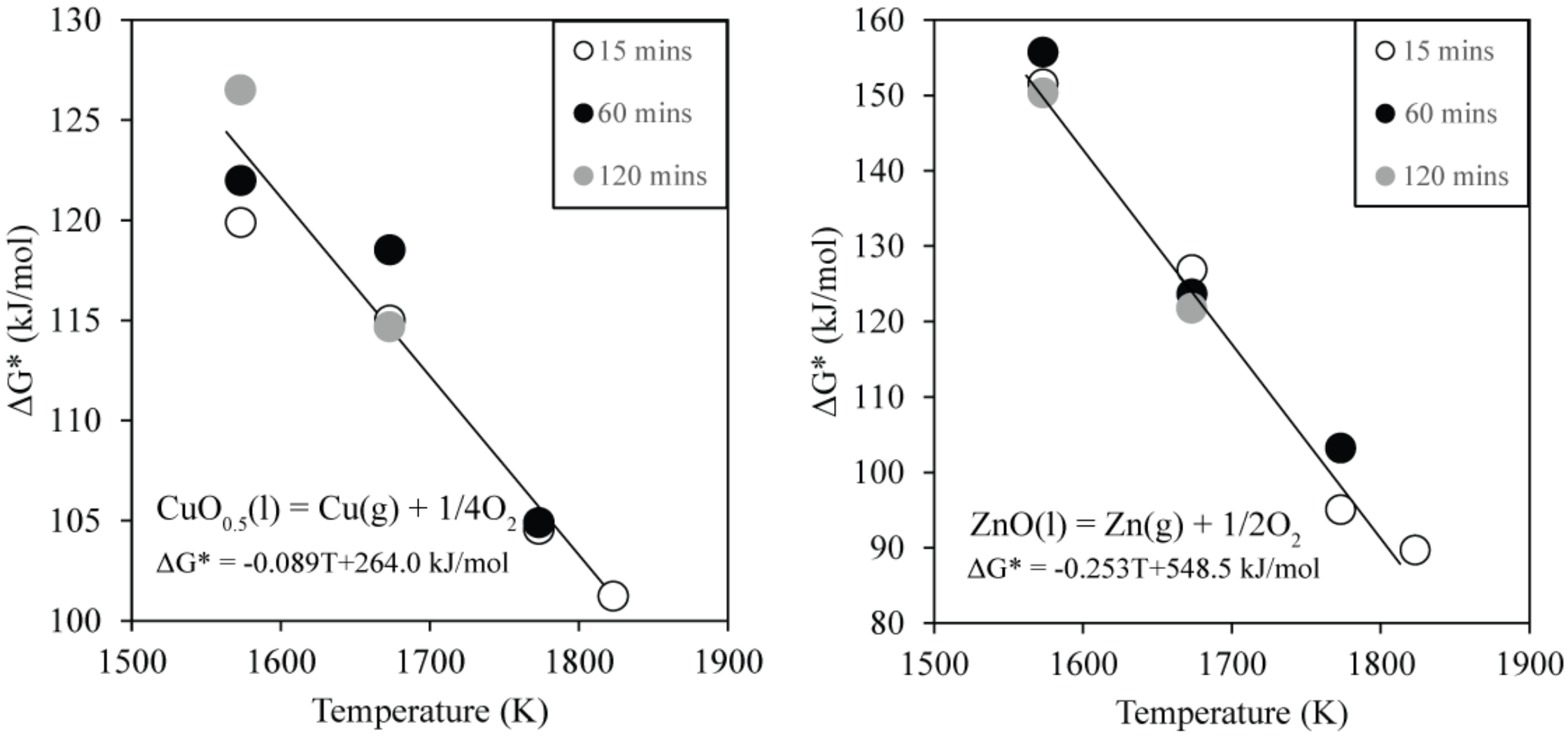

***Fig. 11.*** *The calculated $\Delta G^*_{(r)}$ for (a) Cu and (b) Zn as a function of temperature for given run times (white = 15 minutes; black = 60 minutes; grey = 120 minutes). Data from Sossi et al. (2016).*

Relative volatilities and evaporation temperatures can be calculated from these thermodynamic data; including the activity coefficients of melt oxide species in the silicate melt (Sossi et al. 2016).

Table 3 qualitatively illustrates the differences in the volatility behaviour of the elements calculated for the solar nebula (Lodders, 2003) and observed in Langmuir evaporation experiments of molten basalt (Sossi et al., 2016; Norris and Wood, 2017). Changes in relative element volatility reflect the distinct thermodynamic conditions that typify the two environments; the high pressures (1 bar; though note that Equation (33) is applicable to *any* total pressure, it is the partial pressure of the individual gas species that matter) compared to $10^{-4}$ bar assumed for the solar nebula calculations (see also section 3.0.), in addition to the higher relative $fO_2$ (IW-7 and IW, respectively) and different bulk composition. Low total pressures in the solar nebula promote the formation of solid minerals relative to liquids (Ebel 2004), with many moderately volatile elements condensing into FeS (*e.g.,* Ag, Cd, In, Pb, Zn) or Fe metal (Bi, Ga, Ge, Cu, Sb), whereas the alkalis partition into feldspar aside from Li which, like Mn, condenses into forsterite and enstatite (Lodders 2003; Schaefer and Fegley 2010). By contrast, these elements are probably all dissolved as oxide components in silicate melts. As a result, their activity coefficients will differ from those in solid phases, especially if those solids are metals or sulfides. This phenomenon can explain why In becomes less volatile, and Cu more so in the experiments of Sossi et al. (2016) and Norris and Wood (2017); the $\gamma InO_{1.5}$ in FeO-CaO-MgO-$Al_2O_3$-$SiO_2$ melts is 0.02, while $\gamma CuO_{0.5}$ is 3.5 at 1650°C (see Wood and Wade, 2013 for a discussion). A similar interpretation holds for the decreased volatilities of the alkalis relative to other elements such as Cu and Ag, as the alkalis have very low activity coefficients that stabilise them in silicate melts relative to the gas phase (see Table 2, section 2.2.4.). Additionally, the speciation of the element in the gas may be affected by the more oxidising conditions of the Langmuir evaporation experiments; notably, the Group VI metals Cr, Mo, and W form stable oxide species that enhance their volatilities at oxidising conditions (Fegley and Palme 1985; Sossi et al. 2016; Table 1) and become even more volatile at $fO_2$s > IW. Some other elements are sensitive to the presence of volatiles, such as S, that stabilise $GeS_{(g)}$ and $SnS_{(g)}$ in the solar nebula, but, during evaporation of sulfur-free basaltic magma, are constrained to form oxides (see section 3.0. for a discussion).

**Table 3**. Qualitative order of element volatilities, from least to most, in the solar nebula (IW-7; $10^{-4}$ bar) and from evaporation of silicate melts at 1573 K near the IW buffer at 1 bar.

| | |
|---|---|
| Solar nebula (Lodders, 2003) | Mo < Cr < Mn ≈ Li < Cu < K ≈ Ag < Sb ≈ Ga ≈ Na < Ge < Rb < Bi ≈ Pb ≈ Zn ≈ Sn < Cd < In < Tl |
| Evaporation of ferrobasalt (Sossi et al. 2016) | Mn < Mo < Cr < Li < Ga ≈ Na < K < Cu ≈ Rb < Zn ≈ Ge < Pb < Ag < Cd |
| Evaporation of MORB (Norris and Wood, 2017) | Cr ≈ Ga < In < Zn < Cu ≈ Pb < Sn < Tl ≈ Ag < Bi < Ge < Cd < Sb |

It is emphasised that this behaviour is relevant to volatile-poor systems that lack major gas species found in many terrestrial (and extra-terrestrial) magmatic systems, notably $H_2O$, $SO_2$, $CO_2$, CO, $CH_4$, $H_2S$, in addition to halides such as Cl and F. These compounds can preferentially complex metallic elements, thereby stabilising them in the gas phase relative to the nominally 'dry' case (Symonds and Reed 1993; Matousek 1998; Meschter et al. 2013; Fegley et al. 2016; Renggli et al. 2017). Although these volatiles are readily degassed or lost during high temperature vaporisation processes, such that they may be nearly absent at the temperatures sufficient to vaporise metallic elements, they were also probably extant on primordial planetary atmospheres (section 3.0.). This thermodynamic framework for interactions between gas and condensed phases is then reconciled to provide information as to how planetary bodies acquired (or then subsequently lost) their volatile element budgets during their formation (section 4.0.).

## 3.0. Evaporation in the presence of major volatiles H, C, S and halogens

### *3.1. Overview*

Very few natural materials are truly anhydrous and completely free of other volatiles – even the Moon contains halogens, sulfur, and at least trace water (Tartèse et al. 2013; Hauri et al. 2015; McCubbin et al. 2015) – so the relative importance of $fO_2$ versus the fugacities of these other volatiles for evaporation of the MVE requires consideration.

Henley and Seward discuss volcanic gas chemistry in their chapter in this volume and it is not our purpose to do so here. But to first approximation the observed composition ($H_2O$, $CO_2$, $SO_2$, $H_2S$, $H_2$, CO, HCl, HF) of terrestrial volcanic gases (e.g., Table 6.13 in Lodders and Fegley, 1998; Tables 3-5 in Symonds et al. 1994) is similar to that of BSE – steam atmosphere models for the early Earth (e.g., Figure 15 of Fegley et al. 2016). Thus we use observations and calculations of volcanic gas chemistry to help illustrate important points about the factors controlling volatility of the MVE on the early Earth and its precursor planetesimals.

### *3.2. Cu-bearing gases*

Taking copper as an example, band spectra show CuCl gas in volcanic flames at Kilauea (Murata 1960), and Nyiragongo (Delsemme 1960). These observations agree with equilibrium calculations showing CuCl is the major Cu-bearing species in terrestrial volcanic gases from several different volcanoes (e.g., Symonds et al., 1987 - Merapi; Quisefit et al., 1989 - Momotombo; Symonds and Reed 1993 – Mount St Helens; Brackett et al., 1995; Mather et al., 2012 – Kilauea; Churakov et al 2000, Henley and Seward 2018 – Kudryavy; Zelenski et al., 2013; Renggli et al., 2017 – Erta Ale), volcanic gas with the same composition as the lower atmosphere of Venus (Schaefer and Fegley 2004b), laboratory experiments (Ammann et al. 1993) and primordial lunar volcanic gases with different assumed compositions, pressure, and temperature (Fegley 1991; Renggli et al. 2017).

The formation of CuCl gas from monatomic Cu gas occurs via

$$Cu\ (g) + ½\ Cl_2\ (g) = CuCl\ (g). \tag{36}$$

The equilibrium constant for this reaction, assuming ideality of all gases is

$$K_{(36)} = \frac{p_{CuCl}}{p_{Cu}} \frac{1}{f_{Cl_2}^{1/2}}. \tag{37}$$

One could write this reaction in terms of Cu(g) and HCl(g), which is the major Cl species observed in terrestrial volcanic gases, but there is no guarantee that HCl would be the major Cl-bearing gas on a heated planetary body undergoing outgassing and/or evaporation reactions. For example, as shown in Table 3, the major Cl-bearing gas evolved from H-chondritic material during heating at low pressure is NaCl, not HCl (Schaefer and Fegley 2010). Also, the major Cl-bearing gas changes from HCl to NaCl (at P ~ 3 bar) to Cl (at P ~ $2\times10^{-5}$ bar) at 2000 K and otherwise constant conditions for the BSE (bulk silicate earth) – steam atmosphere model of Fegley et al. (2016). Thus we use the $Cl_2$ fugacity, which is a fundamental variable computed iteratively from the partial pressure sum for all Cl-bearing gases (see below).

However, our use of the $Cl_2$ fugacity as a fundamental variable does not necessarily mean that $Cl_2$ is the major Cl-bearing gas. This is unlikely unless either (1) volcanic HCl is oxidised by the $O_2$ in air to form $Cl_2$ – a reaction known as the Deacon process (Lewis 1906) or (2) chlorine is much more abundant than the elements that usually bond with it – H, the alkalis, and sulfur. Zelenski and Taran (2012) studied the Deacon process and showed it explains the $Cl_2$ emitted in air-rich fumarolic gas at Tolbachik in Kamchatka. Fegley and Zolotov (2000) explored the latter case where Cl/H, Cl/alkali, and Cl/S molar ratios are greater than unity in their calculations of alkali and halogen chemistry in volcanic gases on Io, the volcanically active satellite of Jupiter. However their preferred models – based on elemental abundances deduced from the Io plasma torus and other constraints – show chlorine is emitted as NaCl and KCl in Ionian volcanic gases. Lellouch et al. (2003) subsequently discovered NaCl gas in Io's atmosphere and concluded it was volcanic in origin. Moullet et al. (2013) tentatively observed KCl gas on Io and concluded it was also compatible with a volcanic origin.

Returning to the CuCl formation reaction above, we can rearrange the equilibrium constant expression to show that the CuCl/Cu molar ratio (X denotes mole fraction in the gas) is proportional to the square root of total pressure at otherwise constant conditions

$$\frac{X_{CuCl}}{X_{Cu}} = K_{(36)}\sqrt{X_{Cl_2}}\sqrt{P_T} \tag{38}$$

Thus higher pressure favors CuCl (g) while lower pressure favours $Cu^0$ gas. The tabulated thermodynamic data for gaseous Cu and CuCl (e.g., JANAF or IVTAN) show $K_{eq}$ for the Cu – CuCl reaction decreases (i.e., the ΔG° of reaction is less negative) with increasing temperature; hence at otherwise constant conditions Cu is favoured at higher temperature. Figure 12 shows the pressure dependence of the CuCl/Cu ratio at 2000 K for the BSE – steam atmosphere model of Fegley et al (2016). This model uses a $CuO_{0.5}$ activity coefficient of 3.5 based on Wood and Wade (2013) and Holzheid and Lodders (2001); Fegley et al (2016) describe other details of the calculations, which consider chemical equilibrium in the magma, between gas and magma, and in the gas phase. Figure 12 shows that the 1:1 ratio occurs at ~ 1.25 bar with Cu dominant at lower pressures and CuCl dominant at higher pressures. The position of the 1:1 ratio will be different for this steam atmosphere model at different temperatures. It will also be different in gas mixtures of different composition, e.g., terrestrial volcanic gases, or the outgassed atmosphere on a heated planetesimal, because the Cl-bearing gas speciation and thus the $Cl_2$ fugacity will be different. But the general behaviour – CuCl favoured at higher P and lower T and Cu favoured at lower P and higher T will be the same. Thus we expect Cu volatility at lower pressures/high temperatures is mainly due to its evaporative loss as $Cu^0$ while Cu volatility at higher pressures/low temperatures involves CuCl gas.

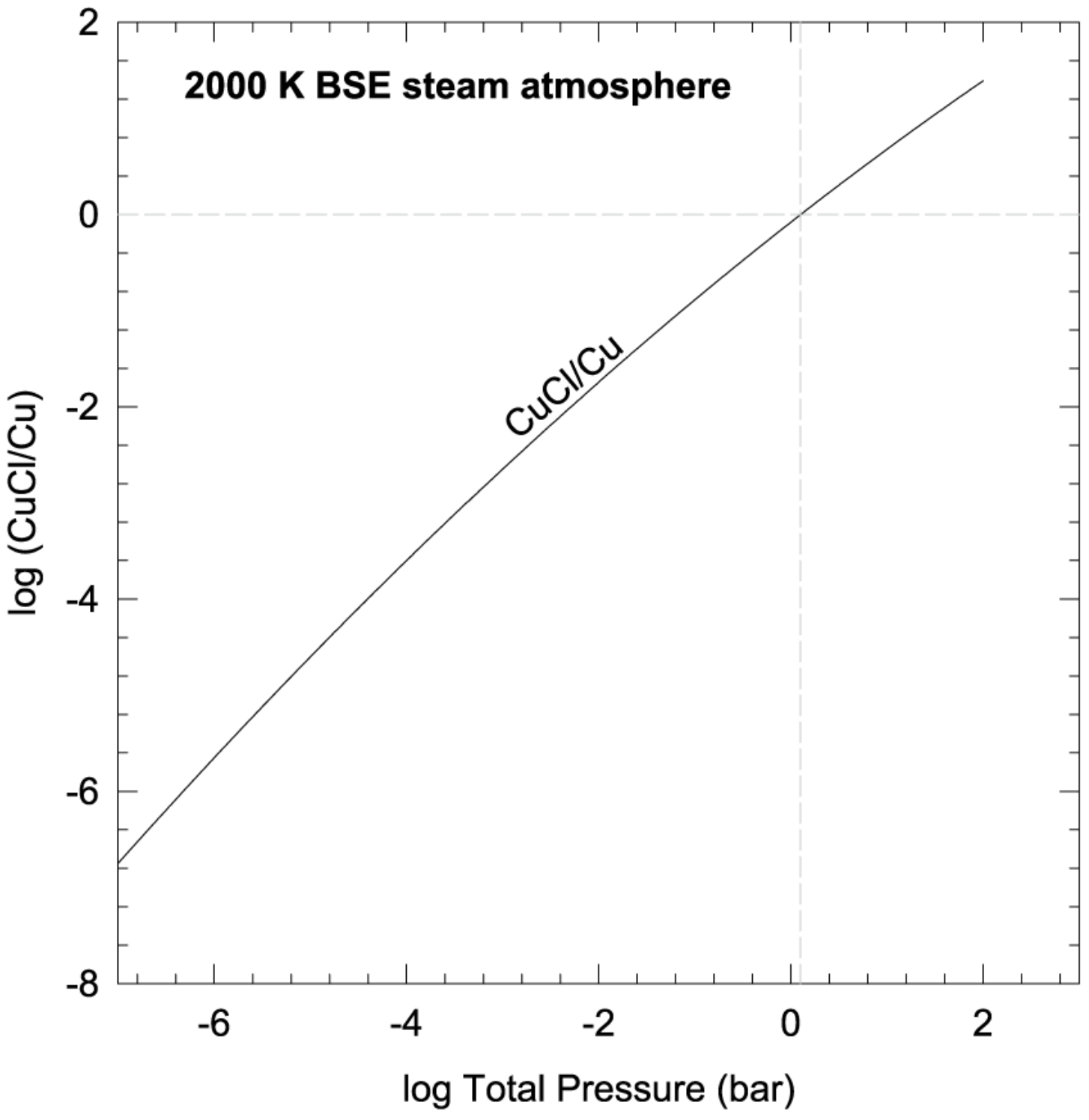

***Fig. 12**. The CuCl/Cu molar ratio in the gas phase is plotted as a function of total pressure at 2000 K for the BSE – steam atmosphere model of Fegley et al (2016). The chemical equilibrium calculations are described in their paper. The CuCl/Cu ratio is unity at ~ 1.25 bar total pressure. Higher pressure and lower temperature favour CuCl while lower pressure and higher temperature favour monatomic Cu gas. See the text for a detailed discussion. Subsequent figures in this section are for the same model.*

### *3.3. Calculations in a multicomponent gas mixture*

Before continuing with another example we return to the $Cl_2$ fugacity and explain what it represents. This discussion is analogous to that given by Lodders (2003) for condensation calculations in the solar nebula. The partial pressure sum for chlorine in a multicomponent gas mixture – be it a volcanic gas, steam atmosphere on the early Earth, or an outgassed atmosphere on a planetesimal – is given by a mass balance equation such as

$$P_{\sum Cl} = X_{\sum Cl} P_T = (P_{Cl} + P_{HCl} + P_{NaCl} + P_{KCl} + P_{CuCl} + \cdots) \quad (39)$$
$$+2\left(P_{Cl_2} + P_{MgCl_2} + P_{FeCl_2} + \cdots\right) + 3\left(P_{CrCl_3} + \cdots\right) + 4\left(P_{Mn_2Cl_4} + \cdots\right) + \cdots$$

This mass balance equation can be rewritten in terms of the $Cl_2$ fugacity and the equilibrium constants for forming each gas from its constituent elements in their reference states, and the thermodynamic activities and fugacities of other elements combined with Cl in gases

$$P_{\sum Cl} = \left(f_{Cl_2}\right)^{1/2} \Big[\Big(K_{Cl} + K_{HCl} f_{H_2}^{1/2} + K_{NaCl} a_{Na} + K_{KCl} a_K + K_{CuCl} a_{Cu} + \cdots\Big) \quad (40)$$
$$+ 2 f_{Cl_2}^{1/2}\left(K_{Cl_2} + K_{MgCl_2} a_{Mg} + K_{FeCl_2} a_{Fe} + \cdots\right) + 3 f_{Cl_2}\left(K_{CrCl_3} a_{Cr} + \cdots\right)$$
$$+ 4 f_{Cl_2}^{3/2}\left(a_{Mn}^2 K_{Mn_2Cl_4} + \cdots\right) + \cdots\Big]$$

The actual mass balance sum contains as many Cl-bearing gases as are in the thermodynamic database of the code used to do the computations; typically several hundred Cl-bearing gases are considered. The equation above shows that chlorine chemistry is coupled to that of all other elements in the gas and the nonlinear system of mass balance equations for Cl and all other elements included in the computations must be solved iteratively. The $Cl_2$ fugacity used to calculate Figure 12 is the value computed from such a mass balance equation and takes into account the distribution of Cl between all Cl-bearing gases in the steam atmosphere model of Fegley et al (2016). For example, Figures 15-24 in their paper show the more abundant Cl-bearing gases, which include NaCl, HCl, KCl, $(NaCl)_2$, $MgCl_2$, $FeCl_2$, $NiCl_2$, and $CaCl_2$; however many more Cl-bearing gases are in the partial pressure sum but are not abundant enough to appear in the figures in Fegley et al (2016).

Analogous considerations apply to the hydrogen, oxygen, fluorine, and sulfur fugacities in complex gas mixtures. In each case the elemental fugacity ($fH_2$, $fO_2$, $fF_2$, $fS_2$) is a fundamental variable that is computed from the mass balance expressions and chemical equilibrium abundances of all gases containing each element. For example, Table 4 below shows elemental fugacities from the computer calculations of Fegley et al (2016) for a BSE – steam atmosphere model (at 2000 K and one bar total pressure). The BSE composition was taken from Palme and O'Neill (2014) in their computations.

**Table 4**. Illustrative elemental fugacities for key volatiles (2000 K, one bar) steam atmosphere

| $fH_2$ | $fO_2$ | $fF_2$ | $fS_2$ | $fCl_2$ |
|---|---|---|---|---|
| $9.02 \times 10^{-2}$ | $4.74 \times 10^{-6}$ | $6.00 \times 10^{-18}$ | $1.29 \times 10^{-3}$ | $4.16 \times 10^{-10}$ |

*3.4. Zn-bearing gases*

Zinc chemistry is analogous to that of copper, but there are no spectroscopic observations of zinc compounds in terrestrial volcanic gases. Chemical equilibrium calculations for terrestrial volcanic gases from several different volcanoes (Symonds et al. 1987; Quisefit et al. 1989; Symonds and Reed 1993; Churakov et al. 2000; Wahrenberger et al. 2002; Renggli et al. 2017) predict Zn is the major species at higher temperatures while $ZnCl_2$ is the major species at lower temperatures. Churakov et al (2000) give the transition temperature between these two species as 1223 K for their calculations modelling Kudryavy fumarolic gases.

The formation of $ZnCl_2$ from monatomic Zn gas occurs via the reaction

$$Zn\ (g) + Cl_2\ (g) = ZnCl_2\ (g) \tag{41}$$

The equilibrium constant for this reaction, again assuming ideality for the gases, is

$$K_{(41)} = \frac{P_{ZnCl_2}}{P_{Zn}} \frac{1}{f_{Cl_2}} \tag{42}$$

Upon rearranging the equilibrium constant expression we see the $ZnCl_2$/Zn molar ratio is proportional to total pressure at otherwise constant conditions,

$$\frac{X_{ZnCl_2}}{X_{Zn}} = K_{(41)} X_{Cl_2} P_T \tag{43}$$

This equation shows higher pressure favours $ZnCl_2$ while lower pressure favours monatomic zinc. The thermodynamic data for gaseous $ZnCl_2$ show it becomes less stable relative to $Zn^0$ at higher temperature. Figure 13 shows that the 1:1 ratio occurs at ~ 890 bar with Zn dominant at lower pressures and $ZnCl_2$ dominant at higher pressures. The behaviour of the Zn – $ZnCl_2$ pair is analogous to that of the Cu – CuCl pair; the chloride is dominant at higher pressure and lower temperature and the monatomic gas dominates at lower pressure and higher temperature. Our calculations, those from the volcanic gas models cited earlier, and the evaporation studies reviewed in section 2 all imply that Zn volatility is governed by its evaporative loss as $Zn^0$.

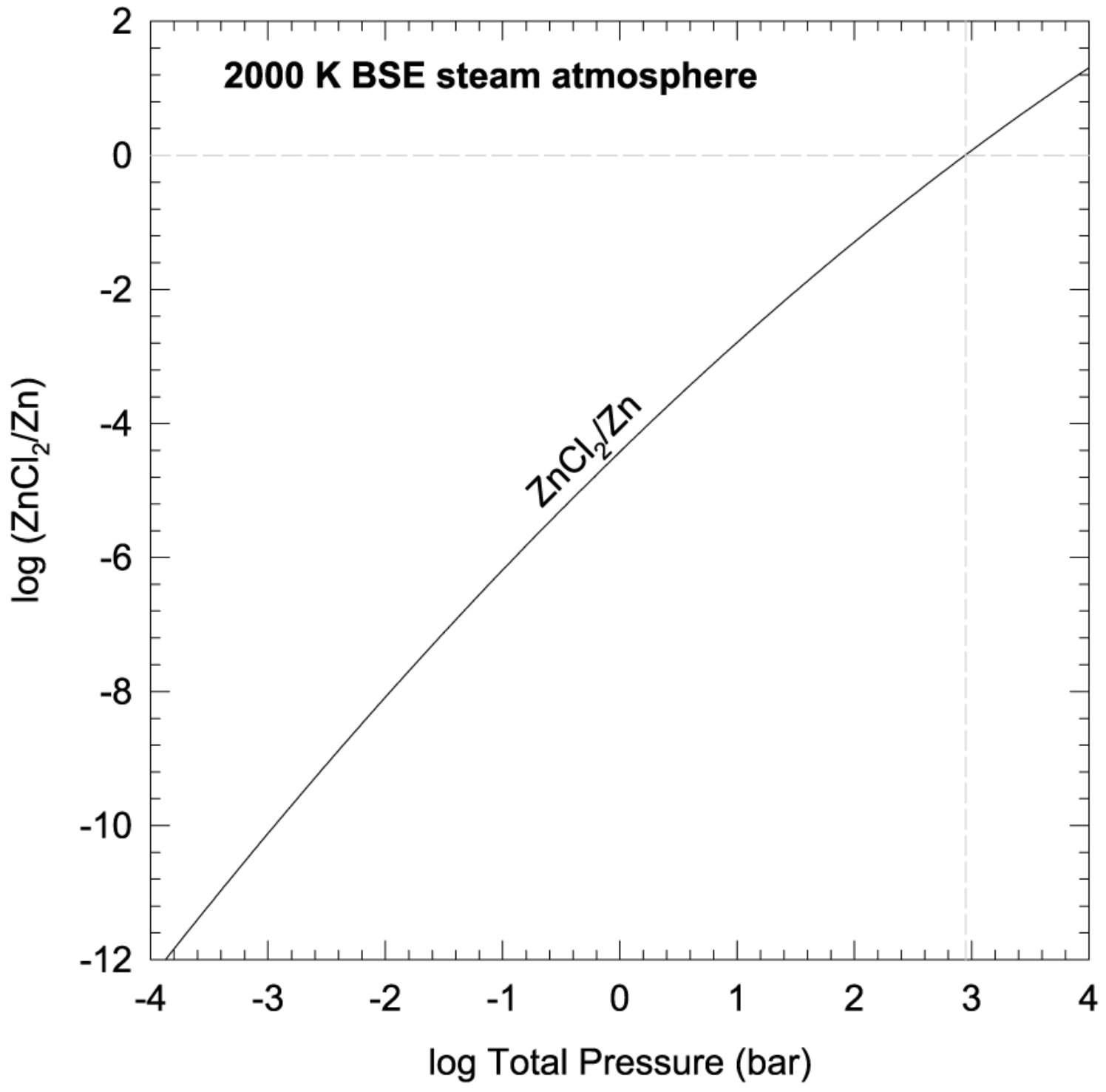

***Fig 13**. The $ZnCl_2/Zn$ molar ratio in the gas phase is plotted as a function of total pressure at 2000 K for the same steam atmosphere model shown in Fig. 9. The $ZnCl_2/Zn$ ratio is unity at ~ 890 bar. Higher pressure and lower temperature favour $ZnCl_2$ while lower pressure and higher temperature favour monatomic Zn gas. See the text for a detailed discussion.*

*3.5. Factors governing the stability of gas species*

Table 5 illustrates several trends as a function of pressure, temperature and composition. The columns for the solar nebula and H-chondrites are taken from our calculations for solar nebula condensation and ordinary chondrite metamorphism (Lodders 2003; Schaefer and Fegley 2010). The volcanic gas column is from published calculations (Symonds et al., 1992 – Augustine, Churakov et al., 2000, Wahrenberger et al., 2002 – Kudravy and Symonds and Reed, 1993 – Mt. St. Helens) and the ferrobasalt column is based on the Langmuir vaporisation experiments of Sossi et al. (2016). As shown in Table 5, the alkalis, Ag, Ga, In and Tl behave analogously to Cu; the chloride is dominant at higher pressure and lower temperature while the monatomic gas is dominant at lower pressure and higher temperature. Figure 14 shows Mn chemistry is like that of Zn but with the interesting complication that with decreasing pressure under otherwise constant conditions the major gas changes from $MnCl_2$ to MnCl to Mn. Volcanic gas chemical equilibrium calculations from the literature show several other elements (Bi, Cr, Fe, Mo, Pb, Sb) may exist as chlorides at one bar pressure and temperatures of 1143 – 1213 K (see Table 5). These trends are easily understood using the same reasoning as described above for the Cu – CuCl and Zn – $ZnCl_2$ equilibria.

Iron gas phase chemistry (see Table 3) exemplifies the competition between halide and hydroxide gases. (Hastie 1975, pp. 68-73) showed the relative abundances of halide and hydroxide gases in steam – rich systems with HCl and HF are controlled by exchange equilibria such as

$$FeCl_2\ (g) + 2\ H_2O\ (g) = Fe(OH)_2\ (g) + 2\ HCl\ (g) \qquad (44)$$

The equilibrium constant expression for this reaction is

$$K_{(44)} = \frac{P_{Fe(OH)_2}}{P_{FeCl_2}} \left(\frac{P_{HCl}}{P_{H_2O}}\right)^2 \tag{45}$$

We can rearrange this to show that the hydroxide/chloride molar ratio is independent of the total pressure but varies with the square of the $H_2O$/HCl molar ratio in the gas

$$\frac{X_{Fe(OH)_2}}{X_{FeCl_2}} = K_{(44)} \left(\frac{X_{H_2O}}{X_{HCl}}\right)^2 \tag{46}$$

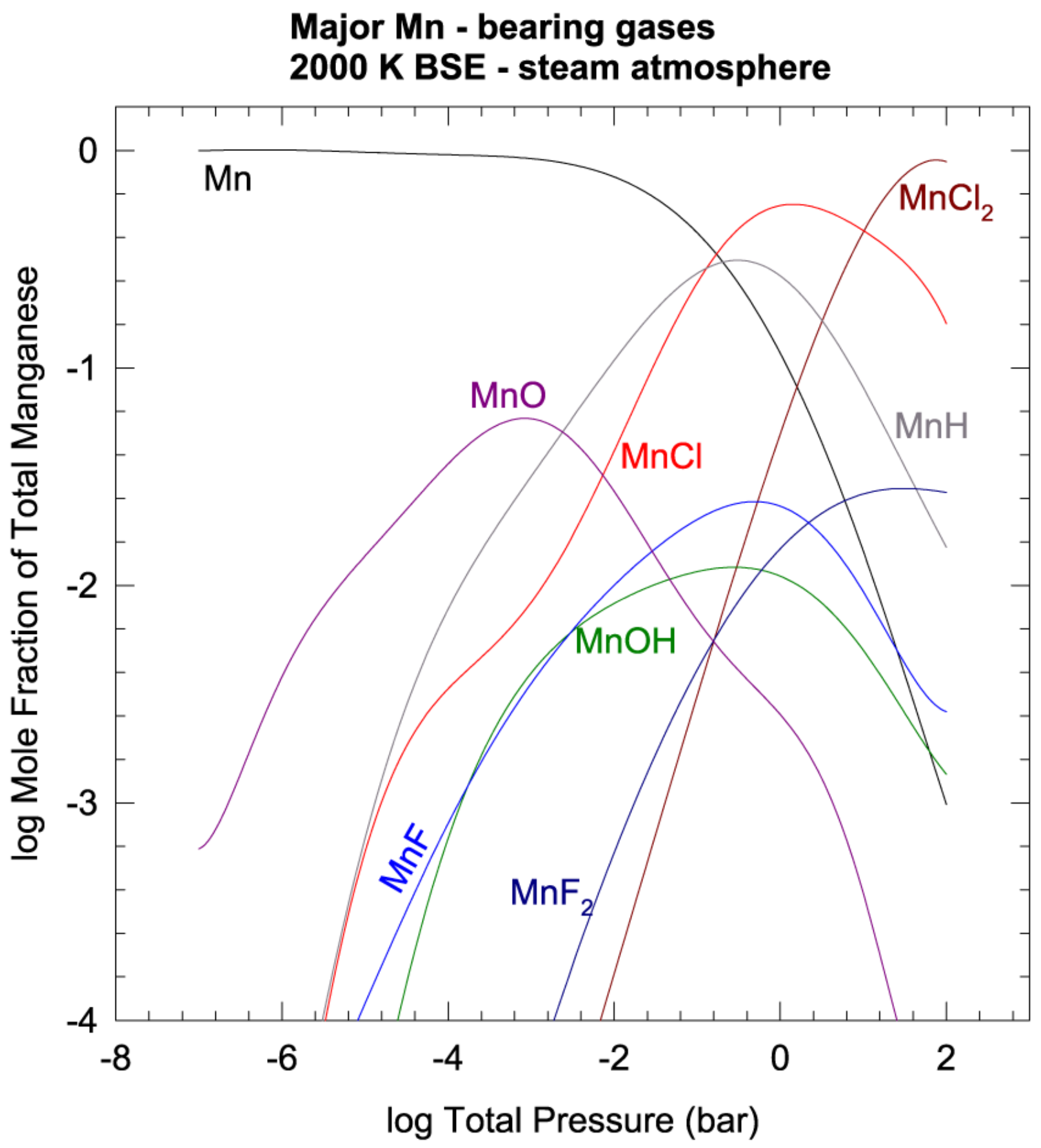

***Fig 14***. *The speciation of manganese is plotted as a function of total pressure at 2000 K for the same steam atmosphere model shown in Fig. 9. The major Mn-bearing gas shifts from $MnCl_2$ to MnCl to Mn with decreasing pressure. Higher pressure and lower temperature favour the Mn chloride gases while lower pressure and higher temperature favour monatomic Mn gas.*

Similar competition between halide and hydroxide gases was calculated to be important for Al, Ca, K, and Na in the steam atmosphere models of Fegley et al (2016). As pointed out by Pokrovski et al. (2013), thermodynamic data may be lacking for trace elements that readily undergo gas phase hydrolysis (notably the Group VI metals), such that the full spectrum of gaseous species may not be represented in these calculations for $H_2O$-rich volcanic vapours. However, the monatomic gases are expected to be the major forms of Fe, K, and Na in the solar nebula (Table 5). Indeed, at high temperatures (>1200 K) even in volcanic vapours with high $fS_2$ and $fCl_2$, the monatomic gases (noted above) and oxide species (*e.g.* Group VI metals, As, Sb) begin to replace their chloride- and hydrated equivalents, respectively (see discussion in Churakov et al. 2000). The experiments of Sossi et al. (2016) notably lack major volatile species, thereby simplifying chemical equilibria in the gas phase by lowering the stability of these –OH, –Cl and S-bearing species observed in volcanic vapours (Table 5). At low pressures (≤1 bar), these molecules are either sparingly soluble in silicate melts (Cl, Carroll and Webster 1995; Webster 1997; $H_2O$, Dixon et al. 1995) and/or highly volatile (S, Cl, F, H) that they may be near-absent in degassing magmas at high temperatures at low pressures, leaving $O_2$ as the major volatile species. Such conditions are likely found on heated planetesimals, on which the escape velocities are too low to retain an atmosphere comprised of light elements, such as H, C, or S, resulting in confining pressures of <1 bar. Under these conditions complex halide or sulfide gases are predicted to be absent; indeed, the major gas species inferred from the experiments of Sossi et al. (2016) are the monatomic gases (the alkalis, Zn, Cd, Ag, Pb, Cu, Fe, Mn and Ga), though notably some have stable di- or trioxide species ($CrO_2$, $MoO_3$), while others form monoxides at $fO_2$ close to air (LiO, GaO, PbO). Their vaporisation behaviour in these volatile-poor systems mirrors that observed for lower-temperature volcanic gases; monatomic gases tend to form chlorides (*e.g.*, alkalis, Cu, Zn), oxide species are hydrated (*e.g.*, $H_2CrO_2$) and gas species of divalent chalcophile elements such as GeO and PbO substitute $O^{2-}$ with $S^{2-}$ (GeS, PbS).

**Table 5**. Speciation of major gases at these conditions.

| Element | Solar nebula, 1500 K, $10^{-4}$ bar | H-chondrite, 1500 K, $10^{-4}$ bar | Volcanic gas 1143 - 1213 K, 1 bar | Anhydrous ferrobasalt (Sossi et al., 2016), 1573-1823 K, 1 bar |
|---|---|---|---|---|
| Ag | Ag | Ag | Ag[8], (AgCl)[8] | Ag |
| As | As | AsS | AsS[5,7,8], AsO[7] | - |
| Au | AuS[1] | AuS | AuS[8] | - |
| Bi | Bi | Bi | Bi[6-8], (BiS[5,6], BiCl[5]) | - |
| Br | HBr | NaBr | HBr[5-8] | - |
| Cd | Cd | Cd | Cd[5,7,8] | Cd |
| Cl | HCl | NaCl | HCl[5-8] | - |
| Cr | Cr | Cr | $CrO_2Cl_2$[5], $CrO_2H_2$[7] | $CrO_2$ |
| Cs | Cs[2] | Cs | CsCl[7,8] | Cs[9] |
| Cu | Cu | Cu | CuCl[5-8] | Cu |
| F | HF | NaF | HF[5-8] | - |
| Fe | Fe | Fe | $Fe(OH)_2$[6-8], ($FeCl_2$[5,7]) | Fe |
| Ga | Ga | GaF[3] | GaCl[6,7] | GaO, Ga |
| Ge | GeS | GeS | GeS[7] | GeO |
| H | $H_2$ | $H_2$ | $H_2O$[5-8] | - |
| I | I | NaI | - | - |
| In | In | In | InCl[6,7] | - |
| K | K | K | KCl[5-8] | K |
| Li | Li | - | LiCl[5,7] | Li, LiO |
| Mn | Mn | Mn | $MnCl_2$[5,7,8] | Mn |
| Mo | Mo | - | $H_2MoO_4$[6-8], ($MoO_2Cl_2$[5]) | $MoO_3$ |
| Na | Na | Na | NaCl[5-8] | Na |
| O | CO[4] | CO | $H_2O$[5-8] | - |
| P | PO | $P_4O_6$ | - | - |
| Pb | Pb | PbS | PbS[6-8], ($PbCl_2$[5]) | PbO, Pb |
| Rb | Rb | Rb | RbCl[7,8] | Rb |
| S | $H_2S$ | $S_2$ | $SO_2$[5-8] | - |
| Sb | Sb | SbS | SbS[5,7,8], SbO[7], ($SbCl_3$[5]) | - |
| Se | Se | SeS | $H_2Se$[8] | - |
| Sn | SnS | SnS | SnS[6,7] | - |
| Te | Te | Te | Te[8] | - |
| Tl | Tl | Tl | TlCl[6,7] | - |
| Zn | Zn | Zn | Zn[7,8], ($ZnCl_2$[5]) | Zn |
| log $fO_2$ | –18.06 | –12.48 | –12.58 to –10.93 | –10 to –0.67 |

[1]Au is about half as abundant as AuS.
[2]If thermal ionisation is considered, $Cs^+$ is major Cs gas.
[3]GaOH is almost as abundant as GaF.
[4]$CO/H_2O$ ~ 0.86 at this P and T.
[5] Symonds et al. 1992, Augustine (log$fO_2$ = -12.58 ; T = 1143 K)
[6] Wahrenberger et al. 2002, Kudryavy (log$fO_2$ = -10.99 ; T = 1173 K)
[7] Churakov et al. 2000, Kudryavy (log$fO_2$ = -10.93 ; T = 1213 K)
[8] Symonds and Reed 1993, Mt. St. Helens (log$fO_2$ = -11.39; T = 1203 K)
[9] Kreutzberger et al., 1986
Gas species that are more stable at lower temperature and/or high $Cl_2$ fugacity are shown in brackets.

## 4.0. Volatility during planetary formation

*4.1. Overview*

Chondritic (undifferentiated) meteorites are currently assumed to provide the best estimate for the bulk composition of the rocky planets. The CI carbonaceous chondrites are particularly useful, as they have relative abundances of rocky elements identical to solar photospheric abundances of these elements. Thus the elemental composition of the solar nebula is given by elemental abundances from CI chondrites and from the Sun (e.g., see Lodders 2003, 2008 and references therein for the details).

However, basaltic rocks and remote sensing measurements show that rocky bodies of the inner solar system, in addition to other chondrites, are depleted in MVEs (Dreibus and Wanke 1985; Palme et al. 1988; Mittlefehldt and Lindstrom 1990; Barrat et al. 2000; Ruzicka et al. 2001; O'Neill and Palme 2008). Element depletion is quantified by normalising a volatile element to a refractory, lithophile element (typically Mg). This ratio is then divided by the same quantity in CI chondrites:

$$f_E = \frac{E_x/Mg_x}{E_{CI}/Mg_{CI}} \tag{47}$$

This exercise is performed for each element, $E$ in a given planetary body, $x$. It has been the convention in cosmochemistry for 50 years (Larimer 1967) to plot these elemental depletion factors ($f_E$) against their nebular half-condensation temperatures[6], $T_c$ (Fig. 15). The mid-plane pressure in the inner solar nebula is $\approx 10^{-4}$ bars and conditions are highly reducing ($fO_2$ is $10^7$ times more reducing than Iron-Wüstite, IW-7; Rubin et al 1988; Grossman et al. 2008). The moderately volatile elements, have $T_c$ between the main components (Mg, Fe, Si, $T_c \approx 1320$ K) and FeS (660 K).

Quantitative condensation of a moderately volatile element occurs over a narrow range of <100 K (*e.g.,* Lodders 2003). Should MVE abundances simply reflect nebular condensation temperatures, then the final temperature of accretion of chondrite components should sharply discriminate between those elements whose $T_c$ is higher than that of last equilibration with the gas phase (fully condensed) from those with lower $T_c$, which reside entirely in the gas phase (as recognised by Larimer, 1967). Rather, MVEs in chondritic meteorites vary monotonically with nebular condensation temperature, and tend to flatten out in the most volatile ($T_c < 650$ K), as noted by Wolf et al. (1980) and expanded upon by Humayun and Cassen (2000). This feature implies two superimposed processes i) the depletion of volatile elements in chondrites according to their condensation temperatures above 650 K and ii) mixing of a second component with its full complement of volatile elements in proportions that are sufficient to elevate the abundances of the highly volatile elements, but not to affect the trend of the moderately volatile elements (see also Anders 1964). The two physical reservoirs for these components are likely manifest in chondrules (volatile-depleted) and matrix (volatile-rich), see

[6] $T_c$ is the temperature at which half the mass of an element is calculated to condense in the solar nebula

Grossman and Larimer (1974). Despite large variations in the chemistry of individual chondrules (*e.g.*, Zanda 2004), the compositions of bulk chondrites are remarkably constant, arguing for complementarity of the two components (Bland et al. 2005; Hezel and Palme 2010; Friend et al. 2017).

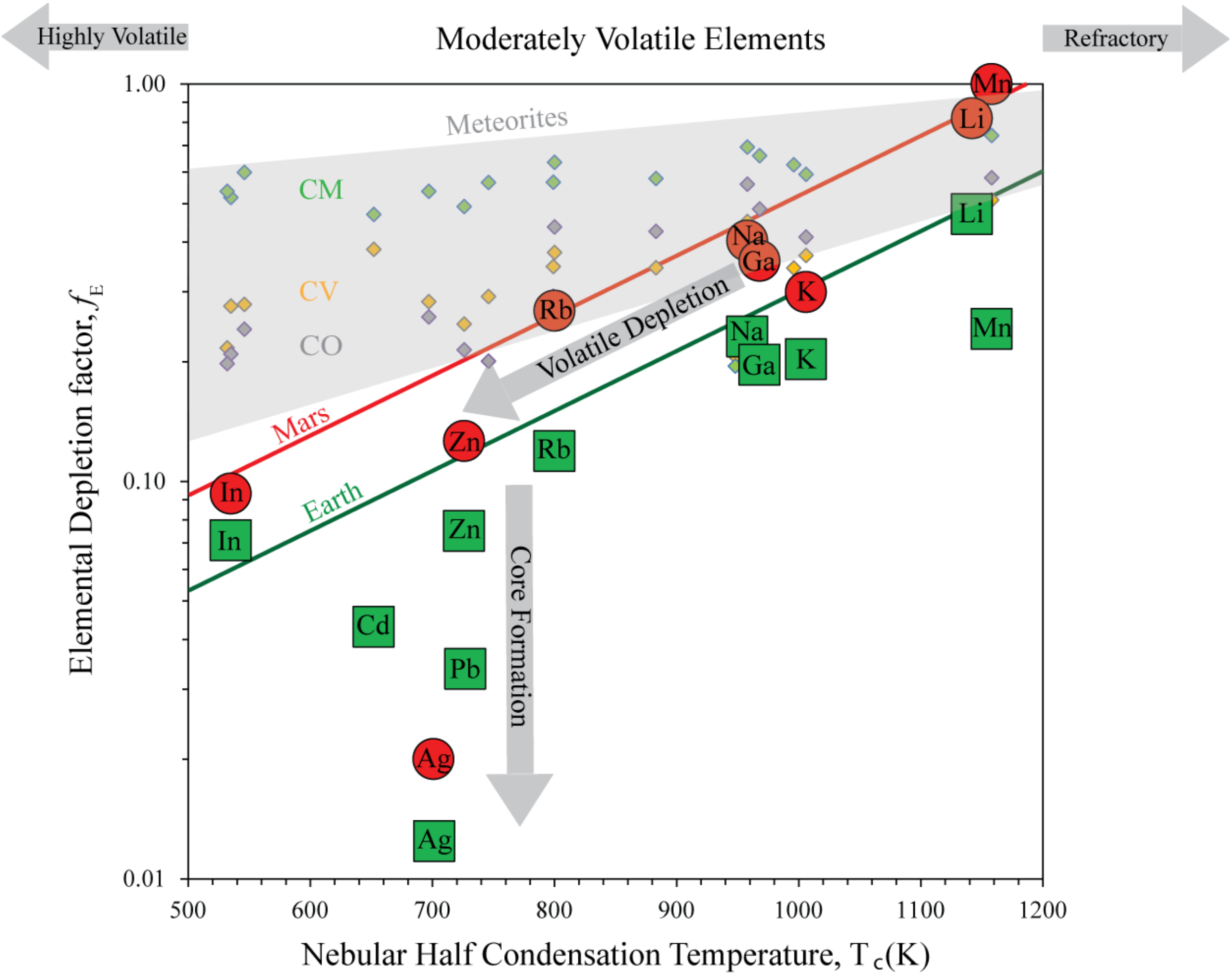

***Fig 15.*** *Depletion factors ($f_E$) of moderately-volatile elements in the Earth (Palme and O'Neill 2014; green squares), Mars (Dreibus and Wänke 1985; red circles) and carbonaceous chondrites; CM, CV and CO classes* (Wasson and Kallemeyn 1988). *Elements are plotted as a function of their nebular half-condensation temperatures, $T_c$ (K) from Lodders (2003) except Ag (Kiseeva and Wood 2015). The trends labelled 'Mars' and 'Earth' are defined by abundances of elements that do not partition into metal, delineated by the 'volatile depletion' arrow. Depletion in Zn, Pb, Cd and Ag, is caused by 'core formation'.*

Although, to a first order, elemental abundances in rocky bodies are correlated with nebular condensation temperatures (*e.g.*, Humayun and Cassen, 2000), in detail this picture becomes blurred. This may be due to each planet accreting different amounts of higher temperature (volatile-depleted) and lower temperature (volatile-rich) material in its feeding zone (e.g., see Barshay, 1981 or the discussion of his modelling in Moynier and Fegley, 2015), though the exact mechanisms remain uncertain. On the planetary scale if we consider only thermal escape (neglecting important non-thermal escape processes (see Tables 7.1, 7.2 of Chamberlain & Hunten 1987) that are very difficult to model), atmospheric loss from a body is described in two regimes; Jeans escape (Jeans, 1916) and hydrodynamic escape (Parker, 1960). Jeans escape assumes a Maxwell-Boltzmann distribution of particles, in which the escape flux is:

$$\left(\frac{dm}{dt}\right)_J = n\left(\frac{2k_BT}{\pi m}\right)^{\frac{1}{2}}(1+\lambda_{esc})e^{-\lambda_{esc}} \tag{48}$$

where $n$ is the particle number density, $m$ the molar mass of the particle, and $\lambda_{esc}$ is the escape parameter, the ratio of the thermal velocity of the particles to the escape velocity of the body, $v_{esc}$ (= $[2GM_{body}/r]^{1/2}$):

$$\lambda_{esc} = \frac{m{v_{esc}}^2}{2k_BT}. \tag{49}$$

When the mean thermal velocity of the Maxwell-Boltzmann distribution of particles begin exceed the escape velocity of the planet (*i.e.,* $\lambda_{esc} < 1$; Genda and Abe, 2003), the condition for hydrodynamic escape is met. Here, Equation (49) simplifies and the flux (kg/s) of atmosphere loss of a particle is given by the product of its density, surface area and the thermal velocity of the particle:

$$\left(\frac{dm}{dt}\right)_H = 4\pi r^2\rho v. \tag{50}$$

The very old Pb and Sr isotope ages for eucrites and angrites (e.g., Amelin 2008; Hans et al. 2013) suggest at least some of the accreted material was differentiated and depleted in MVEs, either in the solar nebula or by these atmosphere escape mechanisms. From a mechanistic standpoint, equilibrium between the gas phase and the liquid is not mandated during gas-condensed phase interaction, and may depend on the scale at which the exchange occurs. Thus, whether the operative process of volatile depletion occurred at equilibrium or was kinetic in nature remains unconstrained. Add to this the difficulty in assessing the degree of volatile depletion experienced by differentiated bodies due to the dual siderophile/chalcophile nature of many MVEs (see Palme et al. 1988), the origin of volatile element depletion is poorly understood. Indeed, a clear departure of chalcophile elements relative to lithophile elements with similar condensation temperatures is observed in terrestrial samples (*e.g.,* Ag and Pb; Schönbächler et al. 2010; Wood and Halliday 2010), implying that core formation also plays a role.

A possible solution to this paradox is the changing locus of volatile depletion with time. The relevance of nebular condensation temperatures is inherently rooted to the conditions of the solar nebula (*i.e.,* that present during the formation of chondrites; and even then there are variations, *e.g.,* Schaefer and Fegley, 2010). However, planetary formation is more protracted (up to 100 My; Kleine et al. 2009; Jacobson and Morbidelli 2014), long after the dispersion of the nebular gas, which likely occurred within 5 My of the solar system's formation (Dauphas and Chaussidon 2011) but may have taken longer, perhaps up to 30 My based on astronomical observations of circumstellar disk dissipation (Cameron 1995). Thereafter, the volatility behaviour of the elements are no longer bound to their nebular condensation temperatures, because the conditions under which evaporation occurs is markedly different in the post-nebular realm.

### *4.2. Planetary Bodies*

Collisions between two embryonic planets, for example between the proto-Earth and a Mars-sized planetesimal thought to have resulted in the formation of the Moon, are highly energetic, reaching temperatures of ≈6000 K (*e.g.* Melosh 1990; Cameron and Benz 1991; Canup 2004; Nakajima and Stevenson, 2014; Lock and Stewart 2017). Such temperatures not only far exceed those present in the solar nebula (<2000 K for any reasonable nebular pressure), but they lead to extensive melting, as they surpass the low-pressure peridotite liquidus (≈ 2000 K) and cause vaporisation of silicate material (e.g., see the discussion in section 6.3.5 of Fegley and Schaefer 2014). Computational models that minimise the Gibbs free energy of both liquid and vapour species for major elements (Fegley and Cameron 1987; Schaefer and Fegley 2004a; Visscher and Fegley 2013) show that the composition of the gas evolved from molten peridotite (>3000 K) is dominated by SiO vapour, though Na predominates between 1800 – 3000 K, yielding oxygen fugacities $10^7$× more oxidising than that of the nebular gas (Visscher and Fegley, 2013; Fig. 16). Furthermore, the total pressure exerted by the gas increases with temperature as the partial pressures of gas species increase, reaching 10 bar, orders of magnitude higher than in the solar nebula, $10^{-4}$ bar (Fig. 16).

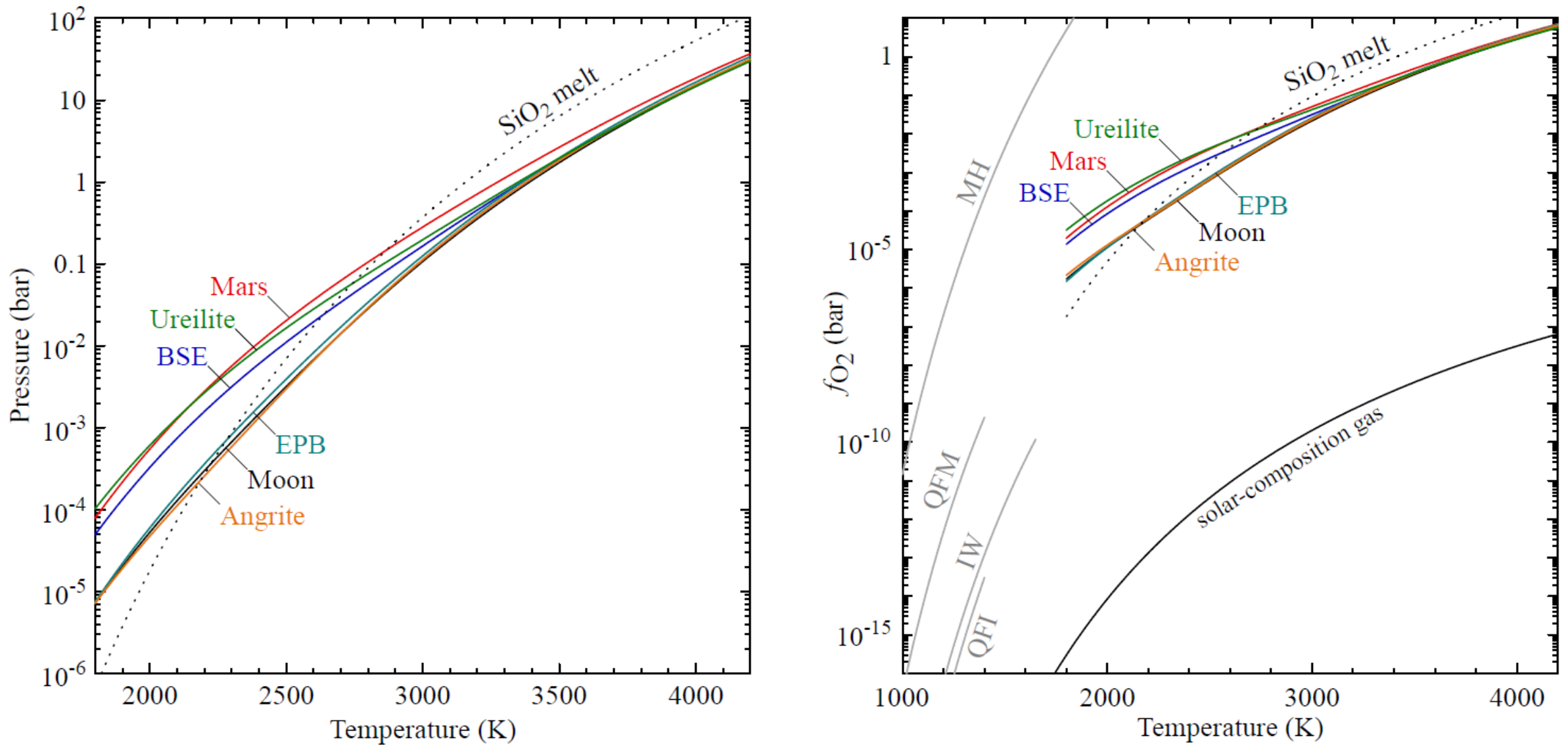

***Figure 16**: Evolution of the a) total pressure and b) oxygen fugacity as a function of temperature (in K) upon evaporation of planetary mantles of their own composition. Reproduced from Visscher and Fegley (2013).*

As illustrated by experimental work (section 2.0.), the order of element volatility must change in response to the different thermodynamic conditions. To this end, a recent push has seen the use of certain element pairs that should be sensitive to volatility under various conditions in order to help to distinguish between nebular and post-nebular volatile loss. Among the first was the study of O'Neill and Palme (2008) who exploited the lithophile behaviour of Mn and Na (though Mn becomes weakly siderophile at high pressures and temperatures; Mann et al. (2009), to demonstrate that all small planetary bodies had undergone post-nebular volatile loss (Fig. 17). This conclusion is reached on the basis of their super-chondritic Mn/Na ratios. Manganese and Na are similarly volatile under nebular conditions (*cf.* the constancy of Mn/Na in chondrites with variable volatile depletion, monitored by the Mn/Mg ratio), but Na becomes relatively more volatile than Mn at more oxidised conditions:

$$MnO_{(l)} + Na_{(g)} = Mn_{(g)} + NaO_{0.5(l)} + \frac{1}{4}O_2 \tag{51}$$

Equation (51) is a combination of the $n = 1$ reaction for Na evaporation, and an $n = 2$ reaction for Mn evaporation. As such, a preferential loss of Na relative to Mn, normalised to chondritic values, must result from an increase in $fO_2$ relative to nebular conditions. Sitting conspicuously on the nebular array is the bulk silicate Earth (Fig. 17). An Mn/Mg ratio that is lower than any chondrite group firstly suggests that Earth is more volatile depleted than any other group (providing only minor Mn was lost to the core; Geßmann and Rubie 1998; Mann et al. 2009; Siebert et al. 2018). A chondritic Mn/Na implies that Mn and Na were similarly volatile during their depletion from Earth. This behaviour only occurs at very low $fO_2$, restricted to those found in the solar nebula, a conclusion supported by short-lived (3.7 My) Mn-Cr isotope system, in which the Earth falls on the isochron defined by chondritic meteorites (Moynier et al. 2007; Trinquier et al. 2008; see Palme and O'Neill 2014 for a discussion).

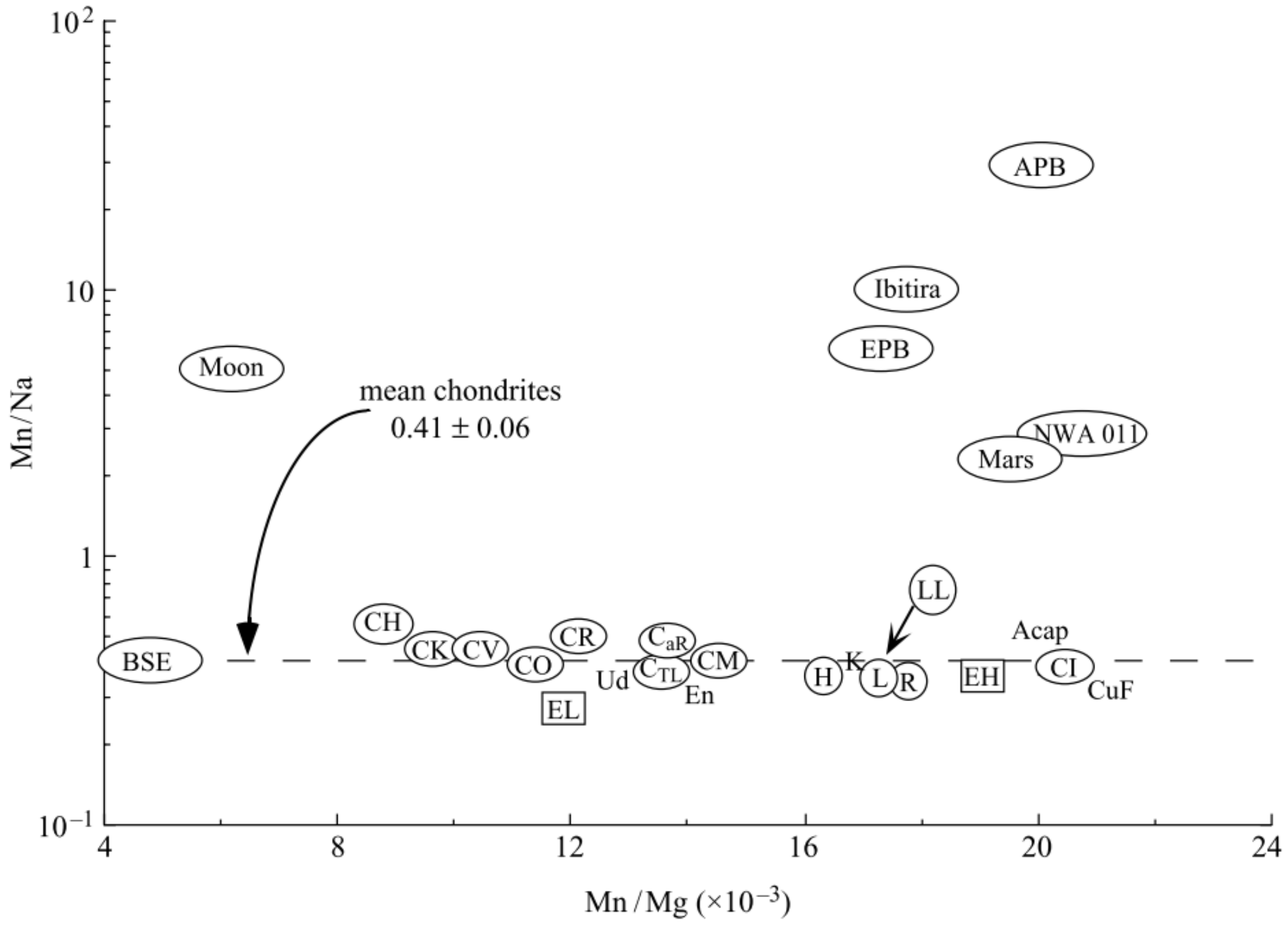

***Fig. 17.*** *The Mn/Na vs. Mn/Mg ratio of planetary bodies (reproduced from O'Neill and Palme 2008). The Mn/Mg ratio tracks the degree of volatile depletion, with no effect of* $fO_2$*. Mn/Na ratios increase as a function of volatile depletion if the* $fO_2$ *is higher than that of the solar nebula. That Mn/Na is constant in chondrites suggests that they were similarly volatile in the solar nebula. As the Earth plots on the chondritic array, its volatile depletion was inferred to have been nebular.*

### *4.3. The Earth*

The view that Earth's moderately volatile element budget was set in the solar nebula has been debated for some time (e.g., Ringwood 1966, 1979) and was recently challenged again by Norris and Wood (2017). They find that the depletion of moderately volatile elements is better correlated with their volatilities during evaporation from a silicate melt at IW-1 than it does with nebular condensation temperatures. Three elements that are notable in this assessment are Zn, In and Cd, as their relative abundances are difficult to reconcile with core formation and nebular condensation temperatures. In an earlier study, this led Wang et al. (2016) to conclude that the Earth is non-chondritic. Their associated $f_E$ are 0.075, 0.071 and 0.043 (Palme and O'Neill, 2014) and calculated $T_c$ are 726 K, 535 K and 650 K, respectively. Extrapolating the trend of $f_E$ vs. $T_c$ for lithophile volatile elements (Fig. 5 of Wang et al., 2016), yields expected $f_E$ for Zn of 0.08; 0.06 for Cd and 0.035 for In. This observation may suggest the calculated condensation temperature of In is too low. Indeed, the CI- and Mg-normalised Zn/In ratio in carbonaceous chondrites is unity (0.96±0.06; Wasson and Kallemeyn 1988), implying that In and Zn have similar volatilities in the solar nebula. This view extends to more oxidising conditions and higher pressures, as supported by the volatilities inferred from Norris and Wood (2017), where Zn is slightly more volatile than In (see Fig. 9), citing the lack of S-bearing species that stabilise $In_2S_{(g)}$ in the solar nebula. The key message is, if Zn and In depletion occurred exclusively by volatility, their depletion factors in the Earth should be very similar, as is observed.

Experimental studies show that element partitioning between silicate liquid and metal is ordered in the sequence $D = 1 \approx Zn < Cd \approx In = 10$ (Ballhaus et al. 2013; Wood et al. 2014; Wang et al. 2016); suggesting that core formation *i)* would deplete In to a significantly greater extent than Zn (which is not observed) and *ii)* would not appreciably fractionate Cd/In ratios. This, together with the observation that Zn and Cd have chondritic stable isotope ratios in the BSE (Wombacher et al. 2008; Sossi et al. 2018) and that they remain relatively impervious to other mass transfer processes during planetary formation such as collisional erosion (Carter et al. 2018) or partial melting (Witt-Eickschen et al. 2009), suggests that the abundances of these three elements should largely reflect fractionation attributed to volatility during Earth's formation, where (least volatile) In ≈ Zn < Cd (most volatile).

A departure from the canonical condensation temperature sequence was noted for the alkali elements by Jones and Palme (2000). In their treatment, a better correlation was found between $f_E$ and the square-root of their masses (Fig. 18a) than with $T_c$ (Fig. 18b). This correlation holds relatively well across other weakly-siderophile MVEs, however, in detail, large discrepancies are apparent. For example, Ag (0.026) is more depleted than In (0.071), and Zn (0.075) and Ga (0.17) despite the heavier mass of the latter element of each pair, especially considering that Ga and In are more siderophile than Zn and Ag, respectively. Jones and Palme (2000) tentatively attributed this broad dependence on the square root of the mass to indicate the controlling influence of a transport process, namely Jeans escape, on the abundances of MVEs. For a Maxwell-Boltzmann distribution of kinetic energies, particles in a hot atmosphere will have mean velocities equal to:

$$\bar{v} = \sqrt{\frac{2RT}{M}}, \quad (52)$$

where $R$ is the gas constant, $T$ the absolute temperature and $M$ the molar mass. Where this velocity exceeds $v_{esc}$, volatile loss occurs. As such, lighter particles should be preferentially lost relative to heavier ones proportional to $M^{-0.5}$, the inverse of what is observed. Rather, the depletions in alkalis are positively correlated with the experimental results of Kreutzberger (1986), where the volatility sequence is, from least to most volatile, Na < K < Rb < Cs (section 2.2.4., Fig. 7). As volatile depletion in their experiments is controlled by their equilibrium partial pressures in the gas phase, so too do these depletions in the Earth likely reflect the same control at some stage of its evolution.

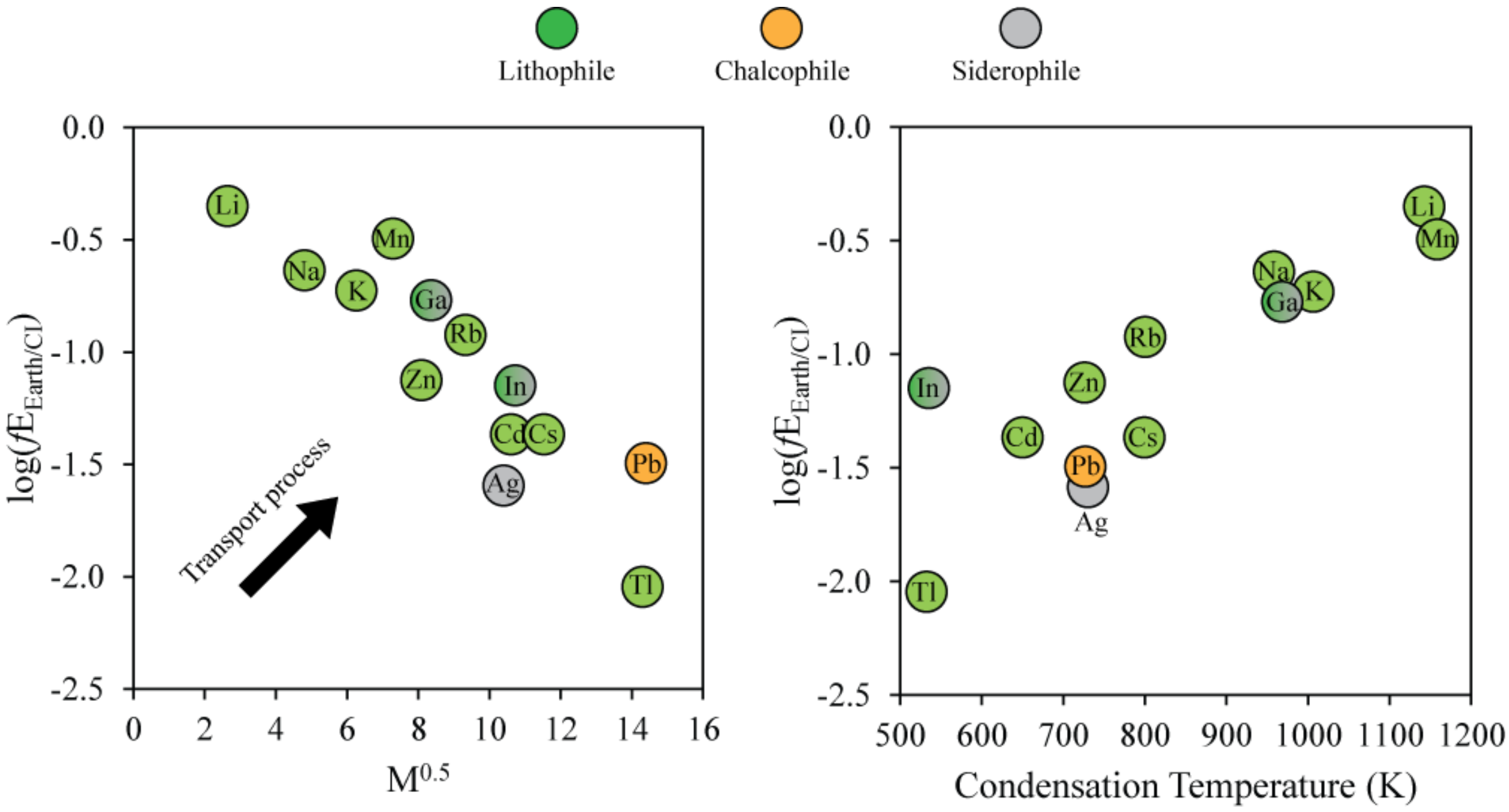

***Fig. 18.*** *CI- and Mg-normalised elemental abundances ($f_E$) in the Earth's primitive mantle (Palme and O'Neill, 2014), as a function of **a)** the square root of their atomic mass (arrow denotes fictive trend for loss controlled by a transport process) and **b)** nebular condensation temperature (Lodders, 2003) and, Ag, (Kiseeva and Wood 2015). Green = lithophile; orange = chalcophile, grey = siderophile.*

Whether volatile depletion on Earth was the result of gas-condensed phase interaction in the solar nebula or later during planetary formation, and to what degree equilibrium was attained, may be edified by ongoing experimental and isotopic work. The emerging picture from numerical modelling suggests that escape velocities for the Earth (11.1 km/s) are prohibitively large for thermal escape to occur. Indeed, even for hydrodynamic escape, which results from dissociation of $H_2O$ in a hot Earth-Moon disk after a giant impact (Genda and Abe 2003; Desch and Taylor 2013), the escape of volatiles is retarded by their necessity to diffuse through a medium dominated by heavier molecules (*e.g.,* $SiO_{(g)}$, $Na_{(g)}$, $NaCl_{(g)}$, see section 3.0.) apparently rendering their loss by this mechanism ineffectual (Nakajima and Stevenson 2018). Nevertheless, proto-lunar disk models are still in flux and Gammie et al. (2016) argue "a hot magnetized disk could drive bipolar outflows that remove mass and angular momentum from the Earth – Moon system". Furthermore, Genda and Abe (2005) argue for enhanced atmospheric loss on the early Earth if oceans were present when impacts took place. Depending on which models are correct, post-nebular volatile loss by evaporation from silicate melts may be restricted to smaller planetesimals where escape velocities are lower (Charnoz and Michaut 2015; Hin et al. 2017). Alternatively, the Earth's MVE depletion was inherited from its constituents that had already lost volatiles in the solar nebula (O'Neill and Palme 2008; Dauphas et al. 2015; Sossi et al. 2016).

*4.4. The Moon and Vesta*

Two such small, volatile depleted bodies are found in the form of the Moon and Vesta, which have radii of 1737 km and 263 km, respectively. Both bodies are volatile depleted and these depletions have some similarities (Fig. 19), but differ in detail (*e.g.,* Dreibus and Wänke 1979, 1980, 2001; O'Neill 1991; Ruzicka et al. 2001). Jochum and Palme (1990) argued that the consistent K/La, Rb/Sm, Cs/U element correlations in eucrites would not be present if evaporative loss of alkalis occurred during volcanism or thermal metamorphism. A similar argument was presented for the lunar mare basalts (see Gooding and Muenow, 1976), meaning that volatile depletion is a widespread characteristic of both bodies and likely occurred during their accretion. Several MVEs are depleted to similar extents in the two bodies (Fig. 19). Although reliable data on the composition of the bulk silicate Moon and Vesta are scarce, $f_E$ appears to exhibit a log-linear decrease with decreasing $T_c$ (Fig. 19). Clues on the nature of volatile metal depletion on Vesta may be found by considering the extent of alkali metal depletion. As per the Earth, alkalis are depleted on Vesta in the sequence Na < K < Rb < Cs; with $f_E$ = 0.07, 0.07, 0.01, and 0.007, respectively. This therefore implies a strong control of equilibrium partial pressures of alkali species in determining the volatile abundances of Vesta, rather than any transport process. This conclusion is in line with the small size of Vesta (263 km radius), and associated escape velocity of 0.36 km/s. Heating of rocky material to minimum temperatures of 1000 K at $10^{-3}$ bar to liberate alkali metals (*e.g.,* Margrave, 1967) yields mean particle velocities, Equation (52), of between 0.96 km/s (Na) to 0.40 km/s (Cs), meaning all alkali thermal velocities exceed the contemporary escape velocity of Vesta. Lodders (1994) discussed thermal loss of alkalis during planetary accretion and modelled alkali loss (as MOH gas, M = Li to Cs) as a function of the ratio of escape/thermal velocities, showing loss by Jeans escape would occur in the sequence of Li > Na > K > Rb > Cs (most volatile but heaviest species), the opposite to that observed. Thus, earlier in its accretion history, heating of the Vestan mantle was likely associated with immediate loss of volatile metals proportional to their partial pressures.

The larger size of the Moon ($r_{Moon}$ = 1737 km; $v_{esc}$ = 2.38 km/s), and its different abundance pattern of volatile metals suggest other processes may have been at play. In detail, lunar volatile metal depletion is characterised by a 'step-wise' pattern, where three groups define plateaus in a $T_c$ vs. $f_E$ plot (Fig. 19). The weakly volatile elements, $T_c > 1000$ K, (Li, Mn) have $f_E \approx 0.4$, the moderately volatile, $700 < T_c$ (K) $< 1000$ with $f_E \approx 0.02$ and the highly volatile $T_c < 700$ K with $f_E \approx 0.0005$. Alkali metals on the Moon show a near-constant CI-, Mg-normalised depletion of 0.02, in disagreement with their partial vapour pressures. This feature is also observed for the so-called 'highly-volatile' elements at $f_E \approx 0.0005$, Ag, Bi, Br, Cd, In, Sb, Sn, Tl and Zn, again with little dependence on their anticipated volatilities (Wolf and Anders 1980; O'Neill 1991; Taylor et al. 2006). Why the Moon should exhibit such a distinctive step-wise depletion pattern has not yet been understood, below several mechanisms are explored.

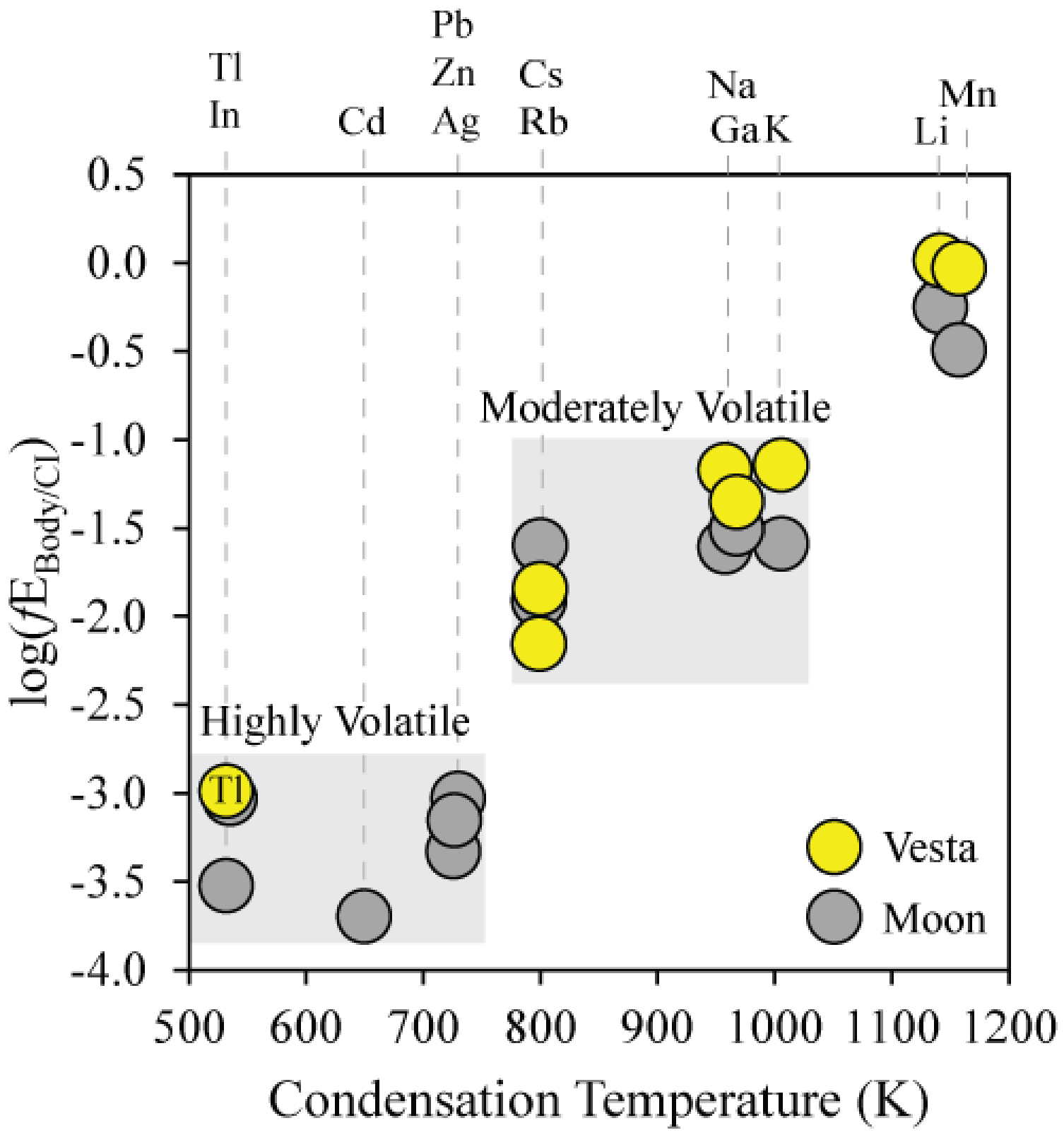

***Fig. 19.*** *CI-, Mg-normalised depletion factors ($f_E$) of volatile metals in the Moon (grey) and Vesta (yellow) as a function of condensation temperature. Trace element abundances from Taylor and Wieczorek (2014), Tl, Cd, Cs, Rb, K and O'Neill (1991), Li, Mn, Na, Ga, Zn, Pb, In for the Moon and Ruzicka et al. (2001) for Eucrites. To calculate mantle composition incompatible element abundances (all except Li, Mn, Zn) were multiplied by $La_{PM}/La_E$, where $La_E$ = 3.35 ppm (Eucrite), with $La_{PM}$ = 0.683 ppm. Condensation temperatures from Lodders (2003) save for Ag (Kiseeva and Wood, 2015).*

That Cs and Rb are not as impoverished in the Moon as might be expected from equilibrium partial pressures relative to other alkali metals has been recognised previously (Kreutzberger et al. 1986; O'Neill 1991; Jones and Palme 2000). O'Neill (1991) considered that this anomalous behaviour may be due to the addition of a small fraction (4%) of H-chondrites to the Moon, which are also relatively rich in alkali metals compared to other MVEs (*e.g.*, Wasson and Kallemeyn). The abundances of HVEs (Zn, Tl, Ag, etc.) are an order of magnitude lower than the alkalis and, in this model, also reflect the H-chondrite contribution. A similar 'chondritic veneer', though this time with C-chondrites, was invoked by Wolf and Anders (1980) to explain the near-chondritic relative proportions of these highly volatile elements.

A 'double-depletion' scenario is favoured by Taylor et al. (2006), in which the timing of the Moon's volatile depletion may have occurred in two stages. This hypothesis is grounded on a striking feature of lunar volatile depletion; it closely parallels that of the Earth (Ringwood and Kesson 1977; Taylor et al. 2006, their Fig. 1). Such depletion is taken as evidence for the inheritance of alkali metals from the Earth's mantle, or from the impactor, characterising the first stage of volatile loss. An early alkali depletion event is signalled by the similar composition of BABI[7] and the earliest lunar rocks (*e.g.*, Carlson and Lugmair 1988; Edmunson et al. 2009; Hans et al. 2013). Volatile depletion was then punctuated by a second event, presumably related to a giant impact, in which vapour loss occurred at a single threshold temperature of ≈1000 K, resulting in minimal accretion of elements present in the vapour (principally the highly volatile elements). A similar scenario had already been envisaged by O'Neill, 1991, exploiting the observation that, relative to the Earth's mantle, Na falls to 1/10$^{th}$ of its initial abundance in the Moon, whereas its neighbouring alkali metal, Li, is essentially undepleted. Satisfying this constraint using thermodynamic data, for which the reaction may be written:

$$LiO_{0.5_{(l)}} + Na_{(g)} = Li_{(g)} + NaO_{0.5_{(l)}} \quad (53)$$

yields temperatures of ≈1400 K, lest Li is lost to greater than 10% for 90% Na loss (see Appendix 2 of O'Neill, 1991). However, this assumes that solid $LiAlSi_2O_6$ (in clinopyroxene) and $NaAlSi_3O_8$ (in ablite), approximate their components in a silicate melt (Ghiorso and Sack, 1995). According to these studies, it is argued that volatile depletion on the Moon was a low-temperature phenomenon occurring at a certain threshold temperature.

Canup et al. (2015), through dynamical and chemical models, provide a mechanism for the Moon's formation that evolves naturally to a state resulting in a wholesale loss of volatiles by incomplete condensation at a certain temperature. In their model, the Moon accretes initially rapidly (<1 yr) from material outside the Roche limit[8] at 2.6 Earth radii. Over the subsequent $10^2$ yr, tidal shearing of hot (>2500 K) silicate material in the inner disk causes it to spread beyond the Roche limit and accrete to the nascent Moon. As such, in the model, the Roche limit represents a natural threshold that dictates the temperature at which volatile elements are lost, though as the study treats only Zn, K and Na, it is unknown as to whether this process results in step-wise lunar MVE depletion. In contrast to the earlier estimates cited above, Canup et al (2015) predict higher temperatures, ≈2500-3000 K, depending on the total atmospheric pressure (typically several bar). Their work used a P – T profile based on the vapour pressure of pure molten silica but also used the MAGMA code, which considers activity coefficients for Zn, Na, K, and major elements in the melt, to compute volatility of Zn, Na, and K from molten peridotite. Further work is needed to include Li, Rb, Cs, other MVE, halogens, water, and sulfur in the MAGMA code. Calculations with other codes give results in the same temperature range for the same total pressures (Petaev et al. 2016; Fegley and Lodders 2017). Furthermore, they show that these complex melts, though their individual components probably evaporate congruently, vaporise incongruently until an azeotropic composition is reached (Heyrman et al 2004; Petaev et al 2016; Fegley and Lodders 2017; Lock et al 2018). Whether these models hold for a non-canonical disk, in which the majority of the disk's mass already lies beyond the Roche limit (Nakajima and Stevenson 2014), or in a magnetised disk (Gammie et al. 2016) is yet to be assessed.

[7] Basalt Achondrite Best Initial
[8] The Roche limit describes the closest distance to a planetary body beyond which its gravitational pull is so weak so as to allow condensed material to coalesce into a second body under its own gravity.

It is clear that the use of a single parameter (*e.g.*, $M^{-0.5}$ or condensation temperature) fails to capture the entire complexity in describing the abundances of MVEs in planetary bodies. Instead, a combined approach is required that involves thermodynamic modelling to constrain temperatures, pressures and fugacities of major volatile species ($O_2$, $Cl_2$, $S_2$) during gas-vapour exchange, together with dynamical modelling that places constraints on the physical processes that moderate volatile escape.

**Conclusions**

Quantitative understanding of the conditions (pressure, temperature, composition) and processes (style of vaporisation/condensation, volatile escape mechanism, equilibrium *vs.* kinetic) that produced the chemical signatures observed in planetary materials necessitates the application of appropriate thermodynamic data. To this end, this chapter describes:

- At equilibrium, evaporation reactions of oxides and oxide components may either occur congruently, in which the M/O ratio in the gas is the same as that in the condensed phase, or incongruently, in which the $M/O_{condensed} \neq M/O_{gas}$. For congruent vaporisation, these reactions may also be associative (with identical speciation in both) or dissociative, where the condensed component decomposes into two or more species. These components in complex silicate melts likely evaporate congruently, though the bulk liquid does so incongruently until reaching an azeotrope.
- A treatment is outlined for equilibrium thermodynamic calculations for evaporation reactions using existing data on pure oxides. Despite the accessibility of thermodynamic data for simple systems, their application to complex, multicomponent systems, requires knowledge of the activity coefficient of the species in the appropriate phase.
- Techniques for measuring equilibrium vapour pressures and thermodynamic quantities are briefly described, and comprise Knudsen cell effusion, torsion effusion, and transpiration methods, in which the partial pressures of vapour species are measured or characterised by mass spectrometry, spectroscopic techniques, or by momentum and/or weight loss or –gain. Newer approaches include thermodynamic modelling and *ab-initio* calculations.

- Table 1 compiles the equilibrium speciation of metal-bearing gas species observed above $MO_x$(s, l) phases for ~ 80 oxides, and serves as an indication for their behaviour in more complex systems.
- Studies reporting thermodynamic data of metal oxides in binary-, ternary- and multicomponent silicate systems, with an emphasis on vapour-pressure measurements are summarised. For reference, a table showing approximate ranges of 28 $MO_x$ activity coefficients in multicomponent (>3) silicate melts is presented.
- Experimental data is reviewed for the evaporation of metal oxides in natural silicate systems, namely basaltic material and olivine, with the aim of quantifying the volatilities of the major silicate components, which are broadly K ≥ Na > Fe, Mn > Cr, Mg, SiO > Al, Ca, Ti. The oxygen fugacity ($fO_2$) of the vapour phase above natural silicate material is close to the Fayalite-Magnetite-Quartz buffer and its influence on element volatility is discussed.
- The condition for kinetic or 'Langmuir' evaporation is presented, and the applicability of the Hertz-Knudsen-Langmuir (HKL) equation to quantify evaporation processes in experiments is evaluated. Providing caveats such as temperature gradients between the gas and the condensed phase are considered, the HKL equation can provide accurate descriptions of vapour pressures above silicate melts at high temperatures, although equilibrium methods are preferred and further testing is required.
- Experimental techniques that deal with kinetic vaporisation are usually performed in a furnace under vacuum or with a continual flow of gases that buffer $fO_2$, and, in contrast to equilibrium techniques, volatilities are calculated from the fraction remaining in the condensed phase or by target collection.
- Evaporation of major elements from solids may be congruent (forsterite) or incongruent (olivine, plagioclase) and is reviewed in more detail in Nagahara (this volume).
- Alkali metals have very low activity coefficients in silicate melts that decreases their equilibrium partial pressures compared to those expected from pure alkali oxides. Their activity coefficients (Li > Na > K > Rb > Cs) are inversely proportional to their volatilities (Cs > Rb > K > Na > Li). Therefore, even though $K_2O$(s) is more volatile than $Na_2O$(s), Na and K are similarly volatile from silicate melts and $\gamma NaO_{0.5}$ and $\gamma KO_{0.5}$ are sensitive to composition.
- Evaporation of elements from a silicate melt at fixed pressure and temperature is controlled by the $fO_2$. This can be understood by the congruent evaporation of metal oxides and their speciation in the gas- and condensed phase. Lower $fO_2$ increases the vapour pressure of the metal-bearing gas species if it is more reduced than in the condensed phase, and vice-versa.

Sossi et al. (2016) show how the speciation of the element in the gas phase can be calculated by analysis of the condensed phase as a function of $fO_2$ in Langmuir evaporation experiments. These authors also describe a method based on the HKL equation that allows for quantitative extraction of free energies of vaporisation and activity coefficients of metal oxides from silicate melts.

- The order of volatility during vaporisation of silicate melts is discussed and contrasted with volatilities expected in the solar nebula. Reasons for differences between the two regimes include *i)* higher total pressures in evaporation experiments such that equilibrium is between liquid-gas rather than solid-gas in the solar nebula; *ii)* differences in phases *e.g.,* silicate melt *vs.* minerals (feldspar, olivine, enstatite and Fe-Ni alloy) in the solar nebula and *iii)* the difference in the fugacity of volatile species ($fO_2$, $fS_2$) in the gas that may stabilise different gas species (*e.g.,* $CrO_2$, $MoO_3$ at high $fO_2$ and SnS and GeS at high $fS_2$).
- The presence of other volatile species, particularly $S_2$, $Cl_2$ and $H_2$ in steam atmospheres above planetary bodies has the potential to complex metal-bearing gases. The effect of pressure is to increase the stability of the chloride species relative to the monatomic gas, where each ratio of Zn/$ZnCl_2$, Cu/CuCl, and Mn/$MnCl_x$ becomes unity at 890, 1.25 and 0.15 bar, respectively.
- The controls on the dominant speciation of 33 elements are contrasted in four different environments, between *i)* the solar nebula *ii)* an ordinary, H-chondrite composition, *iii)* terrestrial volcanic gases and *iv)* evaporation of an anhydrous ferrobasalt.
- Volatile element depletion in planetary materials is quantified by normalisation to a refractory lithophile element (typically Mg) and then to the same ratio in CI chondrites, whose abundances for all but the most volatile elements approximate those of the Sun. Although planetary volatile depletion broadly correlates with nebular half condensation temperatures, this volatility scale is suitable only for understanding condensation as it applies to a gas of solar composition under canonical nebular conditions.
- By contrast, planetary bodies likely experienced volatile depletion at conditions distinct from those extant in the solar nebula. This is expected from results of equilibrium calculations of atmospheric compositions above planetary mantles that are dominated by Na below 3000 K and SiO above (compared to $H_2$ in the solar nebula), and edified by the superchondritic Mn/Na ratios of small planetary bodies that reflect more oxidising conditions, with the notable exception of Earth.
- The Bulk Silicate Earth's abundances of Cd, In and Zn likely reflect their volatility rather than core formation, and suggest that the volatility of In is lower than supposed given its nebular half condensation temperature. A negative correlation between element abundance of moderately volatile elements and $M^{0.5}$ for the Earth show that a transport process (such as Jeans or Hydrodynamic escape) cannot have been responsible for their depletion, likely due to its high escape velocity.

- The Moon and Vesta are more strongly depleted in moderately volatile elements than the Earth, and is likely a global feature of both bodies. The loss of alkalis on Vesta in concert with their equilibrium partial pressures (Cs > Rb > K > Na > Li) points to a quantitative loss of the vapour due to the inability of Vesta to retain an atmosphere due to its small size.
- A step-wise depletion pattern is observed for the Moon, in which weakly volatile, moderately volatile, and highly volatile elements define plateaus of Mg, CI-normalised abundances of 0.4, 0.02 and 0.0005, respectively. Possible explanations for this pattern include inheritance from the Earth and an additional depletion event at 1100 K ('double depletion'); depletion at low temperatures (≈ 1400 K) and addition of a small amount (4%) of H-chondrite; or a high-temperature, high-pressure depletion event at 2500 – 3000 K.

**Acknowledgments**

P.A.S. is grateful to the European Research Council under the Horizon 2020 framework grant #637503 (Pristine), and thanks Hugh O'Neill, Stephan Klemme, Nate Jacobson, Julien Siebert and Frédéric Moynier for discussions. B.F. acknowledges support by NASA Grant NNX17AC02G (XRP Program), NSF Astronomy Program Grants AST-1412175 and AST-1517541, thanks Dick Henley, Nate Jacobson, and Katharina Lodders for useful discussions, and thanks the staff of the Interlibrary Loan Service (ILL) of the Washington University Libraries for their superb support of his research work. We are grateful to Trevor Ireland, Nate Jacobson, Dick Henley, Penny King and an anonymous reviewer for far-reaching and comprehensive reviews on numerous aspects of this chapter that ultimately led to a much-improved contribution.